\documentclass[traditabstract, longauth]{aa} 

\usepackage{graphicx}
\usepackage{txfonts}
\usepackage[breaklinks, colorlinks, citecolor=blue]{hyperref}
\usepackage{natbib}

\bibpunct{(}{)}{;}{a}{}{,} 

\newcommand{\vn}{\vec{\hat{n\,}}}

\newcommand{\vx}{\mbox{\vec{x}}}

\newcommand{\vk}{\mbox{\vec{k}}}

\newcommand{\vrv}{\mbox{\vec{r}}}

\newcommand{\vv}{\mbox{\vec{v}}}

\newcommand{\gsim}{\; ^{>}_{\sim}\;}

\usepackage{multirow}
         
\def\plotancho#1{\includegraphics[width=18cm]{#1}}

\def\setsymbol#1#2{\expandafter\def\csname #1\endcsname{#2}}
\def\getsymbol#1{\csname #1\endcsname}

\def\Planck{{\it Planck\/}}

\newbox\tablebox    \newdimen\tablewidth
\def\leaderfil{\leaders\hbox to 5pt{\hss.\hss}\hfil}
\def\endPlancktable{\tablewidth=\columnwidth 
    $$\hss\copy\tablebox\hss$$
    \vskip-\lastskip\vskip -2pt}

\def\tablenote#1 #2\par{\begingroup \parindent=0.8em
    \abovedisplayshortskip=0pt\belowdisplayshortskip=0pt
    \noindent
    $$\hss\vbox{\hsize\tablewidth \hangindent=\parindent \hangafter=1 \noindent
    \hbox to \parindent{$^#1$\hss}\strut#2\strut\par}\hss$$
    \endgroup}
\def\doubleline{\vskip 3pt\hrule \vskip 1.5pt \hrule \vskip 5pt}

\def\L2{\ifmmode L_2\else $L_2$\fi}

\def\DeltaT{\ifmmode \Delta T\else $\Delta T$\fi}
\def\deltat{\ifmmode \Delta t\else $\Delta t$\fi}
\def\fknee{\ifmmode f_{\rm knee}\else $f_{\rm knee}$\fi}
\def\Fmax{\ifmmode F_{\rm max}\else $F_{\rm max}$\fi}
\def\solar{\ifmmode{\rm M}_{\mathord\odot}\else${\rm M}_{\mathord\odot}$\fi}
\def\Msolar{\ifmmode{\rm M}_{\mathord\odot}\else${\rm M}_{\mathord\odot}$\fi}
\def\Lsolar{\ifmmode{\rm L}_{\mathord\odot}\else${\rm L}_{\mathord\odot}$\fi}

\def\inv{\ifmmode^{-1}\else$^{-1}$\fi}
\def\mo{\ifmmode^{-1}\else$^{-1}$\fi}
\def\sup#1{\ifmmode ^{\rm #1}\else $^{\rm #1}$\fi}
\def\expo#1{\ifmmode \times 10^{#1}\else $\times 10^{#1}$\fi}
\def\,{\thinspace}
\def\lsim{\mathrel{\raise .4ex\hbox{\rlap{$<$}\lower 1.2ex\hbox{$\sim$}}}}
\def\gsim{\mathrel{\raise .4ex\hbox{\rlap{$>$}\lower 1.2ex\hbox{$\sim$}}}}

\def\simprop{\mathrel{\raise .4ex\hbox{\rlap{$\propto$}\lower 1.2ex\hbox{$\sim$}}}}
\def\deg{\ifmmode^\circ\else$^\circ$\fi}
\def\pdeg{\ifmmode $\setbox0=\hbox{$^{\circ}$}\rlap{\hskip.11\wd0 .}$^{\circ}
          \else \setbox0=\hbox{$^{\circ}$}\rlap{\hskip.11\wd0 .}$^{\circ}$\fi}
\def\arcs{\ifmmode {^{\scriptstyle\prime\prime}}
          \else $^{\scriptstyle\prime\prime}$\fi}
\def\arcm{\ifmmode {^{\scriptstyle\prime}}
          \else $^{\scriptstyle\prime}$\fi}
\newdimen\sa  \newdimen\sb
\def\parcs{\sa=.07em \sb=.03em
     \ifmmode \hbox{\rlap{.}}^{\scriptstyle\prime\kern -\sb\prime}\hbox{\kern -\sa}
     \else \rlap{.}$^{\scriptstyle\prime\kern -\sb\prime}$\kern -\sa\fi}
\def\parcm{\sa=.08em \sb=.03em
     \ifmmode \hbox{\rlap{.}\kern\sa}^{\scriptstyle\prime}\hbox{\kern-\sb}
     \else \rlap{.}\kern\sa$^{\scriptstyle\prime}$\kern-\sb\fi}
\def\ra[#1 #2 #3.#4]{#1\sup{h}#2\sup{m}#3\sup{s}\llap.#4}
\def\dec[#1 #2 #3.#4]{#1\deg#2\arcm#3\arcs\llap.#4}
\def\deco[#1 #2 #3]{#1\deg#2\arcm#3\arcs}
\def\rra[#1 #2]{#1\sup{h}#2\sup{m}}

\def\dots{\relax\ifmmode \ldots\else $\ldots$\fi}
\def\WHzsr{\ifmmode $W\,Hz\mo\,sr\mo$\else W\,Hz\mo\,sr\mo\fi}
\def\mHz{\ifmmode $\,mHz$\else \,mHz\fi}
\def\GHz{\ifmmode $\,GHz$\else \,GHz\fi}
\def\mKs{\ifmmode $\,mK\,s$^{1/2}\else \,mK\,s$^{1/2}$\fi}
\def\muKs{\ifmmode \,\mu$K\,s$^{1/2}\else \,$\mu$K\,s$^{1/2}$\fi}
\def\muKRJs{\ifmmode \,\mu$K$_{\rm RJ}$\,s$^{1/2}\else \,$\mu$K$_{\rm RJ}$\,s$^{1/2}$\fi}
\def\muKHz{\ifmmode \,\mu$K\,Hz$^{-1/2}\else \,$\mu$K\,Hz$^{-1/2}$\fi}
\def\MJysr{\ifmmode \,$MJy\,sr\mo$\else \,MJy\,sr\mo\fi}
\def\MJysrmK{\ifmmode \,$MJy\,sr\mo$\,mK$_{\rm CMB}\mo\else \,MJy\,sr\mo\,mK$_{\rm CMB}\mo$\fi}
\def\microns{\ifmmode \,\mu$m$\else \,$\mu$m\fi}

\def\muK{\ifmmode \,\mu$K$\else \,$\mu$\hbox{K}\fi}
\def\microK{\ifmmode \,\mu$K$\else \,$\mu$\hbox{K}\fi}
\def\muW{\ifmmode \,\mu$W$\else \,$\mu$\hbox{W}\fi}
\def\kms{\ifmmode $\,km\,s$^{-1}\else \,km\,s$^{-1}$\fi}
\def\kmsMpc{\ifmmode $\,\kms\,Mpc\mo$\else \,\kms\,Mpc\mo\fi}
\setsymbol{LFI:center:frequency:70GHz:units}{70.3\,GHz}
\setsymbol{LFI:center:frequency:44GHz:units}{44.1\,GHz}
\setsymbol{LFI:center:frequency:30GHz:units}{28.5\,GHz}

\setsymbol{LFI:center:frequency:70GHz}{70.3}
\setsymbol{LFI:center:frequency:44GHz}{44.1}
\setsymbol{LFI:center:frequency:30GHz}{28.5}

\setsymbol{LFI:center:frequency:LFI18:Rad:M:units}{71.7\GHz}
\setsymbol{LFI:center:frequency:LFI19:Rad:M:units}{67.5\GHz}
\setsymbol{LFI:center:frequency:LFI20:Rad:M:units}{69.2\GHz}
\setsymbol{LFI:center:frequency:LFI21:Rad:M:units}{70.4\GHz}
\setsymbol{LFI:center:frequency:LFI22:Rad:M:units}{71.5\GHz}
\setsymbol{LFI:center:frequency:LFI23:Rad:M:units}{70.8\GHz}
\setsymbol{LFI:center:frequency:LFI24:Rad:M:units}{44.4\GHz}
\setsymbol{LFI:center:frequency:LFI25:Rad:M:units}{44.0\GHz}
\setsymbol{LFI:center:frequency:LFI26:Rad:M:units}{43.9\GHz}
\setsymbol{LFI:center:frequency:LFI27:Rad:M:units}{28.3\GHz}
\setsymbol{LFI:center:frequency:LFI28:Rad:M:units}{28.8\GHz}
\setsymbol{LFI:center:frequency:LFI18:Rad:S:units}{70.1\GHz}
\setsymbol{LFI:center:frequency:LFI19:Rad:S:units}{69.6\GHz}
\setsymbol{LFI:center:frequency:LFI20:Rad:S:units}{69.5\GHz}
\setsymbol{LFI:center:frequency:LFI21:Rad:S:units}{69.5\GHz}
\setsymbol{LFI:center:frequency:LFI22:Rad:S:units}{72.8\GHz}
\setsymbol{LFI:center:frequency:LFI23:Rad:S:units}{71.3\GHz}
\setsymbol{LFI:center:frequency:LFI24:Rad:S:units}{44.1\GHz}
\setsymbol{LFI:center:frequency:LFI25:Rad:S:units}{44.1\GHz}
\setsymbol{LFI:center:frequency:LFI26:Rad:S:units}{44.1\GHz}
\setsymbol{LFI:center:frequency:LFI27:Rad:S:units}{28.5\GHz}
\setsymbol{LFI:center:frequency:LFI28:Rad:S:units}{28.2\GHz}

\setsymbol{LFI:center:frequency:LFI18:Rad:M}{71.7}
\setsymbol{LFI:center:frequency:LFI19:Rad:M}{67.5}
\setsymbol{LFI:center:frequency:LFI20:Rad:M}{69.2}
\setsymbol{LFI:center:frequency:LFI21:Rad:M}{70.4}
\setsymbol{LFI:center:frequency:LFI22:Rad:M}{71.5}
\setsymbol{LFI:center:frequency:LFI23:Rad:M}{70.8}
\setsymbol{LFI:center:frequency:LFI24:Rad:M}{44.4}
\setsymbol{LFI:center:frequency:LFI25:Rad:M}{44.0}
\setsymbol{LFI:center:frequency:LFI26:Rad:M}{43.9}
\setsymbol{LFI:center:frequency:LFI27:Rad:M}{28.3}
\setsymbol{LFI:center:frequency:LFI28:Rad:M}{28.8}
\setsymbol{LFI:center:frequency:LFI18:Rad:S}{70.1}
\setsymbol{LFI:center:frequency:LFI19:Rad:S}{69.6}
\setsymbol{LFI:center:frequency:LFI20:Rad:S}{69.5}
\setsymbol{LFI:center:frequency:LFI21:Rad:S}{69.5}
\setsymbol{LFI:center:frequency:LFI22:Rad:S}{72.8}
\setsymbol{LFI:center:frequency:LFI23:Rad:S}{71.3}
\setsymbol{LFI:center:frequency:LFI24:Rad:S}{44.1}
\setsymbol{LFI:center:frequency:LFI25:Rad:S}{44.1}
\setsymbol{LFI:center:frequency:LFI26:Rad:S}{44.1}
\setsymbol{LFI:center:frequency:LFI27:Rad:S}{28.5}
\setsymbol{LFI:center:frequency:LFI28:Rad:S}{28.2}

\setsymbol{LFI:white:noise:sensitivity:70GHz:units}{134.7\muKs}
\setsymbol{LFI:white:noise:sensitivity:44GHz:units}{164.7\muKs}
\setsymbol{LFI:white:noise:sensitivity:30GHz:units}{143.4\muKs}

\setsymbol{LFI:white:noise:sensitivity:70GHz}{134.7}
\setsymbol{LFI:white:noise:sensitivity:44GHz}{164.7}
\setsymbol{LFI:white:noise:sensitivity:30GHz}{143.4}

\setsymbol{LFI:white:noise:sensitivity:LFI18:Rad:M:units}{512.0\muKs}
\setsymbol{LFI:white:noise:sensitivity:LFI19:Rad:M:units}{581.4\muKs}
\setsymbol{LFI:white:noise:sensitivity:LFI20:Rad:M:units}{590.8\muKs}
\setsymbol{LFI:white:noise:sensitivity:LFI21:Rad:M:units}{455.2\muKs}
\setsymbol{LFI:white:noise:sensitivity:LFI22:Rad:M:units}{492.0\muKs}
\setsymbol{LFI:white:noise:sensitivity:LFI23:Rad:M:units}{507.7\muKs}
\setsymbol{LFI:white:noise:sensitivity:LFI24:Rad:M:units}{462.2\muKs}
\setsymbol{LFI:white:noise:sensitivity:LFI25:Rad:M:units}{413.6\muKs}
\setsymbol{LFI:white:noise:sensitivity:LFI26:Rad:M:units}{478.6\muKs}
\setsymbol{LFI:white:noise:sensitivity:LFI27:Rad:M:units}{277.7\muKs}
\setsymbol{LFI:white:noise:sensitivity:LFI28:Rad:M:units}{312.3\muKs}
\setsymbol{LFI:white:noise:sensitivity:LFI18:Rad:S:units}{465.7\muKs}
\setsymbol{LFI:white:noise:sensitivity:LFI19:Rad:S:units}{555.6\muKs}
\setsymbol{LFI:white:noise:sensitivity:LFI20:Rad:S:units}{623.2\muKs}
\setsymbol{LFI:white:noise:sensitivity:LFI21:Rad:S:units}{564.1\muKs}
\setsymbol{LFI:white:noise:sensitivity:LFI22:Rad:S:units}{534.4\muKs}
\setsymbol{LFI:white:noise:sensitivity:LFI23:Rad:S:units}{542.4\muKs}
\setsymbol{LFI:white:noise:sensitivity:LFI24:Rad:S:units}{399.2\muKs}
\setsymbol{LFI:white:noise:sensitivity:LFI25:Rad:S:units}{392.6\muKs}
\setsymbol{LFI:white:noise:sensitivity:LFI26:Rad:S:units}{418.6\muKs}
\setsymbol{LFI:white:noise:sensitivity:LFI27:Rad:S:units}{302.9\muKs}
\setsymbol{LFI:white:noise:sensitivity:LFI28:Rad:S:units}{285.3\muKs}

\setsymbol{LFI:white:noise:sensitivity:LFI18:Rad:M}{512.0}
\setsymbol{LFI:white:noise:sensitivity:LFI19:Rad:M}{581.4}
\setsymbol{LFI:white:noise:sensitivity:LFI20:Rad:M}{590.8}
\setsymbol{LFI:white:noise:sensitivity:LFI21:Rad:M}{455.2}
\setsymbol{LFI:white:noise:sensitivity:LFI22:Rad:M}{492.0}
\setsymbol{LFI:white:noise:sensitivity:LFI23:Rad:M}{507.7}
\setsymbol{LFI:white:noise:sensitivity:LFI24:Rad:M}{462.2}
\setsymbol{LFI:white:noise:sensitivity:LFI25:Rad:M}{413.6}
\setsymbol{LFI:white:noise:sensitivity:LFI26:Rad:M}{478.6}
\setsymbol{LFI:white:noise:sensitivity:LFI27:Rad:M}{277.7}
\setsymbol{LFI:white:noise:sensitivity:LFI28:Rad:M}{312.3}
\setsymbol{LFI:white:noise:sensitivity:LFI18:Rad:S}{465.7}
\setsymbol{LFI:white:noise:sensitivity:LFI19:Rad:S}{555.6}
\setsymbol{LFI:white:noise:sensitivity:LFI20:Rad:S}{623.2}
\setsymbol{LFI:white:noise:sensitivity:LFI21:Rad:S}{564.1}
\setsymbol{LFI:white:noise:sensitivity:LFI22:Rad:S}{534.4}
\setsymbol{LFI:white:noise:sensitivity:LFI23:Rad:S}{542.4}
\setsymbol{LFI:white:noise:sensitivity:LFI24:Rad:S}{399.2}
\setsymbol{LFI:white:noise:sensitivity:LFI25:Rad:S}{392.6}
\setsymbol{LFI:white:noise:sensitivity:LFI26:Rad:S}{418.6}
\setsymbol{LFI:white:noise:sensitivity:LFI27:Rad:S}{302.9}
\setsymbol{LFI:white:noise:sensitivity:LFI28:Rad:S}{285.3}

\setsymbol{LFI:knee:frequency:70GHz:units}{29.5\mHz}
\setsymbol{LFI:knee:frequency:44GHz:units}{56.2\mHz}
\setsymbol{LFI:knee:frequency:30GHz:units}{113.7\mHz}

\setsymbol{LFI:knee:frequency:70GHz}{29.5}
\setsymbol{LFI:knee:frequency:44GHz}{56.2}
\setsymbol{LFI:knee:frequency:30GHz}{113.7}

\setsymbol{LFI:knee:frequency:LFI18:Rad:M:units}{16.3\mHz}
\setsymbol{LFI:knee:frequency:LFI19:Rad:M:units}{15.1\mHz}
\setsymbol{LFI:knee:frequency:LFI20:Rad:M:units}{18.7\mHz}
\setsymbol{LFI:knee:frequency:LFI21:Rad:M:units}{37.2\mHz}
\setsymbol{LFI:knee:frequency:LFI22:Rad:M:units}{12.7\mHz}
\setsymbol{LFI:knee:frequency:LFI23:Rad:M:units}{34.6\mHz}
\setsymbol{LFI:knee:frequency:LFI24:Rad:M:units}{46.2\mHz}
\setsymbol{LFI:knee:frequency:LFI25:Rad:M:units}{24.9\mHz}
\setsymbol{LFI:knee:frequency:LFI26:Rad:M:units}{67.6\mHz}
\setsymbol{LFI:knee:frequency:LFI27:Rad:M:units}{187.4\mHz}
\setsymbol{LFI:knee:frequency:LFI28:Rad:M:units}{122.2\mHz}
\setsymbol{LFI:knee:frequency:LFI18:Rad:S:units}{17.7\mHz}
\setsymbol{LFI:knee:frequency:LFI19:Rad:S:units}{22.0\mHz}
\setsymbol{LFI:knee:frequency:LFI20:Rad:S:units}{8.7\mHz}
\setsymbol{LFI:knee:frequency:LFI21:Rad:S:units}{25.9\mHz}
\setsymbol{LFI:knee:frequency:LFI22:Rad:S:units}{15.8\mHz}
\setsymbol{LFI:knee:frequency:LFI23:Rad:S:units}{129.8\mHz}
\setsymbol{LFI:knee:frequency:LFI24:Rad:S:units}{100.9\mHz}
\setsymbol{LFI:knee:frequency:LFI25:Rad:S:units}{38.9\mHz}
\setsymbol{LFI:knee:frequency:LFI26:Rad:S:units}{58.9\mHz}
\setsymbol{LFI:knee:frequency:LFI27:Rad:S:units}{104.4\mHz}
\setsymbol{LFI:knee:frequency:LFI28:Rad:S:units}{40.7\mHz}

\setsymbol{LFI:knee:frequency:LFI18:Rad:M}{16.3}
\setsymbol{LFI:knee:frequency:LFI19:Rad:M}{15.1}
\setsymbol{LFI:knee:frequency:LFI20:Rad:M}{18.7}
\setsymbol{LFI:knee:frequency:LFI21:Rad:M}{37.2}
\setsymbol{LFI:knee:frequency:LFI22:Rad:M}{12.7}
\setsymbol{LFI:knee:frequency:LFI23:Rad:M}{34.6}
\setsymbol{LFI:knee:frequency:LFI24:Rad:M}{46.2}
\setsymbol{LFI:knee:frequency:LFI25:Rad:M}{24.9}
\setsymbol{LFI:knee:frequency:LFI26:Rad:M}{67.6}
\setsymbol{LFI:knee:frequency:LFI27:Rad:M}{187.4}
\setsymbol{LFI:knee:frequency:LFI28:Rad:M}{122.2}
\setsymbol{LFI:knee:frequency:LFI18:Rad:S}{17.7}
\setsymbol{LFI:knee:frequency:LFI19:Rad:S}{22.0}
\setsymbol{LFI:knee:frequency:LFI20:Rad:S}{8.7}
\setsymbol{LFI:knee:frequency:LFI21:Rad:S}{25.9}
\setsymbol{LFI:knee:frequency:LFI22:Rad:S}{15.8}
\setsymbol{LFI:knee:frequency:LFI23:Rad:S}{129.8}
\setsymbol{LFI:knee:frequency:LFI24:Rad:S}{100.9}
\setsymbol{LFI:knee:frequency:LFI25:Rad:S}{38.9}
\setsymbol{LFI:knee:frequency:LFI26:Rad:S}{58.9}
\setsymbol{LFI:knee:frequency:LFI27:Rad:S}{104.4}
\setsymbol{LFI:knee:frequency:LFI28:Rad:S}{40.7}

\setsymbol{LFI:slope:70GHz:units}{$-1.03$\mHz}
\setsymbol{LFI:slope:44GHz:units}{$-0.89$\mHz}
\setsymbol{LFI:slope:30GHz:units}{$-0.87$\mHz}

\setsymbol{LFI:slope:70GHz}{$-1.03$}
\setsymbol{LFI:slope:44GHz}{$-0.89$}
\setsymbol{LFI:slope:30GHz}{$-0.87$}

\setsymbol{LFI:slope:LFI18:Rad:M:units}{$-1.04$\mHz}
\setsymbol{LFI:slope:LFI19:Rad:M:units}{$-1.09$\mHz}
\setsymbol{LFI:slope:LFI20:Rad:M:units}{$-0.69$\mHz}
\setsymbol{LFI:slope:LFI21:Rad:M:units}{$-1.56$\mHz}
\setsymbol{LFI:slope:LFI22:Rad:M:units}{$-1.01$\mHz}
\setsymbol{LFI:slope:LFI23:Rad:M:units}{$-0.96$\mHz}
\setsymbol{LFI:slope:LFI24:Rad:M:units}{$-0.83$\mHz}
\setsymbol{LFI:slope:LFI25:Rad:M:units}{$-0.91$\mHz}
\setsymbol{LFI:slope:LFI26:Rad:M:units}{$-0.95$\mHz}
\setsymbol{LFI:slope:LFI27:Rad:M:units}{$-0.87$\mHz}
\setsymbol{LFI:slope:LFI28:Rad:M:units}{$-0.88$\mHz}
\setsymbol{LFI:slope:LFI18:Rad:S:units}{$-1.15$\mHz}
\setsymbol{LFI:slope:LFI19:Rad:S:units}{$-1.00$\mHz}
\setsymbol{LFI:slope:LFI20:Rad:S:units}{$-0.95$\mHz}
\setsymbol{LFI:slope:LFI21:Rad:S:units}{$-0.92$\mHz}
\setsymbol{LFI:slope:LFI22:Rad:S:units}{$-1.01$\mHz}
\setsymbol{LFI:slope:LFI23:Rad:S:units}{$-0.95$\mHz}
\setsymbol{LFI:slope:LFI24:Rad:S:units}{$-0.73$\mHz}
\setsymbol{LFI:slope:LFI25:Rad:S:units}{$-1.16$\mHz}
\setsymbol{LFI:slope:LFI26:Rad:S:units}{$-0.79$\mHz}
\setsymbol{LFI:slope:LFI27:Rad:S:units}{$-0.82$\mHz}
\setsymbol{LFI:slope:LFI28:Rad:S:units}{$-0.91$\mHz}

\setsymbol{LFI:slope:LFI18:Rad:M}{$-1.04$}
\setsymbol{LFI:slope:LFI19:Rad:M}{$-1.09$}
\setsymbol{LFI:slope:LFI20:Rad:M}{$-0.69$}
\setsymbol{LFI:slope:LFI21:Rad:M}{$-1.56$}
\setsymbol{LFI:slope:LFI22:Rad:M}{$-1.01$}
\setsymbol{LFI:slope:LFI23:Rad:M}{$-0.96$}
\setsymbol{LFI:slope:LFI24:Rad:M}{$-0.83$}
\setsymbol{LFI:slope:LFI25:Rad:M}{$-0.91$}
\setsymbol{LFI:slope:LFI26:Rad:M}{$-0.95$}
\setsymbol{LFI:slope:LFI27:Rad:M}{$-0.87$}
\setsymbol{LFI:slope:LFI28:Rad:M}{$-0.88$}
\setsymbol{LFI:slope:LFI18:Rad:S}{$-1.15$}
\setsymbol{LFI:slope:LFI19:Rad:S}{$-1.00$}
\setsymbol{LFI:slope:LFI20:Rad:S}{$-0.95$}
\setsymbol{LFI:slope:LFI21:Rad:S}{$-0.92$}
\setsymbol{LFI:slope:LFI22:Rad:S}{$-1.01$}
\setsymbol{LFI:slope:LFI23:Rad:S}{$-0.95$}
\setsymbol{LFI:slope:LFI24:Rad:S}{$-0.73$}
\setsymbol{LFI:slope:LFI25:Rad:S}{$-1.16$}
\setsymbol{LFI:slope:LFI26:Rad:S}{$-0.79$}
\setsymbol{LFI:slope:LFI27:Rad:S}{$-0.82$}
\setsymbol{LFI:slope:LFI28:Rad:S}{$-0.91$}

\setsymbol{LFI:FWHM:70GHz:units}{13\parcm01}
\setsymbol{LFI:FWHM:44GHz:units}{27\parcm92}
\setsymbol{LFI:FWHM:30GHz:units}{32\parcm65}

\setsymbol{LFI:FWHM:70GHz}{13.01}
\setsymbol{LFI:FWHM:44GHz}{27.92}
\setsymbol{LFI:FWHM:30GHz}{32.65}

\setsymbol{LFI:FWHM:LFI18:units}{13\parcm39}
\setsymbol{LFI:FWHM:LFI19:units}{13\parcm01}
\setsymbol{LFI:FWHM:LFI20:units}{12\parcm75}
\setsymbol{LFI:FWHM:LFI21:units}{12\parcm74}
\setsymbol{LFI:FWHM:LFI22:units}{12\parcm87}
\setsymbol{LFI:FWHM:LFI23:units}{13\parcm27}
\setsymbol{LFI:FWHM:LFI24:units}{22\parcm98}
\setsymbol{LFI:FWHM:LFI25:units}{30\parcm46}
\setsymbol{LFI:FWHM:LFI26:units}{30\parcm31}
\setsymbol{LFI:FWHM:LFI27:units}{32\parcm65}
\setsymbol{LFI:FWHM:LFI28:units}{32\parcm66}

\setsymbol{LFI:FWHM:LFI18}{13.39}
\setsymbol{LFI:FWHM:LFI19}{13.01}
\setsymbol{LFI:FWHM:LFI20}{12.75}
\setsymbol{LFI:FWHM:LFI21}{12.74}
\setsymbol{LFI:FWHM:LFI22}{12.87}
\setsymbol{LFI:FWHM:LFI23}{13.27}
\setsymbol{LFI:FWHM:LFI24}{22.98}
\setsymbol{LFI:FWHM:LFI25}{30.46}
\setsymbol{LFI:FWHM:LFI26}{30.31}
\setsymbol{LFI:FWHM:LFI27}{32.65}
\setsymbol{LFI:FWHM:LFI28}{32.66}

\setsymbol{LFI:FWHM:uncertainty:LFI18:units}{0.170\arcm}
\setsymbol{LFI:FWHM:uncertainty:LFI19:units}{0.174\arcm}
\setsymbol{LFI:FWHM:uncertainty:LFI20:units}{0.170\arcm}
\setsymbol{LFI:FWHM:uncertainty:LFI21:units}{0.156\arcm}
\setsymbol{LFI:FWHM:uncertainty:LFI22:units}{0.164\arcm}
\setsymbol{LFI:FWHM:uncertainty:LFI23:units}{0.171\arcm}
\setsymbol{LFI:FWHM:uncertainty:LFI24:units}{0.652\arcm}
\setsymbol{LFI:FWHM:uncertainty:LFI25:units}{1.075\arcm}
\setsymbol{LFI:FWHM:uncertainty:LFI26:units}{1.131\arcm}
\setsymbol{LFI:FWHM:uncertainty:LFI27:units}{1.266\arcm}
\setsymbol{LFI:FWHM:uncertainty:LFI28:units}{1.287\arcm}

\setsymbol{LFI:FWHM:uncertainty:LFI18}{0.170}
\setsymbol{LFI:FWHM:uncertainty:LFI19}{0.174}
\setsymbol{LFI:FWHM:uncertainty:LFI20}{0.170}
\setsymbol{LFI:FWHM:uncertainty:LFI21}{0.156}
\setsymbol{LFI:FWHM:uncertainty:LFI22}{0.164}
\setsymbol{LFI:FWHM:uncertainty:LFI23}{0.171}
\setsymbol{LFI:FWHM:uncertainty:LFI24}{0.652}
\setsymbol{LFI:FWHM:uncertainty:LFI25}{1.075}
\setsymbol{LFI:FWHM:uncertainty:LFI26}{1.131}
\setsymbol{LFI:FWHM:uncertainty:LFI27}{1.266}
\setsymbol{LFI:FWHM:uncertainty:LFI28}{1.287}

\setsymbol{HFI:center:frequency:100GHz:units}{100\,GHz}
\setsymbol{HFI:center:frequency:143GHz:units}{143\,GHz}
\setsymbol{HFI:center:frequency:217GHz:units}{217\,GHz}
\setsymbol{HFI:center:frequency:353GHz:units}{353\,GHz}
\setsymbol{HFI:center:frequency:545GHz:units}{545\,GHz}
\setsymbol{HFI:center:frequency:857GHz:units}{857\,GHz}

\setsymbol{HFI:center:frequency:100GHz}{100}
\setsymbol{HFI:center:frequency:143GHz}{143}
\setsymbol{HFI:center:frequency:217GHz}{217}
\setsymbol{HFI:center:frequency:353GHz}{353}
\setsymbol{HFI:center:frequency:545GHz}{545}
\setsymbol{HFI:center:frequency:857GHz}{857}

\setsymbol{HFI:Ndetectors:100GHz}{8}
\setsymbol{HFI:Ndetectors:143GHz}{11}
\setsymbol{HFI:Ndetectors:217GHz}{12}
\setsymbol{HFI:Ndetectors:353GHz}{12}
\setsymbol{HFI:Ndetectors:545GHz}{3}
\setsymbol{HFI:Ndetectors:857GHz}{4}

\setsymbol{HFI:FWHM:Maps:100GHz:units}{9\parcm88}
\setsymbol{HFI:FWHM:Maps:143GHz:units}{7\parcm18}
\setsymbol{HFI:FWHM:Maps:217GHz:units}{4\parcm87}
\setsymbol{HFI:FWHM:Maps:353GHz:units}{4\parcm65}
\setsymbol{HFI:FWHM:Maps:545GHz:units}{4\parcm72}
\setsymbol{HFI:FWHM:Maps:857GHz:units}{4\parcm39}
\setsymbol{HFI:FWHM:Maps:100GHz}{9.88}
\setsymbol{HFI:FWHM:Maps:143GHz}{7.18}
\setsymbol{HFI:FWHM:Maps:217GHz}{4.87}
\setsymbol{HFI:FWHM:Maps:353GHz}{4.65}
\setsymbol{HFI:FWHM:Maps:545GHz}{4.72}
\setsymbol{HFI:FWHM:Maps:857GHz}{4.39}

\setsymbol{HFI:beam:ellipticity:Maps:100GHz}{1.15}
\setsymbol{HFI:beam:ellipticity:Maps:143GHz}{1.01}
\setsymbol{HFI:beam:ellipticity:Maps:217GHz}{1.06}
\setsymbol{HFI:beam:ellipticity:Maps:353GHz}{1.05}
\setsymbol{HFI:beam:ellipticity:Maps:545GHz}{1.14}
\setsymbol{HFI:beam:ellipticity:Maps:857GHz}{1.19}

\setsymbol{HFI:FWHM:Mars:100GHz:units}{9\parcm37}
\setsymbol{HFI:FWHM:Mars:143GHz:units}{7\parcm04}
\setsymbol{HFI:FWHM:Mars:217GHz:units}{4\parcm68}
\setsymbol{HFI:FWHM:Mars:353GHz:units}{4\parcm43}
\setsymbol{HFI:FWHM:Mars:545GHz:units}{3\parcm80}
\setsymbol{HFI:FWHM:Mars:857GHz:units}{3\parcm67}

\setsymbol{HFI:FWHM:Mars:100GHz}{9.37}
\setsymbol{HFI:FWHM:Mars:143GHz}{7.04}
\setsymbol{HFI:FWHM:Mars:217GHz}{4.68}
\setsymbol{HFI:FWHM:Mars:353GHz}{4.43}
\setsymbol{HFI:FWHM:Mars:545GHz}{3.80}
\setsymbol{HFI:FWHM:Mars:857GHz}{3.67}

\setsymbol{HFI:beam:ellipticity:Mars:100GHz}{1.18}
\setsymbol{HFI:beam:ellipticity:Mars:143GHz}{1.03}
\setsymbol{HFI:beam:ellipticity:Mars:217GHz}{1.14}
\setsymbol{HFI:beam:ellipticity:Mars:353GHz}{1.09}
\setsymbol{HFI:beam:ellipticity:Mars:545GHz}{1.25}
\setsymbol{HFI:beam:ellipticity:Mars:857GHz}{1.03}

\setsymbol{HFI:CMB:relative:calibration:100GHz}{$\lsim 1\%$}
\setsymbol{HFI:CMB:relative:calibration:143GHz}{$\lsim 1\%$}
\setsymbol{HFI:CMB:relative:calibration:217GHz}{$\lsim 1\%$}
\setsymbol{HFI:CMB:relative:calibration:353GHz}{$\lsim 1\%$}
\setsymbol{HFI:CMB:relative:calibration:545GHz}{}
\setsymbol{HFI:CMB:relative:calibration:857GHz}{}

\setsymbol{HFI:CMB:absolute:calibration:100GHz}{$\lsim 2\%$}
\setsymbol{HFI:CMB:absolute:calibration:143GHz}{$\lsim 2\%$}
\setsymbol{HFI:CMB:absolute:calibration:217GHz}{$\lsim 2\%$}
\setsymbol{HFI:CMB:absolute:calibration:353GHz}{$\lsim 2\%$}
\setsymbol{HFI:CMB:absolute:calibration:545GHz}{}
\setsymbol{HFI:CMB:absolute:calibration:857GHz}{}

\setsymbol{HFI:FIRAS:gain:calibration:accuracy:statistical:100GHz}{}
\setsymbol{HFI:FIRAS:gain:calibration:accuracy:statistical:143GHz}{}
\setsymbol{HFI:FIRAS:gain:calibration:accuracy:statistical:217GHz}{}
\setsymbol{HFI:FIRAS:gain:calibration:accuracy:statistical:353GHz}{2.5\%}
\setsymbol{HFI:FIRAS:gain:calibration:accuracy:statistical:545GHz}{1\%}
\setsymbol{HFI:FIRAS:gain:calibration:accuracy:statistical:857GHz}{0.5\%}

\setsymbol{HFI:FIRAS:gain:calibration:accuracy:systematic:100GHz}{}
\setsymbol{HFI:FIRAS:gain:calibration:accuracy:systematic:143GHz}{}
\setsymbol{HFI:FIRAS:gain:calibration:accuracy:systematic:217GHz}{}
\setsymbol{HFI:FIRAS:gain:calibration:accuracy:systematic:353GHz}{}
\setsymbol{HFI:FIRAS:gain:calibration:accuracy:systematic:545GHz}{7\%}
\setsymbol{HFI:FIRAS:gain:calibration:accuracy:systematic:857GHz}{7\%}

\setsymbol{HFI:FIRAS:zero:point:accuracy:100GHz:units}{0.8\MJysr}
\setsymbol{HFI:FIRAS:zero:point:accuracy:143GHz:units}{}
\setsymbol{HFI:FIRAS:zero:point:accuracy:217GHz:units}{}
\setsymbol{HFI:FIRAS:zero:point:accuracy:353GHz:units}{1.4\MJysr}
\setsymbol{HFI:FIRAS:zero:point:accuracy:545GHz:units}{2.2\MJysr}
\setsymbol{HFI:FIRAS:zero:point:accuracy:857GHz:units}{1.7\MJysr}

\setsymbol{HFI:FIRAS:zero:point:accuracy:100GHz}{0.8}
\setsymbol{HFI:FIRAS:zero:point:accuracy:143GHz}{}
\setsymbol{HFI:FIRAS:zero:point:accuracy:217GHz}{}
\setsymbol{HFI:FIRAS:zero:point:accuracy:353GHz}{1.4}
\setsymbol{HFI:FIRAS:zero:point:accuracy:545GHz}{2.2}
\setsymbol{HFI:FIRAS:zero:point:accuracy:857GHz}{1.7}

\setsymbol{HFI:unit:conversion:100GHz:units}{0.2415\MJysrmK}
\setsymbol{HFI:unit:conversion:143GHz:units}{0.3694\MJysrmK}
\setsymbol{HFI:unit:conversion:217GHz:units}{0.4811\MJysrmK}
\setsymbol{HFI:unit:conversion:353GHz:units}{0.2883\MJysrmK}
\setsymbol{HFI:unit:conversion:545GHz:units}{0.05826\MJysrmK}
\setsymbol{HFI:unit:conversion:857GHz:units}{0.002238\MJysrmK}

\setsymbol{HFI:unit:conversion:100GHz}{0.2415}
\setsymbol{HFI:unit:conversion:143GHz}{0.3694}
\setsymbol{HFI:unit:conversion:217GHz}{0.4811}
\setsymbol{HFI:unit:conversion:353GHz}{0.2883}
\setsymbol{HFI:unit:conversion:545GHz}{0.05826}
\setsymbol{HFI:unit:conversion:857GHz}{0.002238}

\setsymbol{HFI:colour:correction:alpha=-2:V1.01:100GHz}{0.9893}
\setsymbol{HFI:colour:correction:alpha=-2:V1.01:143GHz}{0.9759}
\setsymbol{HFI:colour:correction:alpha=-2:V1.01:217GHz}{1.0007}
\setsymbol{HFI:colour:correction:alpha=-2:V1.01:353GHz}{1.0028}
\setsymbol{HFI:colour:correction:alpha=-2:V1.01:545GHz}{1.0019}
\setsymbol{HFI:colour:correction:alpha=-2:V1.01:857GHz}{0.9889}

\setsymbol{HFI:colour:correction:alpha=0:V1.01:100GHz}{1.0008}
\setsymbol{HFI:colour:correction:alpha=0:V1.01:143GHz}{1.0148}
\setsymbol{HFI:colour:correction:alpha=0:V1.01:217GHz}{0.9909}
\setsymbol{HFI:colour:correction:alpha=0:V1.01:353GHz}{0.9888}
\setsymbol{HFI:colour:correction:alpha=0:V1.01:545GHz}{0.9878}
\setsymbol{HFI:colour:correction:alpha=0:V1.01:857GHz}{1.0014}

\begin{document}

\title{\textit{Planck} intermediate results. XIII. Constraints on peculiar
  velocities}

\titlerunning{Peculiar velocity constraints from \Planck\ data}
\authorrunning{\Planck\ Collaboration}

\author{\small
Planck Collaboration:
P.~A.~R.~Ade\inst{77}
\and
N.~Aghanim\inst{51}
\and
M.~Arnaud\inst{65}
\and
M.~Ashdown\inst{62, 7}
\and
J.~Aumont\inst{51}
\and
C.~Baccigalupi\inst{76}
\and
A.~Balbi\inst{33}
\and
A.~J.~Banday\inst{84, 9}
\and
R.~B.~Barreiro\inst{59}
\and
E.~Battaner\inst{86}
% ...
}
\institute{\small
APC, AstroParticule et Cosmologie, Universit\'{e} Paris Diderot, CNRS/IN2P3, CEA/lrfu, Observatoire de Paris, Sorbonne Paris Cit\'{e}, 10, rue Alice Domon et L\'{e}onie Duquet, 75205 Paris Cedex 13, France\\
\and
Aalto University Mets\"{a}hovi Radio Observatory, Mets\"{a}hovintie 114, FIN-02540 Kylm\"{a}l\"{a}, Finland\\
\and
Academy of Sciences of Tatarstan, Bauman Str., 20, Kazan, 420111, Republic of Tatarstan, Russia\\
\and
African Institute for Mathematical Sciences, 6-8 Melrose Road, Muizenberg, Cape Town, South Africa\\
\and
Agenzia Spaziale Italiana Science Data Center, c/o ESRIN, via Galileo Galilei, Frascati, Italy\\
\and
Agenzia Spaziale Italiana, Viale Liegi 26, Roma, Italy\\
\and
Astrophysics Group, Cavendish Laboratory, University of Cambridge, J J Thomson Avenue, Cambridge CB3 0HE, U.K.\\
\and
CITA, University of Toronto, 60 St. George St., Toronto, ON M5S 3H8, Canada\\
\and
CNRS, IRAP, 9 Av. colonel Roche, BP 44346, F-31028 Toulouse cedex 4, France\\
\and
California Institute of Technology, Pasadena, California, U.S.A.\\
\and
Centro de Astrof\'{\i}sica, Universidade do Porto, Rua das Estrelas, 4150-762 Porto, Portugal\\
\and
Centro de Estudios de F\'{i}sica del Cosmos de Arag\'{o}n (CEFCA), Plaza San Juan, 1, planta 2, E-44001, Teruel, Spain\\
\and
Computational Cosmology Center, Lawrence Berkeley National Laboratory, Berkeley, California, U.S.A.\\
\and
Consejo Superior de Investigaciones Cient\'{\i}ficas (CSIC), Madrid, Spain\\
\and
DSM/Irfu/SPP, CEA-Saclay, F-91191 Gif-sur-Yvette Cedex, France\\
\and
DTU Space, National Space Institute, Technical University of Denmark, Elektrovej 327, DK-2800 Kgs. Lyngby, Denmark\\
\and
D\'{e}partement de Physique Th\'{e}orique, Universit\'{e} de Gen\`{e}ve, 24, Quai E. Ansermet,1211 Gen\`{e}ve 4, Switzerland\\
\and
Departamento de F\'{\i}sica, Universidad de Oviedo, Avda. Calvo Sotelo s/n, Oviedo, Spain\\
\and
Department of Astronomy and Astrophysics, University of Toronto, 50 Saint George Street, Toronto, Ontario, Canada\\
\and
Department of Astronomy and Geodesy, Kazan Federal University,  Kremlevskaya Str., 18, Kazan, 420008, Russia\\
\and
Department of Astrophysics/IMAPP, Radboud University Nijmegen, P.O. Box 9010, 6500 GL Nijmegen, The Netherlands\\
\and
Department of Electrical Engineering and Computer Sciences, University of California, Berkeley, California, U.S.A.\\
\and
Department of Physics \& Astronomy, University of British Columbia, 6224 Agricultural Road, Vancouver, British Columbia, Canada\\
\and
Department of Physics and Astronomy, Dana and David Dornsife College of Letter, Arts and Sciences, University of Southern California, Los Angeles, CA 90089, U.S.A.\\
\and
Department of Physics, Gustaf H\"{a}llstr\"{o}min katu 2a, University of Helsinki, Helsinki, Finland\\
\and
Department of Physics, Princeton University, Princeton, New Jersey, U.S.A.\\
\and
Department of Physics, University of California, Santa Barbara, California, U.S.A.\\
\and
Dipartimento di Fisica e Astronomia G. Galilei, Universit\`{a} degli Studi di Padova, via Marzolo 8, 35131 Padova, Italy\\
\and
Dipartimento di Fisica e Scienze della Terra, Universit\`{a} di Ferrara, Via Saragat 1, 44122 Ferrara, Italy\\
\and
Dipartimento di Fisica, Universit\`{a} La Sapienza, P. le A. Moro 2, Roma, Italy\\
\and
Dipartimento di Fisica, Universit\`{a} degli Studi di Milano, Via Celoria, 16, Milano, Italy\\
\and
Dipartimento di Fisica, Universit\`{a} degli Studi di Trieste, via A. Valerio 2, Trieste, Italy\\
\and
Dipartimento di Fisica, Universit\`{a} di Roma Tor Vergata, Via della Ricerca Scientifica, 1, Roma, Italy\\
\and
Dipartimento di Matematica, Universit\`{a} di Roma Tor Vergata, Via della Ricerca Scientifica, 1, Roma, Italy\\
\and
Discovery Center, Niels Bohr Institute, Blegdamsvej 17, Copenhagen, Denmark\\
\and
Dpto. Astrof\'{i}sica, Universidad de La Laguna (ULL), E-38206 La Laguna, Tenerife, Spain\\
\and
European Space Agency, ESAC, Planck Science Office, Camino bajo del Castillo, s/n, Urbanizaci\'{o}n Villafranca del Castillo, Villanueva de la Ca\~{n}ada, Madrid, Spain\\
\and
European Space Agency, ESTEC, Keplerlaan 1, 2201 AZ Noordwijk, The Netherlands\\
\and
Helsinki Institute of Physics, Gustaf H\"{a}llstr\"{o}min katu 2, University of Helsinki, Helsinki, Finland\\
\and
INAF - Osservatorio Astronomico di Padova, Vicolo dell'Osservatorio 5, Padova, Italy\\
\and
INAF - Osservatorio Astronomico di Roma, via di Frascati 33, Monte Porzio Catone, Italy\\
\and
INAF - Osservatorio Astronomico di Trieste, Via G.B. Tiepolo 11, Trieste, Italy\\
\and
INAF/IASF Bologna, Via Gobetti 101, Bologna, Italy\\
\and
INAF/IASF Milano, Via E. Bassini 15, Milano, Italy\\
\and
INFN, Sezione di Bologna, Via Irnerio 46, I-40126, Bologna, Italy\\
\and
INFN, Sezione di Roma 1, Universit`{a} di Roma Sapienza, Piazzale Aldo Moro 2, 00185, Roma, Italy\\
\and
IUCAA, Post Bag 4, Ganeshkhind, Pune University Campus, Pune 411 007, India\\
\and
Imperial College London, Astrophysics group, Blackett Laboratory, Prince Consort Road, London, SW7 2AZ, U.K.\\
\and
Infrared Processing and Analysis Center, California Institute of Technology, Pasadena, CA 91125, U.S.A.\\
\and
Institut Universitaire de France, 103, bd Saint-Michel, 75005, Paris, France\\
\and
Institut d'Astrophysique Spatiale, CNRS (UMR8617) Universit\'{e} Paris-Sud 11, B\^{a}timent 121, Orsay, France\\
\and
Institut d'Astrophysique de Paris, CNRS (UMR7095), 98 bis Boulevard Arago, F-75014, Paris, France\\
\and
Institute for Space Sciences, Bucharest-Magurale, Romania\\
\and
Institute of Astro and Particle Physics, Technikerstrasse 25/8, University of Innsbruck, A-6020, Innsbruck, Austria\\
\and
Institute of Astronomy and Astrophysics, Academia Sinica, Taipei, Taiwan\\
\and
Institute of Astronomy, University of Cambridge, Madingley Road, Cambridge CB3 0HA, U.K.\\
\and
Institute of Theoretical Astrophysics, University of Oslo, Blindern, Oslo, Norway\\
\and
Instituto de Astrof\'{\i}sica de Canarias, C/V\'{\i}a L\'{a}ctea s/n, La Laguna, Tenerife, Spain\\
\and
Instituto de F\'{\i}sica de Cantabria (CSIC-Universidad de Cantabria), Avda. de los Castros s/n, Santander, Spain\\
\and
Jet Propulsion Laboratory, California Institute of Technology, 4800 Oak Grove Drive, Pasadena, California, U.S.A.\\
\and
Jodrell Bank Centre for Astrophysics, Alan Turing Building, School of Physics and Astronomy, The University of Manchester, Oxford Road, Manchester, M13 9PL, U.K.\\
\and
Kavli Institute for Cosmology Cambridge, Madingley Road, Cambridge, CB3 0HA, U.K.\\
\and
LAL, Universit\'{e} Paris-Sud, CNRS/IN2P3, Orsay, France\\
\and
LERMA, CNRS, Observatoire de Paris, 61 Avenue de l'Observatoire, Paris, France\\
\and
Laboratoire AIM, IRFU/Service d'Astrophysique - CEA/DSM - CNRS - Universit\'{e} Paris Diderot, B\^{a}t. 709, CEA-Saclay, F-91191 Gif-sur-Yvette Cedex, France\\
\and
Laboratoire Traitement et Communication de l'Information, CNRS (UMR 5141) and T\'{e}l\'{e}com ParisTech, 46 rue Barrault F-75634 Paris Cedex 13, France\\
\and
Laboratoire de Physique Subatomique et de Cosmologie, Universit\'{e} Joseph Fourier Grenoble I, CNRS/IN2P3, Institut National Polytechnique de Grenoble, 53 rue des Martyrs, 38026 Grenoble cedex, France\\
\and
Laboratoire de Physique Th\'{e}orique, Universit\'{e} Paris-Sud 11 \& CNRS, B\^{a}timent 210, 91405 Orsay, France\\
\and
Lawrence Berkeley National Laboratory, Berkeley, California, U.S.A.\\
\and
Max-Planck-Institut f\"{u}r Astrophysik, Karl-Schwarzschild-Str. 1, 85741 Garching, Germany\\
\and
Max-Planck-Institut f\"{u}r Extraterrestrische Physik, Giessenbachstra{\ss}e, 85748 Garching, Germany\\
\and
National University of Ireland, Department of Experimental Physics, Maynooth, Co. Kildare, Ireland\\
\and
Niels Bohr Institute, Blegdamsvej 17, Copenhagen, Denmark\\
\and
Observational Cosmology, Mail Stop 367-17, California Institute of Technology, Pasadena, CA, 91125, U.S.A.\\
\and
Optical Science Laboratory, University College London, Gower Street, London, U.K.\\
\and
SISSA, Astrophysics Sector, via Bonomea 265, 34136, Trieste, Italy\\
\and
School of Physics and Astronomy, Cardiff University, Queens Buildings, The Parade, Cardiff, CF24 3AA, U.K.\\
\and
School of Physics and Astronomy, University of Nottingham, Nottingham NG7 2RD, U.K.\\
\and
Space Research Institute (IKI), Russian Academy of Sciences, Profsoyuznaya Str, 84/32, Moscow, 117997, Russia\\
\and
Space Sciences Laboratory, University of California, Berkeley, California, U.S.A.\\
\and
Stanford University, Dept of Physics, Varian Physics Bldg, 382 Via Pueblo Mall, Stanford, California, U.S.A.\\
\and
T\"{U}B\.{I}TAK National Observatory, Akdeniz University Campus, 07058, Antalya, Turkey\\
\and
UPMC Univ Paris 06, UMR7095, 98 bis Boulevard Arago, F-75014, Paris, France\\
\and
Universit\'{e} de Toulouse, UPS-OMP, IRAP, F-31028 Toulouse cedex 4, France\\
\and
University Observatory, Ludwig Maximilian University of Munich, Scheinerstrasse 1, 81679 Munich, Germany\\
\and
University of Granada, Departamento de F\'{\i}sica Te\'{o}rica y del Cosmos, Facultad de Ciencias, Granada, Spain\\
% ...
}

\abstract {Using \Planck\ data combined with the Meta Catalogue of X-ray
  detected Clusters of galaxies (MCXC), we address the study of peculiar motions
  by searching for evidence of the kinetic Sunyaev-Zeldovich effect (kSZ).
  By implementing various filters designed to extract the kSZ generated at
  the positions of the clusters, we obtain consistent constraints on the radial
  peculiar velocity average, root mean square (rms), and local bulk flow
  amplitude at different depths. For the whole cluster sample of average
  redshift $0.18$, the measured average radial peculiar velocity with respect to
  the cosmic microwave background (CMB) radiation at that redshift, i.e., the
  kSZ monopole, amounts to
  $72 \pm 60$\,km\,s$^{-1}$. This constitutes less than 1\,\% of the relative
  Hubble velocity of the cluster sample with respect to our local CMB frame.
 While the linear $\Lambda$CDM prediction for the typical cluster radial 
velocity rms at $z=0.15$ is close to 230\,km\,s$^{-1}$, the upper limit imposed by \Planck\ data on the cluster subsample corresponds to 800\,km\,s$^{-1}$ at 95\,\% confidence level, i.e., about three times higher. 
  \Planck\ data also set strong
  constraints on the local bulk flow in volumes centred on the Local Group.
  There is no detection of bulk flow as measured in any comoving sphere
  extending to the maximum redshift covered by the cluster sample. A blind
  search for bulk flows in this sample has an upper limit of
  254\,km\,s$^{-1}$ (95\,\%
  confidence level) dominated by CMB confusion and instrumental noise,
  indicating that the Universe is largely homogeneous on Gpc scales. In this
  context, in conjunction with supernova observations, \Planck\ is able to rule
  out a large class of inhomogeneous void models as alternatives to dark energy
  or modified gravity.  The \Planck\ constraints on peculiar velocities and
  bulk flows are thus consistent with the $\Lambda$CDM scenario.}

\date{Received XXXX; accepted YYYY}

\keywords{cosmology: observations -- cosmic microwave background -- large-scale
  structure of the Universe -- galaxies: clusters: general }

\authorrunning{Planck Collaboration}

\maketitle
%
\section{Introduction}

Today we have a cosmological model that appears to fit all available data.
Nevertheless, it is important to continue to test this picture.  Peculiar
velocities provide an important way to do this.
According to the standard $\Lambda$CDM scenario, gravity drives the growth of
inhomogeneities in the matter distribution of the Universe. After the
radiation-matter equality epoch, fluctuations in the dark matter component were
largely unaffected by the Thomson interaction binding the evolution of baryons
and photons of the cosmic microwave background radiation (CMB).  During that
epoch, the inhomogeneities in the spatial distribution of dark matter kept
growing gravitationally. It was only after the epoch of hydrogen recombination
that the baryons, which had just decoupled from the CMB, could freely fall into
the potential wells created by the dark matter component.

Since then, the gravitational infall of matter into potential wells has been
conditioned by the density field and the universal expansion rate. On large
scales, where baryonic physics and non-linear evolution may be neglected safely,
the continuity equation provides a simple link between the matter density field
and the peculiar velocity field. In particular, in a $\Lambda$CDM scenario, this equation
predicts that peculiar velocities must show typical correlation lengths between 
20 and 40\,$h^{-1}$\,Mpc, and their growth must have practically frozen
since the onset of the accelerated expansion \citep[see the review of,
 e.g.,][]{strauss_willick95}. By averaging the peculiar velocity field on
scales corresponding to galaxy groups and clusters today it is possible to
obtain linear theory predictions for the root mean square (rms) of the radial 
peculiar velocity of
those structures. These predictions typically amount to
about 230\,km\,s$^{-1}$,
\citep[see, e.g.,][]{chmras10}, with a weak dependence on
the galaxy cluster/group mass. If instead one looks at the velocity amplitude
for extended or correlated motion of matter on larger scales, one finds that it
decreases when larger volumes are considered, but should still be at the level
of 50--100\,km\,s$^{-1}$ for radii of a few hundred Mpc \citep[see, e.g., figure
 2 in][]{mak11}. The detection of these large-scale, coherent flows of matter
(hereafter referred to as {\it bulk flows}) has been the subject of active
investigation for several decades \citep[e.g., ][to cite just a
  few]{tonryanddavis81,aaronson82,dressler87,dekel93,lp94,hudson99,willick99,riess00}.
One crucial problem that most of those works encounter is related to the need to
accurately determine distances to galaxies in order to subtract the Hubble
flow-induced velocity.

During the nineties \citet{lp94}, \citet{willick99}, and \citet{hudson99}
claimed that there are large-scale bulk flows with amplitudes of
350--700\,km\,s$^{-1}$ in local spheres of radii
60--150\,$h^{-1}$\,Mpc,
with somewhat discrepant directions.  At the turn of the millennium, however,
\citet{riess00} and \citet{courteau00} reported the lack of any significant
local bulk flow up to depths of about 150\,$h^{-1}$\,Mpc, in apparent
contradiction to the previous works. More recently, claims of the presence of
a large-scale, large amplitude peculiar velocity dipole have been raised again
by various authors. While some works \citep{hudson04,
watkinsetal09,feldmanetal10} find evidence for a peculiar local velocity
dipole of about 400\,km\,s$^{-1}$ (and in tension with $\Lambda$CDM
predictions), others find lower amplitudes for the local bulk flow,
\citep[e.g.,][]{itohyahataandtakada10,nusserdavis11,nusserbranchinianddavis11,mascott12,branchinidavisandnusser12,Courtoisetal2012}.

For greater depths (up to $z\,{\sim}\,0.2$--$0.3$) there are also claims
\citep{kashlinsky08,kashlinsky10,abatefeldman11} of yet higher amplitude bulk
flows (${\sim}\,1000$--4000\,km\,s$^{-1}$).
These cannot be accommodated within a $\Lambda$CDM context,
since the theory predicts that bulk flows are negligible on the very
largest scales. Moreover, these results are in contradiction with other
works, \citep[e.g.,][]{keisler09,osborneetal11,modyandamir}.

Some of the most recent results for bulk flows extending to large distances are
based on the study of the kinetic Sunyaev-Zeldovich effect \citep[hereafter kSZ;
][]{kSZ}. This effect is due to the Doppler kick that CMB photons experience in
Thomson scattering off free electrons moving with respect to the CMB rest
frame. This process introduces intensity and polarization anisotropies in the
CMB along the direction of massive clouds of ionized material, such as galaxy
clusters and groups, but produces no distortion of the CMB spectrum. The kSZ
effect has been theoretically exploited to characterize the growth of velocity
perturbations \citep[e.g.,][]{maandfry02,chm05,zhangetal08}, to search for
missing baryons \citep{dedeo05,chm08, ho09,chm09, shaozhangetal10}, and to study
bulk flows in the local Universe
\citep{kashlinsky08,kashlinsky10,keisler09,osborneetal11, mak11,
  modyandamir,lavaux_afshordi12}.  Very recently, \citet{hand12} have claimed a
detection of the kSZ effect when combining spectroscopic galaxy data from the
Baryonic Acoustic Oscillation Survey (BOSS) with CMB data from the Atacama
Cosmology Telescope (ACT), after searching for the kSZ pairwise momentum
\citep[e.g.,][]{groth89,roman98}. On subcluster scales, as predicted by, e.g.,
\citet{inogamov_ras}, some weak evidence of kSZ has also been reported by the
Bolocam instrument \citep{mustang}.

In this paper we shall focus on the constraints that \Planck\footnote{
\Planck\ (\url{http://www.esa.int/Planck}) is a project of the European Space
Agency (ESA) with instruments provided by two scientific consortia funded by
ESA member states (in particular the lead countries France and Italy), with
contributions from NASA (USA) and telescope reflectors provided by a
collaboration between ESA and a scientific consortium led and funded by
Denmark.} can set on the kSZ-induced temperature anisotropies. These are given
by the line-of-sight integral
\begin{equation}
\frac{\delta T}{T_0}(\vn) = -\int \; dl \, \sigma_{\rm T} \, n_{\rm e}
\frac{\vv_{\rm e}\cdot \vn }{c},
\label{eq:kSZ1}
\end{equation}
where $\sigma_{\rm T}$ is the Thomson scattering cross-section, $n_{\rm e}$ is
the physical electron number density, $\vv_{\rm e}$ denotes the electron
peculiar velocity, $c$ the speed of light, and $\vn$ the direction of
observation on the sky. We are adopting here a reference frame centred on the
observer's position, and hence infalling electrons will have {\it negative}
radial velocities. Note that, unlike in other approaches based upon galaxy
redshift surveys, the distance to the cluster is irrelevant to its peculiar
velocity estimation.  Since the expected kSZ signal coming from an individual
cluster is smaller than the typical level of intrinsic CMB temperature
fluctuations, we shall apply various filters which attempt to minimize the
impact of other signals on the angular positions of a sample of galaxy
clusters, and use these to
extract statistical constraints on the kSZ signal in those sources. 
In the standard $\Lambda$CDM scenario one expects to have matter at rest
with respect to the CMB on
the largest scales, and hence roughly the same number of clusters with
positive and negative radial velocities. This means that the mean or
{\em monopole\/} of kSZ estimates should be consistent with zero, although
there are inhomogeneous scenarios (addressed in Sect.~\ref{sec:inhomo}) in
which the average velocity of clusters may differ from zero. Likewise it is
possible to set constraints on the kSZ-induced variance in the CMB temperature
anisotropies measured along the direction of galaxy clusters. 
This is a direct probe of the rms peculiar velocity of those objects with
respect to the CMB, and can be compared to theoretical predictions. In this
context, it has been mentioned above that the motion of matter is predicted to
occur in bulk flows with coherence on scales of
about $30\,h^{-1}$Mpc. If these
bulk flows are local and the observer is placed inside them, then they should
give rise to a dipolar pattern in the kSZ measurements of individual clusters
\citep{kashlinsky00}. If they are instead distant, then the projection of the
coherence length of kSZ measurements on the sky should shrink down to a few
degrees \citep{chm05}. Therefore it is possible to use the set of
{\em individual\/} kSZ estimates from galaxy clusters to place constraints on
the monopole (mean), variance, and dipole of the peculiar velocities of the
cluster population. While some of our statistical tools target the kSZ signal
in each cluster separately, others are particularly designed to probe the
local bulk blow and set constraints on the kSZ dipole at the positions of
clusters, as will be shown below.

This paper is organized as follows. In Sect.~\ref{sec:data} we describe the data
used, both for the CMB and large-scale structure. The statistical tools we use
for the kSZ detection are described in Sect.~\ref{sec:filters}, and the results
obtained from them are presented in Sect.~\ref{sec:results}. We examine the
robustness of our results in Sect.~\ref{sec:robust}. Finally, in
Sect.~\ref{sec:discussion}, we discuss the cosmological implications of our
findings and conclude. Throughout this paper, we use a cosmological parameter
set compatible with {\it WMAP}-7 observations \citep{komatsu2010}:
density parameters $\Omega_{\rm M} = 0.272$ and $\Omega_{\Lambda} = 0.728$;
Hubble parameter $h = 0.704$; $8\,h^{-1}$Mpc normalization $\sigma_8=0.809$;
and $n_{\rm S} = 0.96$ for the spectral index of scalar perturbations.

\section{Data and simulations}
\label{sec:data}
\subsection{\Planck\ data}

\Planck\ \citep{tauber2010a, Planck2011-1.1} is the third generation space
mission to measure the anisotropy of the CMB. It observes the sky in nine
frequency bands covering 30--857\,GHz with high sensitivity and angular
resolution from 31\arcm\ to 4.39\arcm.  The Low Frequency Instrument (LFI;
\citealt{Mandolesi2010, Bersanelli2010, Planck2011-1.4}) covers the 30, 44, and
70\,GHz bands with amplifiers cooled to 20\,\hbox{K}. The High Frequency
Instrument (HFI; \citealt{Lamarre2010, Planck2011-1.5}) covers the 100, 143,
217, 353, 545, and 857\,GHz bands with bolometers cooled to 0.1\,\hbox{K}.
Polarization is measured in all but the highest two bands \citep{Leahy2010,
Rosset2010}. A combination of radiative cooling and three mechanical coolers
produces the temperatures needed for the detectors and optics
\citep{Planck2011-1.3}.  Two data processing centres (DPCs) check and calibrate
the data and make maps of the sky \citep{Planck2011-1.7, Planck2011-1.6}.
\Planck's sensitivity, angular resolution, and frequency coverage make it a
powerful instrument for Galactic and extragalactic astrophysics as well as
cosmology. Early astrophysics results are given in Planck Collaboration
VIII--XXVI 2011, based on data taken between 13~August 2009 and 7~June 2010.
Intermediate astrophysics results are now being presented in a series of papers
based on data taken between 13~August 2009 and 27~November 2010.

Although the 70\,GHz LFI channel was included initially in the analysis, it was
found that constraints were practically identical when using HFI frequency maps
alone (see details in Table~\ref{tab:hfi_table}).  Measuring the kSZ effect
requires avoiding, in the best possible way, contamination by the much stronger
thermal Sunyaev-Zeldovich effect \citep[hereafter tSZ; ][]{tSZ}. While in theory
observations at 217\,GHz, close to the zero of the tSZ emission, should not
suffer much from tSZ contamination, it is necessary to account for the broad
spectral band of each detector and each channel, \citep{Planck2011-1.7}.
 In terms of the tSZ effect, the
`effective' frequencies of the HFI channels (i.e., those frequencies at which
the tSZ emission is equal to its integral over the frequency band) are listed in
the second column of Table~\ref{tab:hfi_table}. Raw HFI frequency maps are
useful for testing for systematic effects associated with foreground emission,
tSZ spectral leakage, or FWHM characterization errors.  Note also that HFI
frequency maps are produced in thermodynamic temperature units, so that both
primary CMB and kSZ emission have constant amplitude across frequency channels.
In the third column we display the corresponding tSZ Comptonization parameter
($y_{\rm SZ}$) to $\Delta T$ conversion factors.
The Comptonization parameter $y_{\rm SZ}$ is a dimensionless
 line-of-sight integral of the gas pressure,
\begin{equation}
y_{\rm SZ} = \int dl\, \sigma_{\rm T} n_{\rm e}
 \frac{k_{\rm B}T_{\rm e}}{m_{\rm e} c^2},
\label{eq:ysmall}
\end{equation}
with $T_{\rm e}$ and $m_{\rm e}$ the electron temperature and rest mass,
and $k_{\rm B}$ the Boltzmann constant, \citep{tSZ}.

Two different strategies are used in this paper to measure the kSZ effect at the
positions of the cluster catalogue. The first consists of estimating directly
the kSZ signal at MCXC cluster positions from the original HFI frequency maps,
using both aperture photometry and matched multi-band filtering.
The second consists of first
producing a map of the CMB and kSZ effect that is nearly free from tSZ
contamination before estimating the kSZ emission from MCXC clusters
using the aperture photometry
and single frequency matched filtering. As described below,
this map makes use of both HFI and LFI data.

\begin{table}[tmb] 
\begingroup 
\newdimen\tblskip \tblskip=5pt
\caption{Nominal and tSZ-effective frequencies, $\Delta T$ to $y_{\rm SZ}$
  conversion factors and FWHMs for each HFI channel used in this paper. The
  second column provides the effective frequency of the channels after
  considering the spectral dependence of the non-relativistic tSZ effect and the
  finite response of the HFI detectors.  The third column displays the
  conversion factor between SZ Comptonization parameter ($y_{\rm SZ}$) and
  thermodynamic temperature (in K). The fourth column provides, for each HFI channel, 
  the average FWHM value of the effective Gaussian beam at map level, as described in \cite{Planck2011-1.7}.
}
\label{tab:hfi_table}
\vskip -3mm
\footnotesize 
\setbox\tablebox=\vbox{ %
\newdimen\digitwidth 
\setbox0=\hbox{\rm 0}
\digitwidth=\wd0
\catcode`*=\active
\def*{\kern\digitwidth}
\newdimen\signwidth
\setbox0=\hbox{+}
\signwidth=\wd0
\catcode`!=\active
\def!{\kern\signwidth}
\halign{\hfil#\hfil\tabskip=0.2cm& \hfil#\hfil\tabskip=0.2cm&
 \hfil#\hfil\tabskip=0.2cm& \hfil#\hfil\tabskip=0cm\cr
\noalign{\doubleline}
\noalign{\vskip -2pt}
HFI nominal& HFI effective&  $y_{\rm SZ}/\Delta T$& FWHM\cr
frequency& frequency& &\cr
[GHz]& [GHz]& [K$^{-1}_{\rm {CMB}}$]& [arcmin]\cr
\noalign{\vskip 3pt\hrule\vskip 3pt}
100& 103.1& $-0.2481$&*9.88\cr
143& 144.5& $-0.3592$&*7.18\cr
217& 222.1& !$5.2602$&*4.87\cr
353& 355.2& !$0.1611$&*4.65\cr
545& 528.5& !$0.0692$&*4.72\cr
857& 775.9& !$0.0380$&*4.39\cr
\noalign{\vskip 3pt\hrule\vskip 3pt}}}
\endPlancktable 
\endgroup
\end{table}

\subsubsection{The two-dimensional internal linear combination map} 
  
In the absence of a fully reliable model of foreground emission (including
number of foregrounds, emission laws, and coherence of their emission across
\Planck\ frequencies), a minimum variance map of CMB emission can be obtained by
the so-called ``Internal Linear Combination'' approach (hereafter ILC). The CMB
map is obtained from a linear combination of input observations, subject to the
constraint that the CMB is preserved. I.e., for CMB-calibrated maps (in
thermodynamic units) $x_i$, the CMB is obtained as
$\sum_i w_i x_i$ with $\sum_i w_i = 1$,
the latter condition guaranteeing the preservation of the CMB
signal.  This obviously also preserves the kSZ signal, which has the same
frequency dependence.

The minimum variance map, however, is not necessarily that of minimum
contamination by any particular foreground. In our present analysis, the
measured map of CMB+kSZ will be further processed, first being filtered on the
basis of predicted kSZ cluster shapes and locations, to suppress contamination
by the larger scale primary CMB, and then stacked to combine the
measurements of all individual clusters. While this filtering and stacking will
reduce contamination of the measurement by independent foregrounds such as
Galactic dust emission, as well as by detector noise,
tSZ residuals are likely to
add-up coherently and contaminate the measurement significantly. The ILC must
then be modified to ensure that instead of the total variance of the map being
minimized, the contamination by tSZ must be minimized instead.

It is possible to extend the ILC method to add a constraint to reject the tSZ
effect specifically, and thus make sure that the CMB+kSZ map is completely free
from tSZ contamination. The idea is similar to that used in the unbiased
Multifrequency Matched Filter approach \citep{herranz05,mak11},
i.e., a constraint is added to null the tSZ contribution to the output map. The method is described
and validated on realistic simulations in \citet{2011MNRAS.410.2481R}.

Note that this method is a special case of a multi-dimensional generalization of
the ILC \citep{2011MNRAS.418..467R}, in which several components of interest
with known emission laws, can be recovered simultaneously with vanishing
contamination from each other. Here we consider two specific components,
one with the CMB emission law, which comprises both primary CMB and kSZ, and one
with the emission law of the tSZ effect (neglecting relativistic
corrections). We refer to the map obtained by this method as a two-dimensional
ILC (hereafter 2D-ILC).

In detail, the 2D-ILC map used in the present analysis is obtained from all LFI
and HFI maps as follows. For each frequency band, point sources detected
by a Mexican Hat Wavelet filter at more than $5\,\sigma$ at that frequency are
masked. The masked region has a radius of three times the standard deviation of
the Gaussian beam (i.e., $1.27 \times$\,FWHM). The masked regions are filled in
by interpolation using neighbouring pixels. Maps are then analysed on a frame of
spherical needlets for implementation of the ILC in needlet space, in a very
similar way to what has been done on \textit{WMAP\/} data by
\citet{2009A&A.493.835D}. However, the covariance matrices associated with
the filter, instead of being computed using average
covariances of needlet coefficients over
HEALPix\footnote{\url{http://healpix.jpl.nasa.gov}}
superpixels, are computed from products of maps of needlet coefficients,
smoothed using a large Gaussian beam, similarly to what was done
by \citet{2012MNRAS.419.1163B} on \textit{WMAP\/}
7-year data. The constrained ILC filter implemented is that of Eq.~20 of
\citet{2011MNRAS.410.2481R}. Thus, the exact linear combination used to
reconstruct the CMB+kSZ map depends both on the sky region and on the angular
scale. In particular, on scales smaller than some of the \Planck\ LFI and HFI
beams, the relative weights of the corresponding lower frequency channels become
negligibly small, due to their low resolution. The final CMB+kSZ map is
reconstructed at $5\arcm$ resolution. In order to carry this out, at the very
smallest scales the CMB+kSZ map is reconstructed mostly from observations in
the frequency channels at 217\,GHz and above. At intermediate scales (of order
$10\arcm$), however, measurements from all HFI channels are used to reconstruct
the final map.

The ILC (classical or 2D version) assumes the emission law of the component of
interest to be known. This knowledge is necessary to ensure the preservation of
the signal of interest (here, the kSZ effect) and, for the 2D-ILC, to reject
the contaminating signal (here, the tSZ). As discussed in
\citet{2010MNRAS.401.1602D}, imperfect knowledge of the emission law can
result in a significant loss of CMB power.  In practice, the effective emission
laws, as observed by the detectors, depend on the calibration of the
observations in each frequency channel. For the CMB and kSZ signals, it hence
depends on the absolute calibration of all the \Planck\ channels used in the
analysis (here, the HFI channels). For the tSZ, it also depends on the accuracy
of the knowledge of the spectral bands, and on the validity of the
non-relativistic approximation for tSZ emission.

For \Planck\ HFI channels, the absolute calibration error is estimated
(conservatively) to be less than about 0.1\,\% in the channels calibrated on
the CMB itself with the CMB dipole, and on CMB anisotropies themselves for
cross-calibration, and less than a few percent on channels calibrated on the
dust emission measured by FIRAS
\citep[545 and 857\,GHz channels,][]{Planck2011-1.7}. 
Small uncertainties on the frequency dependence of the (CMB+ kSZ)
signal may induce a large bias in the calibration of the output of the ILC. 
This effect can be strong in the high SNR regime \citep{2010MNRAS.401.1602D}, 
which is the case here
because the strong CMB anisotropy signal itself contributes to the total
signal. With CMB calibration errors of 0.1\,\%, we check on simulated data
sets generated with the Planck Sky Model \citep{jacques12} that the
corresponding error on the final map is small (less than 1\,\%).
This is also confirmed on the actual \Planck\ data by comparing the power spectrum of
the CMB+kSZ map with the current CMB best fit $C_\ell$, since any serious
loss of power would be immediately visible in the power spectrum of the
reconstructed CMB+kSZ map.

Errors in the assumed tSZ emission law (by reason of relativistic corrections,
colour correction, or mis-calibration), can also potentially result in residual
contamination by tSZ in the CMB+kSZ ILC map. Note, however, that the
two-dimensional ILC does not amplify the contamination by a mis-calibrated
tSZ component. Uncertainties of a few percent on the tSZ frequency dependence
(the typical size of relativistic corrections to the thermal SZ effect)
will hence not impact the reconstruction of the CMB+kSZ signal by more
than a few percent of the original tSZ.

\subsection{Tracers of moving baryons: X-ray MCXC clusters}
\label{sec:clusters}

We use the Meta Catalogue of X-ray detected Clusters of galaxies (MCXC), a
compilation of all publicly available {\it ROSAT} All Sky Survey-based samples
(NORAS, \citealt{boh1}; REFLEX, \citealt{boh2}; BCS, \citealt{eb1},
\citealt{eb2}; SGP, \citealt{cru}; NEP, \citealt{hen}; MACS, \citealt{eb3}; and
CIZA, \citealt{eb4}, \citealt{koc}), and serendipitous cluster catalogues
(160SD, \citealt{mul}; 400SD, \citealt{bure}; SHARC, \citealt{rom},
\citealt{burk}; WARPS, \citealt{per}, \citealt{hor}; and EMSS, \citealt{gio}
\citealt{hen2}). The information was systematically homogenized and duplicate
entries were carefully handled, yielding a large catalogue of approximately 1750
clusters. The MCXC is presented in detail in \cite{pif}, and has been used
in previous \Planck\ studies, \citep[e.g.,][]{planck2011-5.2a}.

For each cluster the MCXC provides, among other quantities, the position,
redshift, and mass of each cluster.
The masses are estimated from the luminosities thanks to the
REXCESS $L_{500}$--$M_{500}$ relation of \cite{prat}. Hereafter, all cluster
quantities with the subscript ``$500$'' are evaluated at the radius ($R_{500}$)
at which the average density equals 500 times the critical density at the
cluster's redshift. In this way, $M_{500}$ is defined as $M_{500} = (4\pi/3)
\,500 \, \rho_{\rm c}(z) \, R_{500}^{3}$, where $\rho_{\rm c}(z)$ is the
critical density at the cluster redshift $z$.

For the measurement of velocities, we also need cluster optical depths. Our
approach here is based upon the study of \citet{arnaudetal10}: using REXCESS
data, we either use the universal pressure profile and then divide by the
average temperature profile to estimate a density profile, or fit directly an
average density profile. For this purpose, we make use of $Y(x)$,
the volume integral 
of gas pressure up to a radius given by $x\equiv r/ R_{\rm 500}$:
\begin{eqnarray}
Y_{\rm sph}(x) & = & \int_0^{r=x\,R_{\rm 500}} d\vrv' \, 
 \sigma_{\rm T} n_{\rm e} (\vrv')
 \frac{k_{\rm B}T_{\rm e}(\vrv')}{m_{\rm e} c^2}; \\
Y_{\rm cyl}(x) & = & \int_0^{r=x\,R_{\rm 500}} d\vx' \,
 \int_{-\infty}^{+\infty} dz'
 \sigma_{\rm T} n_{\rm e} (\vrv')
 \frac{k_{\rm B}T_{\rm e}(\vrv')}{m_{\rm e} c^2} .
\label{eq:Ydefs}
\end{eqnarray}
These equations describe the spherical and cylindrical integrals of
pressure, respectively.  The vector centred on the cluster $\vrv'$ is
decomposed into a vector on the plane of the sky, $\vx'$, and a
vector normal to this plane (given by the component $z'$),
$\vrv' = (\vx',z')$.
With this, we use the $Y (x)$ vs.\ $M_{\rm500}$
relations in \citet{arnaudetal10}, i.e.,
\begin{eqnarray}
Y_{\rm sph} (x) & = & Y_{\rm 500} I(x), \\
\label{eqs:yarray1}
Y_{\rm cyl} (x) & = & Y_{\rm 500} J(x),
\label{eqs:yarray2}
\end{eqnarray}
in which
\[
Y_{\rm 500} = 1.38\times 10^{-3} \,E^{2/3}(z)
 \,\biggl( \frac{M_{\rm 500}}{B_{\rm 500}}\biggr)^{\alpha_{\rm Y}} \, \times \]
\begin{equation}
\phantom{xxxxxxxxxxxxxxx}
\biggl( \frac{D_{\rm A}(z)}{500\,h_{\rm 70}^{-1}{\rm Mpc}}
 \biggr)^{-2}h_{70}^{-1}
\,\,{\rm arcmin}^2,
\label{eq:y500}
\end{equation}
and $I(x)$ and $J(x)$ are functions expressing the spherical/cylindrical
integrals of pressure around the cluster's centre, respectively. The factor $h_{\rm 70}$
denotes the Hubble reduced parameter in units of $70$\,km\,s$^{-1}$\,Mpc$^{-1}$.  
An observer's angular aperture $\theta$ translates into an
effective cluster radius $r = \theta\,D_{\rm A}(z)$, with $D_{\rm A}(z)$ the
angular diameter distance to redshift $z$. Note that, as expressed above,
the cylindrical case
considers a full integration along the line of sight up to a given aperture on
the plane of the sky, as is the case for real observations. The spherical case,
instead, integrates out to a given radius in all directions,
and differs from the
cylindrical case in a geometric factor. The constants in Eq.~\ref{eq:y500} are
$B_{500} = 3 \times 10^{14}\,h_{70}^{-1}\Msolar$ and $\alpha_{\rm Y} =
1.78$, while $E(z)$ is the Hubble parameter normalized to its current value,
\begin{equation}
E(z) = \sqrt{\Omega_{\rm M} (1+z)^3 + \Omega_{\Lambda}}.
\end{equation}
The functions $I(x)$ and $J(x)$ depend on the particular model adopted for the
pressure profile, which, in our work, is taken to follow the universal scaling
provided in Eqs.~11 and 12 of \citet{arnaudetal10}, for which $I(x=1)=0.6541$
and $J(x=1)=0.7398$. As will be addressed in Sect.~\ref{sec:robust}, results
do not change significantly when adopting different choices for
\citet{arnaudetal10} type pressure profiles, 
but their uncertainty is dominated by our ignorance of the gas density profile
in the clusters' outskirts.

\begin{figure}
\centering
\includegraphics[width=8.cm]{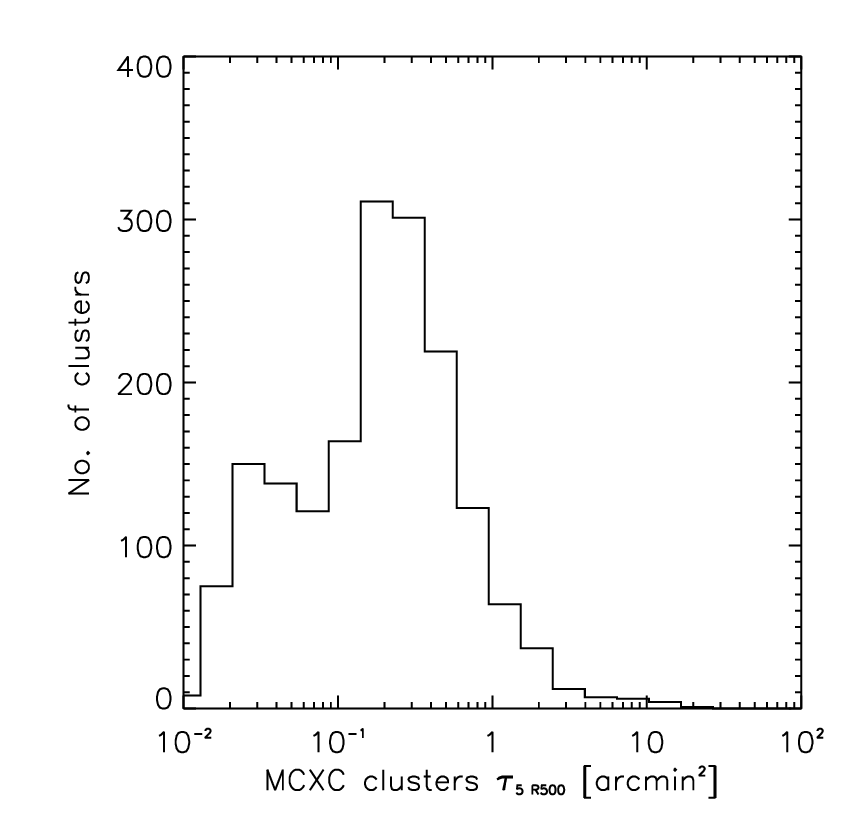}
\caption[fig:histtaus]{Histogram of predicted values of $\tau_{5\,R_{500}}$,
the cylindrical optical depth times solid angle out to $R = 5 \, R_{500}$.}
\label{fig:histtaus}
\end{figure}

In the isothermal case, the clusters' optical depth integrated up to
$x R_{\rm 500}$ is equal to the corresponding $Y(x)$,
modulo a $k \bar{T}/(m_{\rm e} c^2)$ factor, where $\bar{T}$ is the average spectroscopic
temperature measured in a fraction of the volume enclosed by $R_{\rm 500}$.
To account for this, we use the $\bar{T}$--$M_{\rm 500}$ relation
given in \cite{arn2}:
\begin{equation}
\label{eq:TvsM}
E(z)\, M_{500} = A_{500} \left(\frac{k \bar{T}}{5\,{\rm keV}} \right)^{
  \alpha_{\rm T}},
\end{equation}
with $\alpha_{\rm T} = 1.49$ and $A_{500}=4.10 \times
10^{14}\,h_{70}^{-1}\,\Msolar$. In this simple case, it is possible to
derive an expression for the optical depths $\tau_{\rm sph} (x) = \tau_{\rm 500}
I(x)$ and $\tau_{\rm cyl} (x) = \tau_{500} J(x)$, with
\[
\tau_{\rm 500} = 1.3530\,\times 10^{-5}\, E^{2/3-1/\alpha_{\rm T}}(z) \,\times
\]
\begin{equation}
\phantom{xx} \biggl( \frac{D_{\rm A}(z)}{500\,h_{\rm 70}^{-1}
 {\rm Mpc}}\biggr)^{-2} \,\biggl( \frac{511\,\rm{keV}}{k \bar{T}}\biggr)
 \,\biggl(\frac{M_{\rm 500}}{C_{\rm 500}} \biggr)^{\frac{1}{\alpha_{\rm
      Y}}-\frac{1}{\alpha_{\rm T}}}\,h_{70}^{-1}\,\,\rm{arcmin}^2,
\label{eq:tau500}
\end{equation}
and 
\begin{equation}
C_{\rm 500} = \biggl( B_{\rm 500}^{1/\alpha_{\rm Y}}
 A_{\rm 500}^{-1/\alpha_{\rm T}}\biggr)^{\alpha_{\rm Y}\alpha_{\rm T}/
 (\alpha_{\rm T}-\alpha_{\rm Y})}.
\label{eq:c500}
\end{equation}
For the non-isothermal case, we use the average $T(x)/\bar{T}$ scaling obtained
from the middle panel of figure~3 of \citet{arnaudetal10}. This scaling is only
applied for $x<1$, and divides the pressure profile to obtain the density, which
becomes the integrand in $I(x)$ and $J(x)$. Since the $T(x)/\bar{T}$ scaling has
only been measured for $r<R_{\rm 500}$ ($x<1$), a different approach is followed
for $r>R_{\rm 500}$.  In this radius range, we express the electron density in
terms of the pressure and the entropy, $K(r)= k T(r) /n_{\rm e}^{2/3}(r)$, and
adopt the relation $K(r)\propto r^{0.5}$. This defines the scaling of density
versus radius that enters the outer parts ($r>R_{\rm 500}$) of the integrals
$I(x)$ and $J(x)$. This constitutes our best guess of the radial dependence of
density in clusters, although in Sect.~\ref{sec:robust} we discuss the
motivation and limitations of this approach.  Fig.~\ref{fig:histtaus} displays a
histogram of the estimated values of the cylindrical optical depth integrated
out to a radius of $R = 5 \, R_{500}$, $\tau_{5\,R_{500}}$, for the
non-isothermal case.

In the analyses described below, we exclude clusters located at less than
$1.5\,$FWHM from point sources detected at more than $5\,\sigma$ in any of the
single frequency \Planck\ maps. 
This is done in order to remove any spurious signals caused by point sources
associated with clusters.
We also mask clusters lying in regions with high
Galactic emission, and with estimated masses below $10^{13}$\,\Msolar. This
leads to a basic mask that leaves 1405 clusters on the sky (out of the initial
1743 clusters). However, the 2D-ILC has its own (and slightly more conservative)
mask, which leaves only 1321 clusters for analysis on this map.

\begin{figure}
\centering
\includegraphics[width=6.cm,height=9.cm, angle=90]{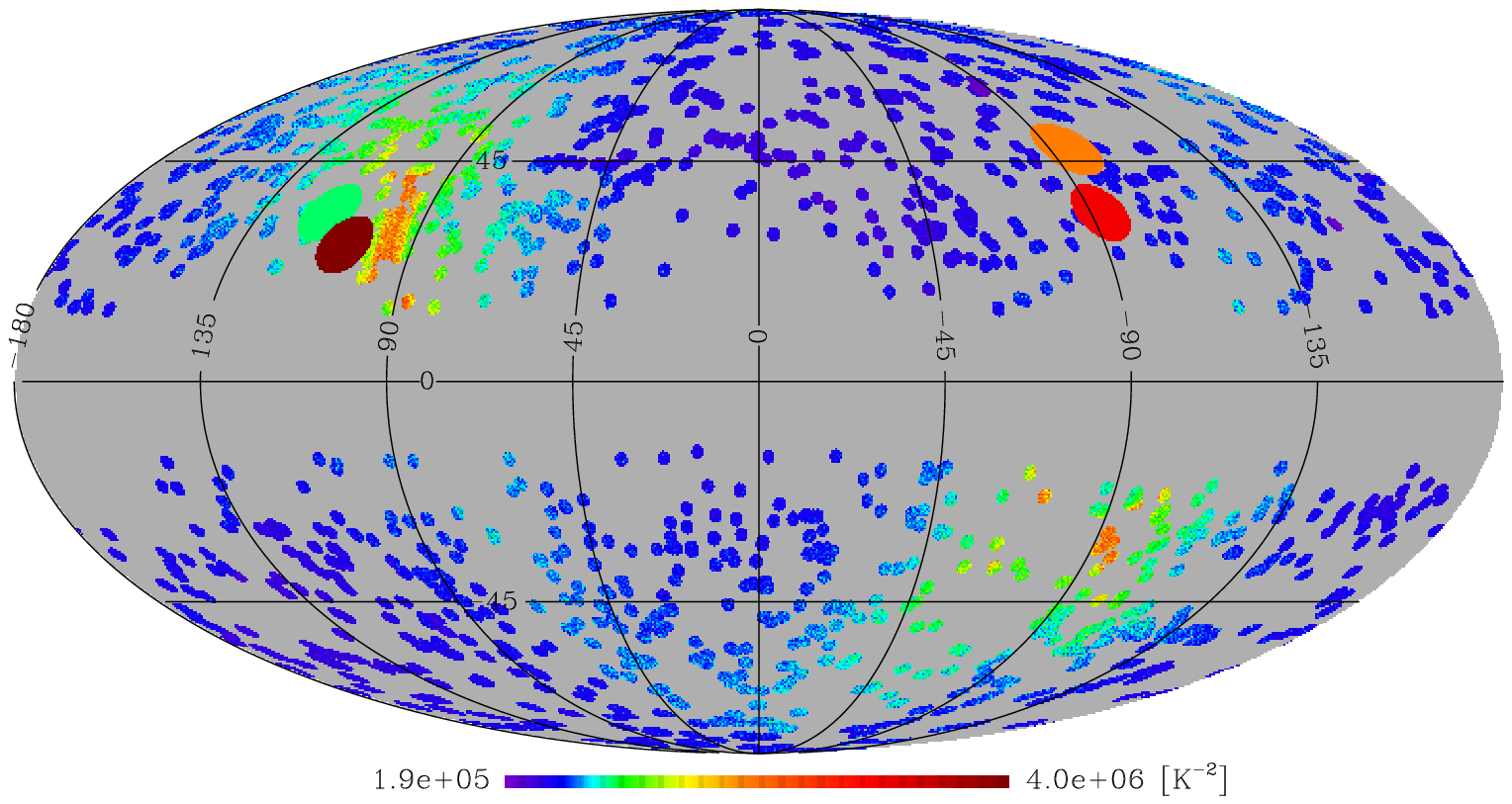}
\caption[fig:clunoise]{Locations of MCXC clusters outside the masked region.
  The colour scale indicates inverse noise squared at the positions of the
  MCXC clusters. The large coloured circles
  indicate directions for various dipole determinations: HFI dipole from MCXC
  cluster locations (green); HFI dipole from shifted positions (brown); CMB
  dipole (orange); \citet{kashlinsky10} dipole (red). The HFI dipole from MCXC cluster locations is compatible with CMB and Galactic residuals.}
\label{fig:clunoise}
\end{figure}

In Fig.~\ref{fig:clunoise} we show the spatial distribution of the surviving
clusters.  The spatial distribution is quite uniform, except for some areas
where deeper X-ray observations have allowed for more detections.

\subsection{Simulations}
\label{sec:simul}
In order to test and disentangle the effects of instrumental noise, CMB,
Galactic and extragalactic foregrounds on our results, we make use of
simulations. Specifically, we use
\begin{enumerate}
\item[a)]Simulations of clusters with similar characteristics to the MCXC
  sample. We simulate SZ clusters at the actual MCXC clusters' locations, using
  the \citet{arnaudetal10} pressure profile, and the cluster mass and size
  values obtained from X-ray observations. Clusters are assumed to be
  isothermal.  The Comptonization $Y$ parameter of each cluster is
  then computed using the cluster mass, as outlined in
  Sect.~\ref{sec:clusters}. We include in the simulations a scatter in the
  parameters of the scaling relations with a normal distribution, which results
  in an averaged scatter in the $Y$ parameter of 10\,\%. We generate a set of
  1000 such simulations at the effective frequencies in
  Table~\ref{tab:hfi_table} in order to assess the effect of that scatter on our
  derived results.
\item[b)] CMB realizations for the current {\it WMAP}-7 best-fit model (1000 of
  them).
\item[c)] 1000 noise realizations with noise variance estimated from
 the difference between the first and second halves of the \Planck\ rings for
 a given position of the satellite and divided by \Planck's appropriate hit map
 (or exposure map).
 In doing so, we disregard noise correlations between pixels.
 Such simulations take into account the non-uniform sky coverage. The
 (non-uniform) noise level in the direction of the MCXC clusters is visible in
 Fig.~\ref{fig:clunoise}.
\end{enumerate}


Additionally, we make use of the simulations developed by the
\Planck\ collaboration \citep[the Planck Sky Model,][]{jacques12}
in order to assess the Galactic contribution to the
bulk flow measurement on the whole sky. These contain our current best
knowledge of the diffuse Galactic component (``PSM diffuse'').
Since \Planck\ is most sensitive at
frequencies above 100\,GHz where the dust emission dominates,
we only consider this component. We use model number 7
of \citet{Finkbeiner99} which extrapolates the IRAS 100\,$\mu$m data
to lower frequency using a modified blackbody frequency
dependence and a spatially varying dust grain equilibrium
temperature based on the DIRBE 240/140\,$\mu$m maps.

\section{Statistical methods}
\label{sec:filters}
Here, we outline two different statistical approaches implemented when searching
for the kSZ signal in \Planck\ data. The first one (the aperture photometry
method) is applied on a cluster by cluster basis and makes no assumption
about the gas distribution within a given radius
where most of the cluster signal must be generated. It nevertheless has to make
assumptions about the amount of signal that is present in the outskirts. This
method is quick and easy to implement, in particular when performing checks for
systematic errors. The second approach
(the unbiased Matched Multi-frequency Filter)
makes use of information related to the expected spatial distribution of gas and
the scale dependence of all sources that can be regarded as noise (including the
CMB).  The use of these two, different filters is motivated by the consistency
requirement of having independent algorithms that should provide compatible kSZ
estimates.  The unbiased Multi-frequency Matched Filter was implemented
independently by four different teams, two implementations being on square
patches centred on clusters, and the other two on the whole celestial sphere.
The first two implementations target the determination of {\em individual\/}
cluster peculiar velocities, while the latter two are specifically developed
to derive local bulk flows, since they focus on the
dipole pattern of the kSZ effect in clusters on the celestial sphere.
Implementations targeting clusters individually allow constraints to be placed
on the mean cluster velocity (or monopole), the rms or variance of cluster
peculiar velocities, and also on local bulk flows, i.e., the kSZ dipolar
pattern, as will be shown below.
Results from different codes confirm the robustness of our results.

\subsection{The aperture photometry method}

The aperture photometry method (hereafter AP) computes the average temperature
within an input radius $R$, and subtracts from it the average temperature
computed in a surrounding ring of inner and outer radii $R$ and $f R$ ($f > 1$),
respectively \citep[see, e.g.,][]{2004MNRAS.347..403H}. In this work we use
$f=\sqrt{2}$, so that the outer ring has the same area as the inner circle. This
is a compromise between having too few pixels in thin rings (yielding noisy
estimates of the average) and being insensitive to local background fluctuations
of typical size just above $R$ (that are washed out for choices of
$f$ which are too large).
This filter constitutes a simple approach to enhance the signal coming
from a region of size $R$ against the background. In our analyses, the AP
procedure was applied in the direction of each galaxy cluster, separately in
each HFI frequency map. When looking at clusters, the filter scale $R$ was taken
equal to either $k\theta_{500}$ ($k$ times the angle subtended by the radius at
which the cluster's density equals 500 times the critical density) or the FWHM
of the beam, depending on whether the object is resolved
($k \theta_{500} > {\rm FWHM}$ )
or not.  Values of $k$ ranging from $0.25$ up to $2$ showed that the
strongest constraints were obtained for $k=0.25$.  Yet smaller values of $k$ do
not yield significant differences, since for such low $k$ practically all
clusters become unresolved. We shall describe results with $k=0.25$, unless
other values are explicitly quoted.

The subtraction of the average temperature in the outer ring from the average of
the inner circle also removes some fraction of the object's flux, which must be
accounted for. This results in a correction factor of the order of 12\,\% if
clusters are smaller than the beam size (${\rm FWHM} > \theta_{500}$).  If,
instead, ${\rm FWHM} < \theta_{500}$, this correction must make use of some
model for the cluster gas density profile, as we address next.

In order to obtain velocity estimates, it is necessary to divide the filter
outputs (in units of temperature) by the CMB temperature monopole and a
prediction of the clusters' optical depths integrated out to the radial
aperture. This then provides an estimate of the entire cluster's peculiar
velocity. The amount of kSZ signal that is subtracted by the removal of the
outer ring has to be accounted for by the same model which provides the optical
depth versus aperture radius. As explained in more detail in
Sect.~\ref{sec:robust}, we use the REXCESS observations provided in
\citet{arnaudetal10} to infer an analytic fit to the average density profile in
clusters within $R_{500}$, and use arguments on the behaviour of gas entropy at
$r > R_{500}$ to extrapolate the density profile at larger radii. The adopted
model for density provides velocity amplitudes that are about 28\,\% higher
than those obtained under the assumption of isothermal clusters, although we
expect clusters to be closer to our adopted profile than to an isothermal
one. Nevertheless, it is our ignorance of the clusters' density profiles which
drives most of the uncertainties in the velocity constraints.

When testing for systematic effects, this same filter can easily be applied at
displaced positions on the sky (that is, positions on the sky separated from the
real cluster positions by a known angle). In the absence of sources and
clusters, the average of the outputs of this filter at those displaced positions
should be compatible with zero, and their rms provides an error estimate for the
AP filter output at the real cluster's position.

\subsection{The unbiased Multifrequency Matched Filter (uMMF) on patches}
\label{sec:uMMF}

The unbiased Multifrequency Matched Filter \citep[][hereafter uMMF]{herranz05,mak11} 
is a linear multi-frequency filtering technique
that is specifically tailored to deal with signals that have the same spatial
template but different frequency dependence. A good example of this is the
imprint on CMB photons caused by the tSZ and kSZ effects.  The uMMF can be
considered as a modification of the Multi-frequency Matched Filter \citep[MMF,
][]{herr02,schaf06,melin06} that optimally enhances one of the two superimposed
signals while cancelling out the other. As demonstrated in \cite{herranz05}, it
is possible to devise a uMMF that detects the tSZ effect and estimates its
intensity without the bias produced by the kSZ effect, or a different uMMF that
extracts the kSZ signal and removes the bias caused by the tSZ effect. In this
paper we are interested in the latter option. In thermodynamic units, the uMMF
for optimal detection and estimation of the kSZ effect is given, in Fourier
space, by
\begin{equation} 
\label{eq:UMMF}
\tens{\Phi} = \frac{1}{\Delta} \tens{P}^{-1} \left( - \beta \vec{F} + \alpha
\vec{\tau} \right),
\end{equation}
where the constants $\alpha$, $\beta$ and $\Delta$ are given by
\begin{eqnarray}
\label{eq:abcd}
\alpha &=& \int d\vec{k} \  \vec{F}^{\rm T}    \tens{P}^{-1} \vec{F},
 \nonumber \\
\beta  &=& \int d\vec{k} \  \vec{\tau}^{\rm T} \tens{P}^{-1} \vec{F},
 \nonumber \\
\Delta &=& \alpha \gamma - \beta^2, \\
{\rm with\ }\gamma &=& \int d\vec{k}\ \vec{\tau}^{\rm T}\tens{P}^{-1}\vec{\tau},
 \nonumber
\end{eqnarray}
and where $\tens{P}$ is the cross-power spectrum matrix of the
\emph{generalized noise\/} (CMB plus foregrounds plus instrumental noise),
$\vec{\tau}=[\tau_{\nu}\left(\vec{k}\right)]$ is a vector containing the spatial
profile of the optical depth of the cluster (obtained from the universal profile
of \citet{arnaudetal10} after dividing by the {\em constant\/} temperature and
convolving by the beam that corresponds to each channel) and $\vec{F}=[f_{\nu}
  \tau_{\nu}\left(\vec{k}\right)]$ is the vector obtained by multiplying,
element by element, the profile $\tau$ by the well-known tSZ frequency
dependence $f_{\nu}$. Thus this method observes only isothermal profiles for
clusters, an assumption which results in a roughly 5\,\% low bias in the radial 
velocity amplitude, as shown in
Sect.~\ref{sec:robust}. The power spectrum matrix $\tens{P}$ is computed from
real data in patches surrounding the sources. Once we have the filters
(Eq.~\ref{eq:UMMF}), the filtered image
\begin{equation}
\label{eq:ffield}
w \left( \vec{x} \right) = \sum_i \int d\vec{y} \ d_i\left(\vec{y}\right) \Phi_i
\left( \vec{x} - \vec{y} \right)
\end{equation}
is optimal for the detection of the kSZ effect and has no trace of the tSZ
effect. In this equation, $d_i$ represents the unfiltered map in the $i$th
frequency channel. The filters are normalized so that $w(\vec{x}_0)$, where
$\vec{x}_0$ is the location of the centre of the cluster, is an unbiased
estimator of the kSZ signal due to the cluster. An estimation of the error of
this is given by the square root of the variance $\sigma_w^2(\vec{x}_0)$, which
can be directly obtained from the filtered map or calculated through
\begin{equation}
\sigma_w^2 = \frac{\alpha}{\Delta},
\label{eq:errmmf}
\end{equation}
where $\alpha$ and $\Delta$ have the same meaning as in Eq.~\ref{eq:abcd}.

In this work, two different uMMF implementations on square patches were used,
confirming the robustness
of the results. As mentioned above, the two implementations
assume that the spatial distribution of the thermal and kinetic signals follows
the pressure, for which we adopt the universal pressure profile from
\citet{arnaudetal10}. For each cluster, the profile is scaled with
$R_{500}$. The two implementations mainly differ in the size of the patches used
to estimate the background around each cluster and the details of the
cross-power spectrum estimation on the data. In both cases, we apply the
resulting uMMF to the patches and directly obtain the estimated velocity at the
centre of the filtered patch. The rms of the filtered patch outside the centre
region occupied by the cluster gives an estimation of the velocity
error. This leads to a good statistical match between velocities measured by the
two implementations, but not detailed agreement on a cluster by cluster basis.
This is expected, since the peculiar velocity estimate per cluster is dominated
by noise, and the actual noise component present in each estimate is dependent
on the details of each specific implementation.  The method provides estimates 
of the kSZ flux integrated over the cluster profile;
these are translated into velocity estimates for each cluster after dividing
by the integrated optical depth. Errors 
in these estimates of the optical depth will lead to errors in the velocity
estimates, but, as will be discussed below, these should have little
impact on estimates of ensemble quantities like
velocity averages, dipoles, and rms estimates. More important error offsets (at
the 5--25\,\% level) are expected from inaccuracies associated with the gas
density profile in clusters (see Sect.~\ref{sec:robust} for details).

The uMMF method may also be applied to a single map (as is the case for the
2D-ILC map), a situation in which the uMMF becomes a simple {\it Matched Filter}
(MF).

\subsection{Constraining kSZ-induced rms in AP/uMMF measurements}
\label{sec:spatial_app}

Since the signal-to-noise ratio for the kSZ on a typical MCXC cluster
is very small (see, e.g., \citep{nabila01} 
for forecasts on the analysis of bulk flows and the kSZ effect), we attempt to
set constraints on the kSZ signal by performing statistical analyses
on the entire MCXC
cluster sample. We next describe our approach to set constraints on the kSZ
contribution to the variance of a set of AP/uMMF outputs. This method relies on
the fact that the kSZ contribution to our AP or uMMF measurements is
uncorrelated with the dominant noise sources (CMB residuals, instrumental noise
and dust emission). In practice this reduces to searching for a kSZ-induced
{\em excess\/} variance, and this demands a good knowledge of the variance
of the variance of AP/uMMF measurements, as we next describe.

In this work, we set constraints on the variance of the cluster
radial peculiar velocities by looking at the variance of our filter outputs. 
For both AP and uMMF filters, the data consist of a radial velocity component
($v_{||}$) plus a number of noise sources (CMB
anisotropies, instrumental noise, Galactic and extragalactic emissions not
associated with the clusters, etc., here denoted by $n$):
\begin{equation}
d_i = v_{||,i} + n_i,
\end{equation}
where $i$ is the index of each cluster in our sample of
$N=1405$ objects ($N=1321$ under the strict mask). Sufficiently distant from
such locations, the data $d$ and the noise $n$ coincide.

\begin{figure*}
\centering
\includegraphics[width=18.cm]{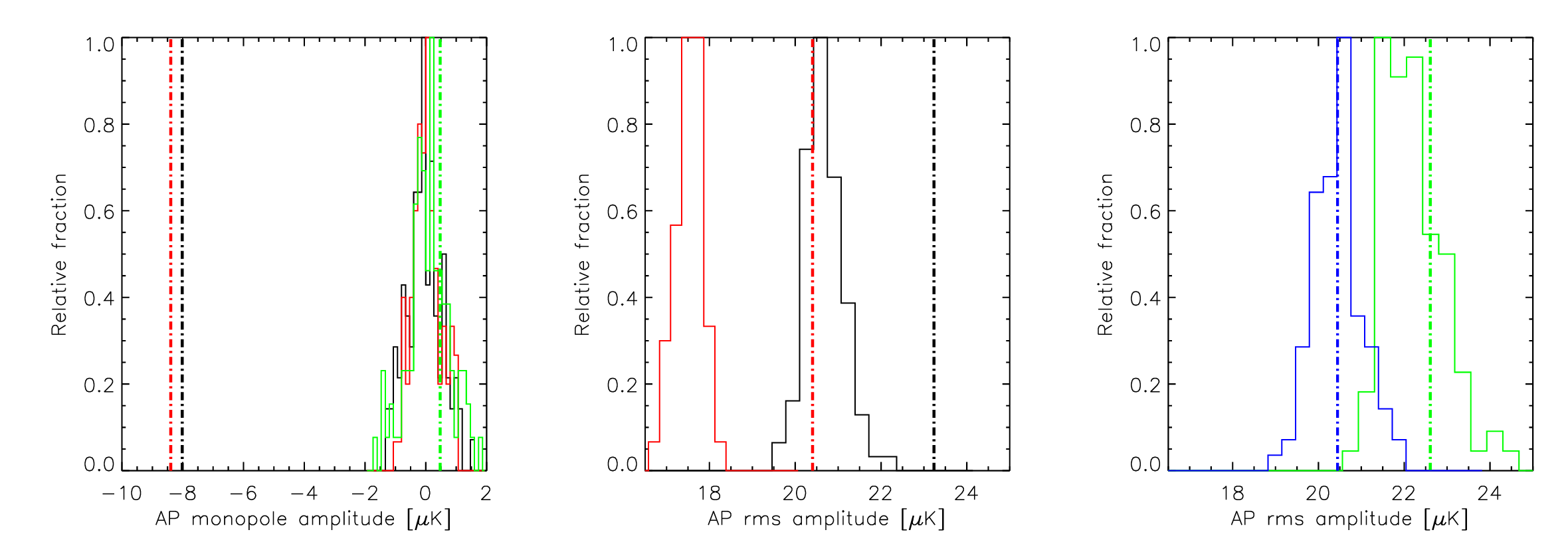}
\caption[fig:AP_hist]{Colour coding common for all panels: black corresponds to
  the 100\,GHz channel; red to 143\,GHz; green to 217\,GHz; and blue to the
  2D-ILC map. Histograms are obtained from AP output at 100 displaced
  positions, while vertical dot-dashed lines correspond to AP outputs obtained
  on the clusters.  The left panel refers to the AP output monopole/average,
  whereas the middle and right one display histograms built upon rms estimates.
  }
\label{fig:AP_hist}
\end{figure*}

We therefore want to measure $ \sigma_{\rm kSZ}^2 = {\rm Var}(v) = \sum_{i=1}^N
{(v_{||,i}-v_{\rm m})^2/(N-1)}$, where $v_{\rm m}$ is the mean velocity. The
variance of the data at cluster locations reads:

\[
\sigma_{\rm clusters}^2= {\rm Var}(d)=  \frac{\sum_{i=1}^N
 (v_{||,i}-v_{\rm m})^2}{N-1} \,+  
\]
\begin{equation}
\phantom{xxxxxxxxx}
\frac{\sum_{i=1}^N (n_i-n_{\rm m})^2}{N-1} +
 \frac{\sum_{i=1}^N 2(v_{||,i}-v_{\rm m})(n_i-n_{\rm m})}{N-1},
\label{eq:err0}
\end{equation}
where $n_{\rm m}$ is the mean noise. Assuming that noise terms and cluster
velocities are uncorrelated, for the large number of clusters considered here
we expect the last term to be subdominant with respect to the first two.  We
therefore interpret the variance of the data in Eq.~\ref{eq:err0} as the sum of
variances of the velocity and noise terms:
\begin{equation}
\sigma_{\rm clusters}^2 = \sigma_{\rm kSZ}^2 + \sigma_{\rm noise}^2.
\label{eq:rms1}
\end{equation}
We then estimate the noise variance by looking at 100 locations near to
clusters where noise properties will be similar.
By doing so in each of the 100 nearby locations, we can obtain 100 estimates of
the noise variance and hence construct a histogram representing its probability
distribution.  Note that this distribution is, in general, not Gaussian. An
example from the derived noise rms distributions from the AP and uMMF filters
are provided in the right panels of Figs.~\ref{fig:AP_hist} and
\ref{fig:uMMF_spatial}, respectively.

Given the probability distribution of the noise and our measured variance at
the cluster locations, we can deduce upper limits for the clusters' velocity
variance. Because the variance velocity term is positive and added in quadrature
to the noise, as in Eq.~\ref{eq:rms1}, we can conclude that at 95\,\%
confidence limit (C.L.) the
kSZ contribution should be below the following value:
\begin{equation}
\sigma_{\rm kSZ}^2 (95\,\%) = \sigma_{\rm clusters}^2
 - \sigma_{\rm noise}^2 (5\,\%).
\label{eq:rms2}
\end{equation}
Here $\sigma_{\rm noise}^2 (5\,\%)$ is the noise variance amplitude limiting
the lowest 5\,\% of the noise variance distribution. In practice, since our
histogram is based upon 100 different variance estimates, we write:
\begin{equation}
\sigma_{\rm kSZ}^2 (95\,\%) = \sigma_{\rm clusters}^2
 - \sigma_{\rm noise}^2 ({\rm 5th}).
\label{eq:var_diff}
\end{equation}
In this equation, $\sigma_{\rm noise}^2 ({\rm 5th})$ denotes the fifth lowest
AP/uMMF output variance estimate picked from the 100 variance estimates making
the histogram. The quantity $\sigma_{\rm kSZ} (95\,\%) $ constitutes our limit
of the peculiar radial velocity rms at the 95\,\% confidence level. Such a
constraint is, however, obtained from histograms built upon only 100
measurements. Using the histograms built upon the filter outputs in blank
positions we have run Monte Carlo simulations and studied the uncertainty on the
lower 5\,\% limit on $\sigma_{\rm noise}^2 $ if estimated as outlined above. 
We find that these uncertainties lie
typically below the 5\,\% level when only a subsample of 100 clusters are used,
and below 1\,\% when using the entire cluster sample (around 1400 objects).


\subsection{All-sky bulk flow with the unbiased Multifrequency Matched Filter (uMMF)}
\label{sec:bfmak}

In order to evaluate the bulk flow in \Planck\ data, we adopted the procedure
previously used on simulations for forecasting \Planck's performance, as
detailed in \cite{mak11}.  We briefly summarize the approach here, and we refer
the reader to \cite{mak11} for further information. In this procedure, we do
not focus on the velocities of individual clusters,
but rather fit for both amplitude and
direction of the bulk velocity for the whole ensemble. The first step of the
procedure consists of filtering the observed maps with a whole-sky version of
the uMMF that adopts the universal pressure profile from \citet{arnaudetal10}
convolved with the beam profile of a given frequency 
as in the case of uMMF on patches. For this whole-sky version, instead of
designing the filters individually for each cluster (that match its
size), we construct one single filter for all clusters, with a characteristic
scale of $\theta_{500} = 8\arcm$. This choice is motivated by the fact that the
average size of the MCXC sample is $ \langle \theta_{500} \rangle=7.8\arcm$.
The filtering procedure combines maps at different frequencies into a cleaned
temperature map that is then used to fit for the real spherical harmonic
coefficients of the dipole terms ($v_{{\rm x}}$, $v_{{\rm y}}$ and
$v_{{\rm z}}$).  In doing so, we adopt the effective frequencies listed in
Table~\ref{tab:hfi_table}.

We fit the dipole terms of the filtered map using a
weighted least squares fit that is based on the HEALPix \citep{gorski2005}
IDL procedure
{\tt remove\_dipole}. We weight the central pixels of the clusters that are
outside the masked region with inverse noise variance weights,
i.e., $W_i=1/\sigma_{N,i}^2$ where $\sigma_{N,i}$ is the $i$th pixel
noise variance calculated from filtered
CMB and noise realizations.  In such realizations, the instrumental noise is
white and spatially uncorrelated, with a variance estimated from the half ring
maps and divided by the hit maps appropriate for \Planck\ data in a given pixel.
We convert the dipole from temperature units ($\vec{\delta T}$)
to velocity ones ($\vec{v}$) by means of a conversion matrix
$\tens{M}$ previously constructed using simulations 
of clusters with the same characteristics as the sample in hand, i.e.,
$\vec{v}=\tens{M}\, \vec{\delta T}^{\rm T}$.  We then evaluate the error on
the bulk flow dipole coefficients by fitting dipoles to sets of simulations
of CMB anisotropies, instrumental noise, and the tSZ effect.
In order to do this, we assume that these
sources of errors are uncorrelated, but we consider potential correlations in
the errors for the dipole coefficients.  The magnitude of the dipole velocity
follows a $\chi^2$ probability distribution with three
degrees of freedom that can be computed as follows:
\begin{equation}
\chi^2 = (\vec{v}-\vec{v}_{\rm m})^{\rm T} \tens{N}^{-1}
 (\vec{v}-\vec{v}_{\rm m}), 
\label{eq:chi}
\end{equation}
where $\vec{v}$ is the variable of the distribution, $\vec{v}_{\rm m}$ is
the mean of the velocity as estimated from simulations, and $\tens{N}$ is the
noise covariance matrix under consideration. We compute the covariance matrix
by passing 1000 simulations of the noise components 
(CMB and/or instrumental noise and/or tSZ) through our pipeline
and performing the dipole fit on them. The scatter in dipole coefficients
provides an estimate of the noise correlations between the dipole directions,
i.e., $\tens{N}=\left \langle \vec{v}^{\rm noise}\vec{v}^{\rm T,noise} \right
\rangle$.  The 95\,\% upper limit is then determined to be the velocity at which
$\chi^2 = 7.8$, which is the 95\,\% upper limit for a $\chi^2$ distribution with
three degrees of freedom. Errors on the bulk flow measurements are therefore
computed on the basis of simulations, and include sources of uncertainties in
the mass--observable relation as well as in the residual contamination
from thermal SZ, CMB and instrumental noise.

We verified that the most stringent constraints are obtained when only the
central pixel in the direction of cluster's location is considered after
filtering the map (and since the data have already been matched filtered,
applying an aperture would not be valid).
For the frequency maps used, we present results based on the
four lowest HFI channels, i.e., 100, 143, 217, and 353\,GHz. We verified that
extending the analysis to the two highest LFI channels, 44 and 70\,GHz
\citep[as in][]{mak11} gives consistent results, but
does not significantly improve the constraints.

Finally, before we end this section we stress the difference existing
between methods working on patches (such as the ones described in previous
sections) and this method, which works on the entire celestial sphere.
The former methods are insensitive to scales larger than the patch size,
unlike all-sky methods for which filtering is implemented on all angular
scales.

\section{Analyses and results}
\label{sec:results}

This section contains the entire set of results in this paper, and is divided
into three sections. The first one (Sect.~\ref{sec:kszmonorms}) addresses
the constraints on the kSZ monopole and the rms of the cluster peculiar
velocities, and is divided into two parts, devoted to the results obtained
with the AP and uMMF filters. The second section
(Sect.~\ref{sec:bulkflows}) studies the constraints on bulk flows, and is
divided into four parts. The first three outline the constraints obtained with
the three filters defined in Sect.~\ref{sec:filters}. The fourth part revisits
the specific filter implemented by \citet{kashlinsky08}. Finally, the third
section (Sect.~\ref{sec:inhomo}) sets constraints on inhomogeneous
cosmological models.

As mentioned above, the MCXC catalogue consists of a sample of massive clusters
of galaxies hosting large reservoirs of hot gas, where the CMB is distorted by
means of the tSZ effect \citep{planck2011-5.2a,planck2011-5.2b}. We target this
cluster sample in our attempt to detect or put constraints on peculiar motions
in the local Universe. Provided that the expected typical correlation length for
peculiar velocities is of order 20--40\,$h^{-1}$\,Mpc, we do not expect MCXC
clusters to show any significant level of coherent motion. Clusters in the MCXC
catalogue cover a wide redshift range and distances between them are far larger
than the velocity coherence length. However, in the last few years there have
been several works \citep{kashlinsky08,kashlinsky10,abatefeldman11} claiming the
presence of extremely large-scale bulk flows, and such scenarios can be tested
with the MCXC cluster sample.

The linear theory expectation for the line of sight peculiar velocity
variance can easily be derived from the continuity equation in terms of the
matter density power spectrum:
\begin{equation}
\sigma_v^2 (M)= \frac{1}{3} \int \frac{d\vk}{(2\pi)^3} H^2(z) \biggl| \frac{d{\cal D}_{\delta}}{dz}\biggr|^2 \frac{P_{\rm m}(k)}{k^2} |W(kR[M])|^2.
\label{eq:v2}
\end{equation}
This equation refers to the radial velocity rms of a cluster of mass $M$.
The symbol $W(kR[M])$ corresponds to the Fourier window function associated
with a top hat filter of size given by the linear scale corresponding to the
cluster mass $M$, $R = [3\rho_{\rm m}/(4\pi)]^{1/3}$, where $\rho_{\rm m}$ the
average matter comoving density. The linear matter power spectrum is given by
$P_{\rm m}(k)$, ${\cal D}_{\delta}(z)$ denotes the density linear growth
factor, and $H(z)$ corresponds to the Hubble parameter. The rms inferred from
this expression at $z=0$ is about 230\,km\,s$^{-1}$ for a
$2\times 10^{14}\,h^{-1}$\,M$_{\odot}$ cluster.
Note that the linear theory $\Lambda$CDM
predictions for the peculiar velocities of the clusters are supported by the
output of numerical simulations, although clusters and groups may show biases
depending on their environment, with higher velocities in overdense regions,
and non-Maxwellian tails \citep{shethdiaferio01}. In any case, the velocity rms
expectation, when translated into temperature fluctuations via
Eq.~\ref{eq:kSZ1}, yields too small a signal to be detected on an individual
basis (typical velocity estimate errors lie at the level of thousands of
km\,s$^{-1}$). This motivates a statistical approach which targets ensemble
properties of the cluster peculiar velocities.

We first apply the AP filter to raw HFI frequency maps. Since this filter is
applied independently on different frequency bands, it permits us to track
separately the impact of other contributions like the tSZ effect or dusty point
sources. When imposing constraints on the cluster peculiar velocities, we also
use the cleaned 2D-ILC CMB map. Likewise, the use of the uMMF on raw HFI
frequency maps allows us to test for dust contamination, tSZ spectral leakage,
or errors in the cluster size determination. However, the most restrictive
velocity constraints are usually obtained from the 2D-ILC map.   

\subsection{Constraints on kSZ monopole and rms}
\label{sec:kszmonorms}

In this section we present the constraints that \Planck\ sets on the amplitude
of the peculiar velocity monopole (average) and rms in our cluster sample.

\subsubsection{Constraints from the AP filter}

For all MCXC clusters outside the joint HFI mask, an AP estimate is provided for
each frequency band. In order to test for systematic errors, this filter is
applied not only at the cluster positions, but also on 100 other positions
displaced from the real ones in either Galactic or equatorial latitude. For each
position, the amount of displacement is an integer multiple of three times the
FWHM of the beam corresponding to the frequency map under study.
For a {\em fixed\/} cluster {\it i},
the AP output rms from the displaced positions provide
an estimate of the rms of the AP output at the $i$th cluster's real position
($\sigma_{{\rm AP},\,i}$).  For each displacement, we consider only positions
outside the effective mask, and compute both the average (or monopole) of the AP
outputs, and their rms, as we run over different clusters. The left panel of
Fig.~\ref{fig:AP_hist} displays the histograms of the AP outputs for the 100
displaced positions at 100\,GHz (black solid line), 143\,GHz (red solid line)
and 217\,GHz (green solid line). The vertical, dot-dashed lines correspond to
the AP outputs at the real cluster positions (zero angular displacement).

Note that for each displacement some of the real MCXC cluster positions
may fall in masked pixels. In those cases, the AP filter outputs are
ignored, that is, for each set of displaced positions, the number of
{\it useful\/} AP estimates equals the number of clusters under consideration
minus the number of times that the ``displaced'' AP filter centres falls on a
masked pixel. We hence do not consider AP outputs whenever the filter is centred
on masked pixels.  The left panel of Fig.~\ref{fig:AP_hist} shows that the AP
approach is sensitive to the tSZ-induced decrements at 100 and 143\,GHz, since
the AP output monopoles at cluster positions fall in the negative temperature
range, far from the histograms coming from displaced positions (which are
centred near zero). The observed monopoles in this panel are less
negative (by about 20\,\%) than predictions based upon the universal
pressure profile of \citet{arnaudetal10}. Given measurement errors, this low
bias is marginally significant (around $3\,\sigma$) and is probably due to
residual point source emission and/or inaccuracies in the modelling of the
beam impact on our predictions. For 217\,GHz, however, the AP monopole falls
on the positive part of the histogram, possibly indicating traces of
tSZ-induced emission (since the effective frequency of this channel is above
the tSZ null). This histogram can be converted into velocity units
through dividing by each cluster's estimated
optical depth (see Sect.~\ref{sec:clusters}). After averaging over the full MCXC
un-masked cluster sample one can compute the conversion factor from
thermodynamic temperature fluctuations ($\delta T$) to peculiar radial velocity
($v_{||}$) for this sample,
\begin{equation}
v_{||} = f_{\rm T2v} \, \delta T.
\label{eq:conv_T2vel}
\end{equation}
This conversion factor however depends on the AP radius applied.
We obtain values for
$f_{\rm T2v}$ of 172\,km\,s$^{-1}$\,$\mu$K$^{-1}$ and
203\,km\,s$^{-1}$\,$\mu$K$^{-1}$, for the AP radius choices of
$0.25\,\theta_{500}$ and $\theta_{500}$, respectively.  After weighting the AP
velocity estimate of each cluster by its variance
($w_i = 1/\sigma_{{\rm AP},\,i}^2$\,), we obtain an estimate for the kSZ
average/monopole from HFI
217\,GHz data: $-212 \pm 80$\,km\,s$^{-1}$. If instead we use the 2D-ILC map,
the constraint becomes $-1 \pm 73$\,km\,s$^{-1}$. The HFI data in the channel
near the tSZ null seem to show
some residual tSZ contamination (as expected from the effective frequency of
this channel quoted in Table~\ref{tab:hfi_table}), but the 2D-ILC result is
consistent with zero.

The middle and right panels
of Fig.~\ref{fig:AP_hist} display the histograms of the
rms values obtained from the displaced positions. As for the left panel, black
and red colours refer to the 100 and 143\,GHz channels, respectively.
In this case, and given the measurement uncertainties, predictions from our
adopted pressure profile are in good agreement with our rms-excess
measurements. The AP
output rms estimates determined at the positions of clusters are displayed by
the vertical, dot-dashed lines: they fall clearly off the histograms obtained
from displaced positions, showing an excess rms, which is however not seen at
217\,GHz (green curve in the right panel). In this panel, the AP output rms
estimated at the cluster positions falls in the middle of the histogram obtained
from displaced positions. The same occurs for the 2D-ILC map, denoted by blue
lines. This suggests that the rms excess found in the middle panel has come from
tSZ, since it does not show up in the 217\,GHz channel.  In ideal
conditions, with identical beams throughout channels and an absence of noise and
foregrounds, the histograms in the middle and right panels would be identical;
the differences among them are reproduced when performing the analysis on
\Planck\ simulated maps, which account for different noise levels and beam
sizes.

Applying the procedure outlined in Sect.~\ref{sec:spatial_app} on the AP rms
distribution displayed in the right panel of Fig.~\ref{fig:AP_hist}, we set
constraints on the kSZ-induced contribution to the total measured rms. We find
that radial velocity rms constraints for the whole cluster sample are, at the
95\,\% confidence level, 2017\,km\,s$^{-1}$ and 1892\,km\,s$^{-1}$,
for the raw 217\,GHz and 2D-ILC maps, respectively.
These upper limits are about a factor of 8
above theoretical predictions, and can be only slightly improved by looking at
subsets of the cluster sample.
We have checked that kSZ errors decrease with cluster mass
and angular distance, since the more massive clusters provide higher kSZ signals
and the CMB contamination is less important on smaller angular scales. However,
constraints on radial velocities do not improve significantly.  By using only an
un-masked cluster subsample containing the first 1000 clusters, which have
larger values of mass times angular distance ($M_{500} \times D_{\rm A}$,
$\langle M_{500}\rangle_{\rm subsample} = 2.3 \times 10^{14}\,\Msolar$, $\langle
z\rangle_{\rm subsample} = 0.18$), the 95\,\% confidence level constraints from
HFI 217\,GHz and 2D-ILC data become 1806\,km\,s$^{-1}$ and 1229\,km\,s$^{-1}$,
respectively. These are still a factor 5--7 above theoretical predictions
for $\Lambda$CDM.

\begin{figure}
\centering
\includegraphics[width=8.cm]{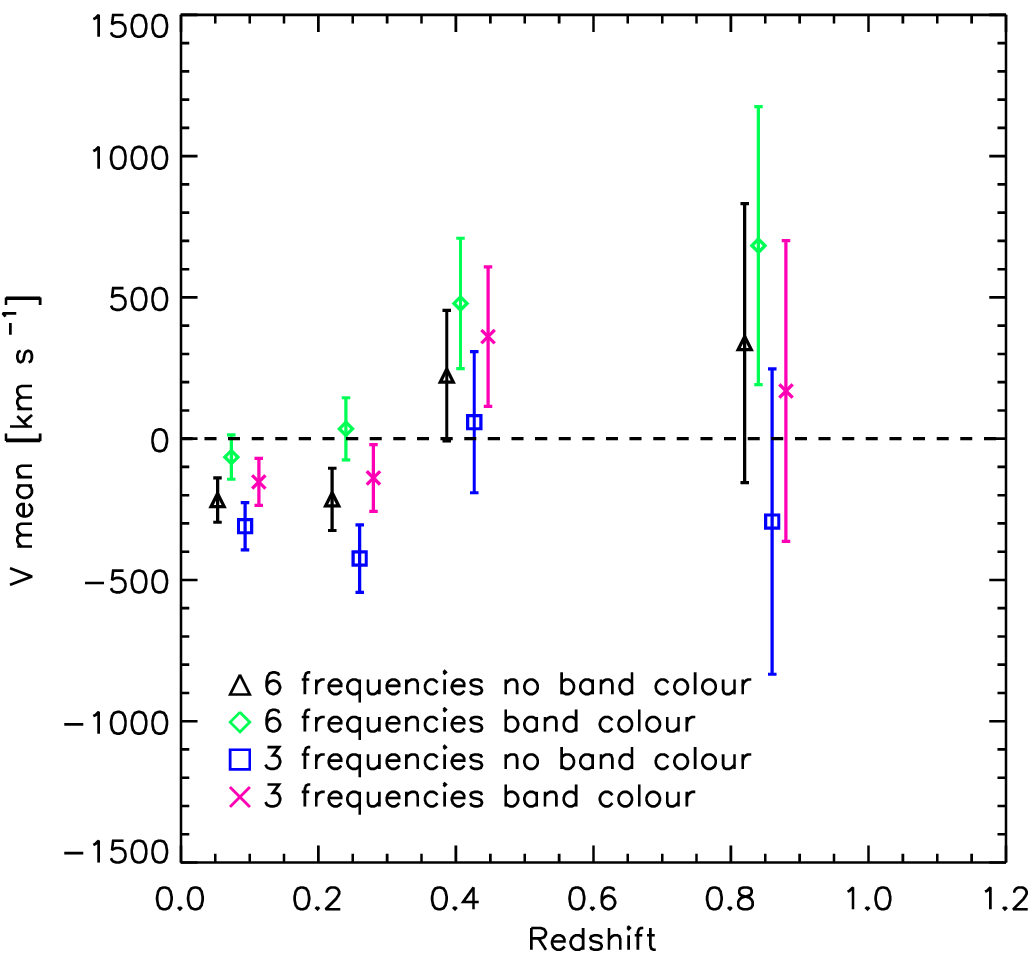}
\caption[fig:kSZmono_vs_z]{Dependence of the average radial velocity of clusters
  for different redshift bins.  Note that the ``no band colour'' symbols are
  known to give unreliable results.  For clarity, symbols within the same redshift bin
  have been slightly shifted horizontally.}
\label{fig:kSZ_mono_vs_z}
\end{figure}

\subsubsection{Constraints from the uMMF/MF filters}

The uMMF approach
can provide accurate kSZ amplitude estimates under the assumption that
clusters, as well as being isothermal, follow the universal pressure profile of
\citet{arnaudetal10}. In Sect.~\ref{sec:robust} we address the bias that this
assumption introduces when clusters show different density profiles, finding a
roughly 5\,\% low bias in the velocity amplitude estimates.  On an individual
basis, the uMMF provides velocity errors that depend on the mass and the
size of each cluster
on the sky, and lie at the level of a few thousand km\,s$^{-1}$,
with an average value of about 4100\,km\,s$^{-1}$ for the un-masked MCXC sample
and the six HFI channels.  When properly accounting for the finite bandwidth of
the HFI channel spectral responses, the average peculiar radial velocity of MCXC
clusters is compatible with zero ($15 \pm 60$\,km\,s$^{-1}$).  The 2D-ILC map
provides $72 \pm 60$\,km\,s$^{-1}$.

When binning the cluster sample in
redshifts, we again find no evidence for any statistically significant average
peculiar velocity (see red and green symbols in
Fig.~\ref{fig:kSZ_mono_vs_z}). Estimates of the kSZ monopole for clusters
belonging to different redshift bins are shown in
Fig.~\ref{fig:kSZ_mono_vs_z}.

\begin{figure}
\centering
\includegraphics[width=8.cm]{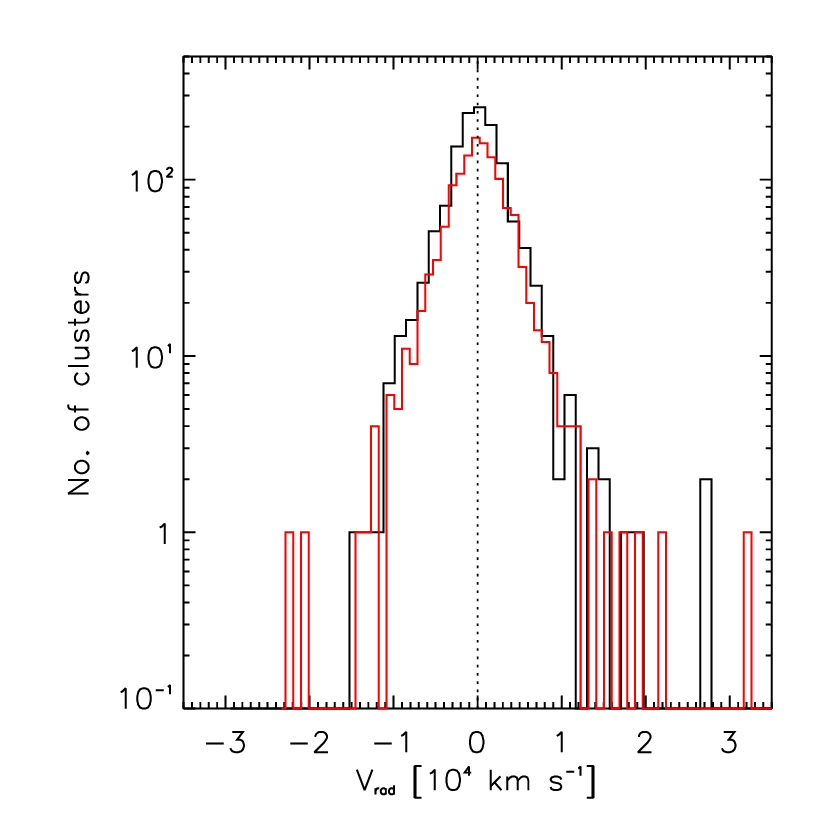}
\caption[fig:kSZ_velhist]{ Histogram of recovered radial peculiar velocities as
  estimated by the uMMF (black) and AP (red) implementations on HFI frequency maps.  }
\label{fig:kSZ_velhist}
\end{figure}

Bear in mind that if no colour correction is taken into account,
and use is made of the nominal
HFI frequencies instead of the effective ones (see Table~\ref{tab:hfi_table}),
then effects associated with the finite spectral response in HFI channels become
of relevance. In Fig.~\ref{fig:kSZ_mono_vs_z} blue squares and black triangles
display the average radial velocity estimates (within different redshift bins)
as inferred by the uMMF when colour correction is ignored, using the three
lowest frequency or six frequency channels, respectively. In those cases,
average velocity estimates lie a few $\sigma$ below the zero level for several
redshift bins, pointing to some {\it positive} residual temperature fluctuations
at the cluster positions (which is expected since the effective frequency for
the third HFI channel is about 222\,GHz, i.e., above the tSZ null frequency).  
Colour corrections must hence be made when interpreting kSZ measurements.
However, the fact that velocity estimates for three and six HFI bands are
compatible suggests
that dust contamination is not important. We also checked that no significant
differences are found when introducing changes of $\pm 1 \sigma$ in the adopted
HFI Gaussian fit FWHM values.
 
In Fig.~\ref{fig:kSZ_velhist} we display the histograms of radial peculiar
velocities as estimated, for MCXC clusters, by the uMMF (black) and AP (red) implementations. For the uMMF case, the distribution is almost symmetric around zero and has a standard deviation of
$4100$\,km\,s$^{-1}$. Both histograms show non-Gaussian tails, presumably due to the
presence of un-resolved point sources and other non-Gaussian signals in
some galaxy clusters.

\begin{figure*}
\centering
\includegraphics[width=18.cm,height=6.cm]{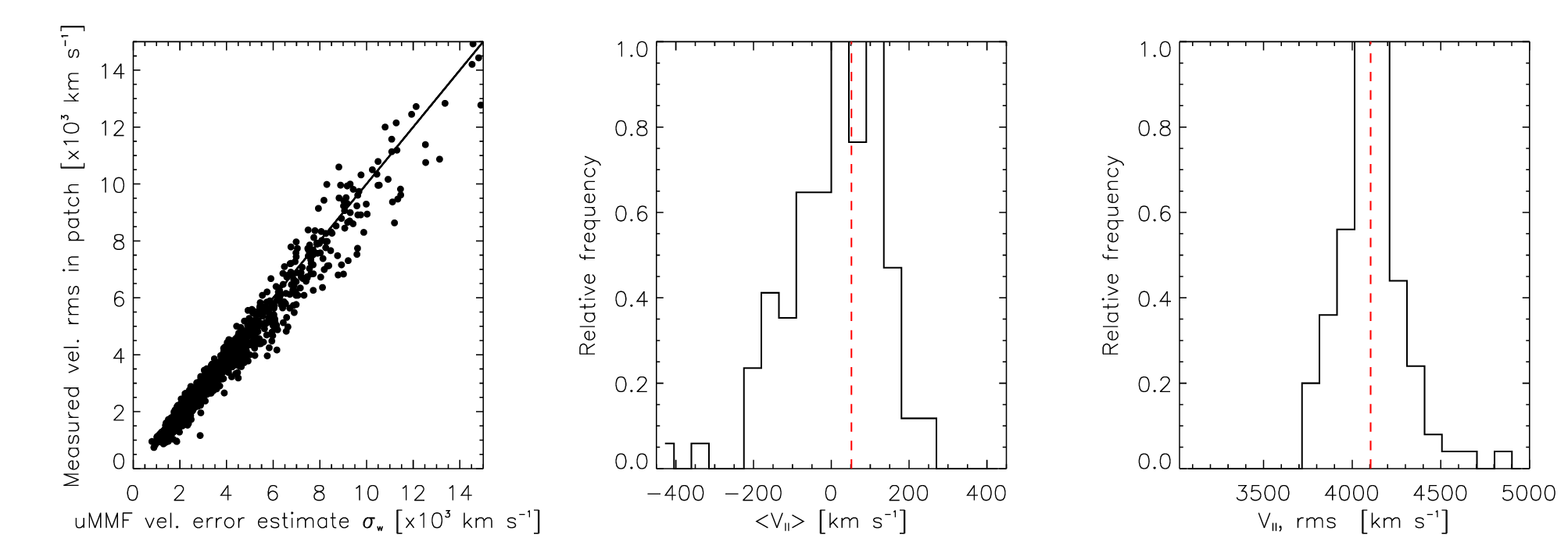}
\caption[fig:uMMF_spatial]{{\it Left panel:} Estimated rms of uMMF outputs at
  100 random positions within the patch surrounding each MCXC cluster versus the
  corresponding peculiar velocity error as predicted by the uMMF. {\it Middle
    panel:} Histogram of mean kSZ radial velocity estimated using all un-masked
  MCXC clusters, for the 101 positions considered in each cluster patch. The
  vertical, red dashed line corresponds to the average kSZ estimate when
  averaging throughout the patch {\em centres}, i.e., at the real cluster
  positions.  {\it Right panel:} uMMF velocity rms histogram obtained for the
  MCXC cluster set, at different positions within the patches, just as for the
  middle panel. Again, the vertical red dashed line corresponds to the kSZ
  velocity rms estimated at real cluster positions (on patch centres).  }
\label{fig:uMMF_spatial}
\end{figure*}

We next conduct a {\it spatial} analysis of the Matched Filter outputs, just as
done for the AP approach above, and again analysing both the set of HFI
frequency maps and the 2D-ILC map. In this analysis we first apply the filter on
100 {\it blank, random} positions on the sky where no MCXC clusters are found,
and compare the resulting rms of the filter velocity estimates with the rms of
the estimates obtained for the real cluster positions. Contrary to the displaced
positions for the AP filter, in this case the random positions on which the
uMMF/MF filters are evaluated are not displaced in either equatorial or ecliptic
latitude with respect to the real cluster, but simply randomly placed within a
$10\degr \times 10\degr$ patch centred on the real object. If the kSZ signal
generated for those ``sources''
leaves a measurable effect, then the rms at the real positions
of the clusters must be larger than the rms obtained at blank positions.

In each patch centred on each MCXC cluster, we record the uMMF/MF
outputs for 100
random positions not coincident with the centre. This provides peculiar velocity
estimates for positions where we expect the kSZ effect to be zero while, at the
same time, having similar levels of instrumental noise and foreground
contamination as the positions corresponding to real MCXC clusters. For each set
of 100 positions, it is possible to compute the rms and compare it to the
expected value predicted by the uMMF/MF method, as provided by
Eq.~\ref{eq:errmmf}.
This is shown in the left panel of Fig.~\ref{fig:uMMF_spatial} for HFI frequency
maps: the solid line displays a one to one relation, and it is roughly followed
by the recovered rms from the random estimates within the patch (vertical axis)
versus the predicted velocity errors (horizontal axis). Here we neglect all
uMMF/MF outputs within the patch that fall in masked pixels.
After fixing one of the 101 positions within each patch, we compute an average
velocity by considering velocity estimates in all patches at that particular
position.
The histogram of these average velocities computed in the 100 displaced
positions is shown in the middle panel. The vertical, red dashed line
corresponds to the average velocity as computed from the velocity estimates at
the patch centres, that is, at the real MCXC cluster positions. As expected, the
middle panel of Fig.~\ref{fig:uMMF_spatial} shows that the entire ensemble of
MCXC clusters exhibits average peculiar velocities that are compatible with
zero. The right panel displays the histogram of the velocity rms estimates,
computed exactly in the same way as for velocity averages. For each of the 101
positions within a patch, we calculate a velocity rms after considering the
filter outputs for the whole set of patches at that position.
We end up with 100 velocity rms estimates in displaced positions, and the
resulting histogram in the right panel is compared to the velocity rms estimated
at the cluster positions, again shown as a vertical, red dashed line. From this
distribution, and after following Eq.~\ref{eq:var_diff}, the uMMF/MF can set the
following upper limits at 95\,\% C.L. on the cluster radial peculiar velocity
rms: 1514\,km\,s$^{-1}$ for HFI frequency maps; and 987\,km\,s$^{-1}$ for ILC
data. As for the AP filter, when restricting ourselves to the cluster subsample
which maximizes the product $M_{500}\times D_{\rm A}$, we obtain improvements on
these constraints: for the 1000 clusters maximizing $M_{500}\times D_{\rm A}$,
upper limits become 798 and 754\,km\,s$^{-1}$ for the raw HFI and 2D-ILC data,
respectively.  If we instead choose the top 100 clusters in the sorted list of
the previous subsample, the constraints change, but only slightly: 794 and
614\,km\,s$^{-1}$ for raw HFI and 2D-ILC maps, respectively.  These limits
have systematic uncertainties at the level of a few percent, but are
nevertheless a factor of about 3 above the level
of $\Lambda$CDM predictions.


\subsection{Constraints on bulk flows and the kSZ dipole }
\label{sec:bulkflows}

In this section we describe the constraints that
\Planck\ sets on the existence of bulk flows at different scales and the
measurement of the kSZ dipole in our cluster sample.

\subsubsection{Constraints from individual cluster velocities}

Extensive efforts have been made in recent years to try to set constraints
on the local bulk flow \citep[][]{hudson04, watkinsetal09, feldmanetal10,
nusserdavis11,mascott12}, without reaching full agreement so far.
The kSZ estimates from \Planck\ provide a different approach to the question
of the local bulk flow: if clusters embedded
in structure around the Local Group are comoving with it towards a nearby
overdensity, then the kSZ measured for those sources should show a dipolar
pattern. By looking at clusters within spheres of different radii
from us it is possible to
set constraints on the local bulk flow within different volumes. This provides a
direct test on the studies of
\citet[][]{kashlinsky08,kashlinsky10,kashlinsky12}, which claim that clusters
extending at least up to 800\,$h^{-1}$\,Mpc are part of a bulk flow of amplitude
about 1000\,km\,s$^{-1}$.

\begin{figure}
\centering
\includegraphics[width=8.cm]{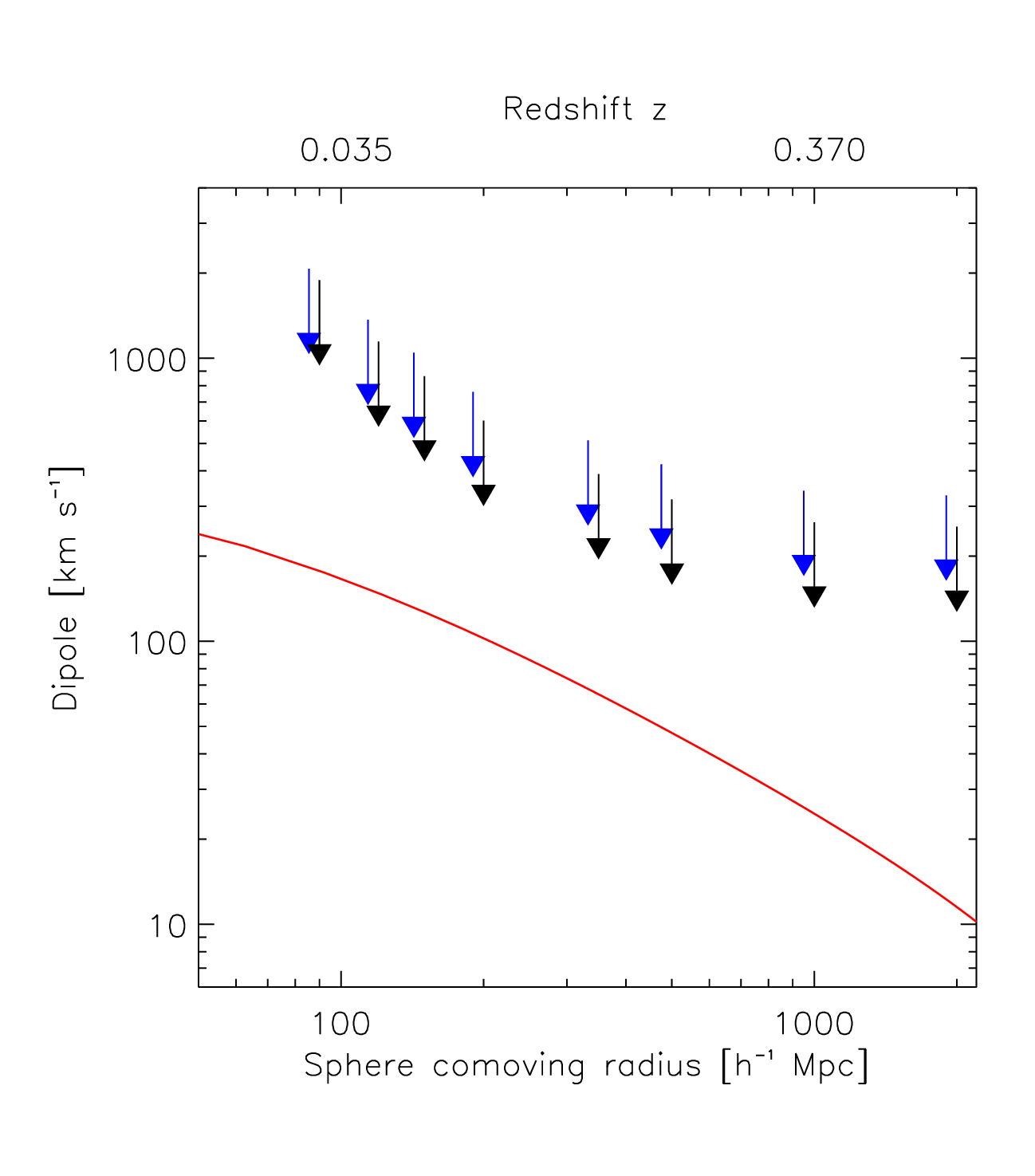}
\caption[fig:dip_vs_z]{ Upper limits at the 95\,\% confidence level for the
  dipole amplitude from MCXC clusters contained in local spheres of varying
  radii. Blue and black arrows denote limits for AP and uMMF
  methods, respectively. The upper limits are indicated by the tails of the arrows. 
  The red solid line depicts the $\Lambda$CDM prediction.}
\label{fig:dip_vs_z}
\end{figure}

Given any MCXC cluster subsample, we compute the amplitude of the kSZ dipole
along a given direction $\vn_{\rm dip}$ by minimizing $\chi^2 = \sum_j (v_{||,j}
- \alpha\, ( \vn_{\rm dip} \cdot \vn_j) )^2 / \sigma^2_{\rm v_j}$. Here,
$v_{||,j}$ denotes the AP/uMMF/MF radial velocity estimate of the $j$th cluster
(which is located in the direction $\vn_j$), and $\sigma_{\rm v_j}$ is its
associated error. After assuming uncorrelated errors (from cluster to cluster),
this minimization yields both an estimate of $\alpha$ and a formal associated
error:
\begin{eqnarray}
\hat{\alpha} &=& \frac{ \sum_{j} v_{||,j}\,( \vn_{\rm dip} \cdot \vn_j) /
  \sigma^2_{\rm v_j} } {\sum_{j} ( \vn_{\rm dip} \cdot \vn_j)^2 /
  \sigma^2_{\rm v_j}}; \nonumber \\ \sigma_{\hat{\alpha}} &=& \sqrt{
  \frac{1}{\sum_{j} ( \vn_{\rm dip} \cdot \vn_j)^2 / \sigma^2_{\rm v_j}}
}.
\label{eq:dipfit1}
\end{eqnarray}
We compute the kSZ dipole for cluster sub-samples contained within increasing
radii from the Local Group.  For each cluster sub-sample and associated kSZ
estimates, we fit a dipole along every direction in the sky, i.e., we sweep in
$\vn_{\rm dip}$, and retain the direction which yields the highest uncertainty
(i.e., highest $\sigma_{\hat{\alpha}}$ value) in order to find
the corresponding upper limit.  Figure~\ref{fig:dip_vs_z} displays the
corresponding upper limits (at the 95\,\% confidence level, calculated as
2$\sigma$, assuming a Gaussian distribution) on the dipole values for different
radii and both AP and uMMF methods on HFI frequency maps: for both methods, the
95\,\% C.L. upper limit for radii of $90\,h^{-1}$\,Mpc amounts to
${\sim}\,2000\,$km\,s$^{-1}$, but it decreases rapidly as the volume increases.
For spheres of radius around $350$\,$h^{-1}$\,Mpc, the uMMF limits fall to
about 390\,km\,s$^{-1}$, and the corresponding AP limit is just slightly higher
(520\,km\,s$^{-1}$). In the largest volume probed by the MCXC clusters, the
95\,\% C.L. upper limits become 329 and 254\,km\,s$^{-1}$ for the AP and uMMF filters,
respectively. All these limits are well above the $\Lambda$CDM prediction, displayed by the
red solid line. Despite being very different in their definition, the two
methods give rise to a very similar pattern in the bulk flow constraints inside
different volumes, and in all cases the measured dipoles are compatible with
zero.

\begin{figure}
\centering
\includegraphics[width=9.cm,height=6.cm]{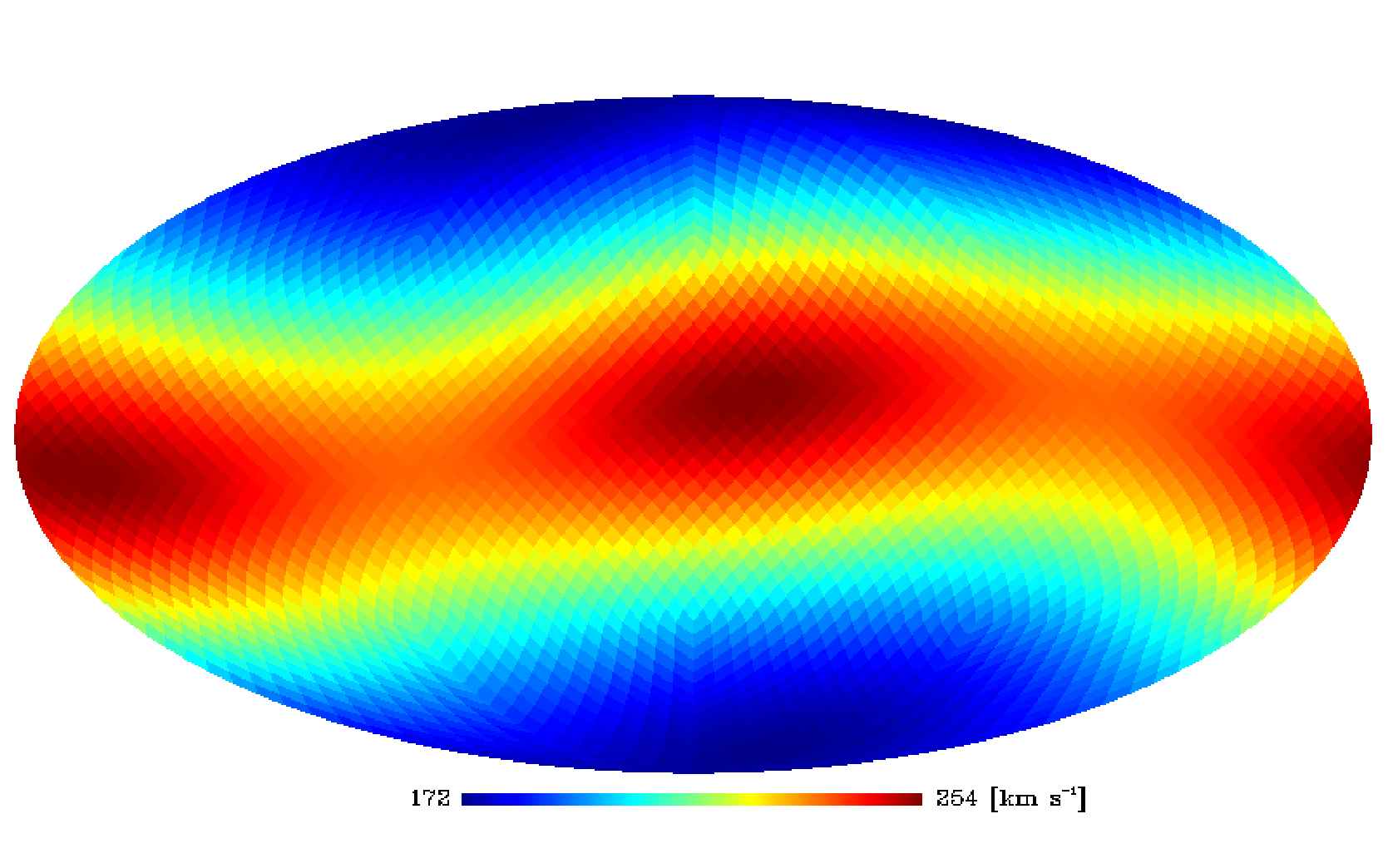}
\caption[fig:s2n_uMMF_dip]{Mollweide projection in Galactic coordinates of the
  upper limit (at 95\,\% C.L.) of the kSZ dipole
  amplitude from applying the uMMF approach to HFI frequency maps using
  the whole MCXC cluster
  sample. In no direction is the dipole detected at more than 2\,$\sigma$.}
\label{fig:s2n_uMMF_dip}
\end{figure}

In Fig.~\ref{fig:s2n_uMMF_dip} we display the 95\,\% upper limit on the kSZ
amplitude from the uMMF filter using
the whole MCXC cluster set (for which $\langle z \rangle = 0.18$)
and HFI frequency maps. In no direction does the measured
dipole exceed $2\sigma$, and the direction with the highest
$\hat{\alpha}/\sigma_{\hat{\alpha}}$ value is close to the Galactic plane.
This is to
be expected if the errors in the Galactic $x$- and $y$-components of the dipole are
larger than the $z$-component, due to the lack of clusters at low Galactic
latitudes. When restricting ourselves to clusters below $z=0.25$, the dipole
amplitude along the CMB dipole direction $(l,b) = (264\deg, 48\deg)$
\citep{hinshaw09}\footnote{We use the CMB dipole as measured by
{\it WMAP}, since \Planck\ has not yet provided a measurement of the CMB
dipole.} amounts to $80 \pm 150$\,km\,s$^{-1}$, and limits to $50 \pm
160$\,km\,s$^{-1}$ along the direction of apparent motion of the Local Group
with respect to the CMB, $(l,b) = (276\deg, 30\deg)$ \citep{kogut93}. This
result is in clear contradiction with \citet{kashlinsky10}, who find
a bulk flow of about 1000\,km\,s$^{-1}$ amplitude within radii of
$300$--$800\,h^{-1}$\,Mpc at ${\sim}\,3\,\sigma$ C.L. Since our error
bars are a factor of about 2 smaller, this suggests that we test the
outcome of the filter used by those authors on our data
(see Sect.~\ref{sec:kash} below).

\subsubsection{Constraints from all-sky method}

We can also compute the bulk flow according to the procedure outlined in
Sect.~\ref{sec:bfmak}. We filter the observed HFI maps and fit monopole and
dipole velocity coefficients to the filtered data, as well as to simulations of
the data (PSM diffuse component, tSZ and CMB plus instrumental noise
simulations; see Sect.~\ref{sec:simul}). Simulations provide one way of
estimating uncertainties which allow one to propagate errors on cluster-derived
measurements, mainly induced by dispersion in the scaling relations,
throughout the whole pipeline. An alternative procedure to derive uncertainties
consists of randomizing the positions of the clusters on the sky and computing
the monopole and dipole from these random directions, adopting
the same procedure
used on the real data. We use the latter to show the typical variations of the
diffuse component's contribution for small displacements around cluster
locations. Specifically, we consider directions displaced by $30^\prime$
to $1^\circ$ from the cluster
nominal locations, while also avoiding mask boundaries
(these are the ``shifted positions'' in
Table~\ref{t:dipoles}).
 
Values for the resulting velocity dipole coefficients are presented in
Table~\ref{t:dipoles}.  The main result is that \Planck\ data give dipole
coefficient amplitudes consistent with those expected from the $\Lambda$CDM
scenario, once one has taken into account the contamination from Galactic
foregrounds and other signals.
The apparent bulk flow measured is 614\,km\,s$^{-1}$.  However, with this
particular configuration for cluster positions, the diffuse Galactic
component provides a
non-negligible contribution to the dipole signal, 529\,km\,s$^{-1}$, as
measured in the PSM simulations.  The errors on the diffuse component,
as estimated by
randomizing the cluster directions on the PSM diffuse component simulations, are
smaller than those induced by the thermal SZ and CMB plus instrumental noise
simulations (see Table~\ref{t:dipoles}).

Simulations of the tSZ component, which account for uncertainties in the SZ
signal for clusters with a given temperature, induce a 1$\sigma$ uncertainty on
the bulk velocity of 40\,km\,s$^{-1}$, and an overall bias in the velocity
estimation of the order of 400\,km\,s$^{-1}$.

Uncertainties from CMB confusion and instrumental noise (140--290\,km\,s$^{-1}$
in the different directions) are dominant over tSZ ones.  The fraction of the
observed bulk flow not accounted for by Galactic foregrounds (by subtracting the
dipole as a vector, this amounts to 350\,km\,s$^{-1}$) is within 95\,\% of the
error on bulk flows induced by the tSZ, CMB and instrumental noise
(893\,km\,s$^{-1}$) and below the 95\,\% level of CMB plus instrumental noise
alone (543\,km\,s$^{-1}$).

\begin{figure}
\centering
\includegraphics[width=8.cm]{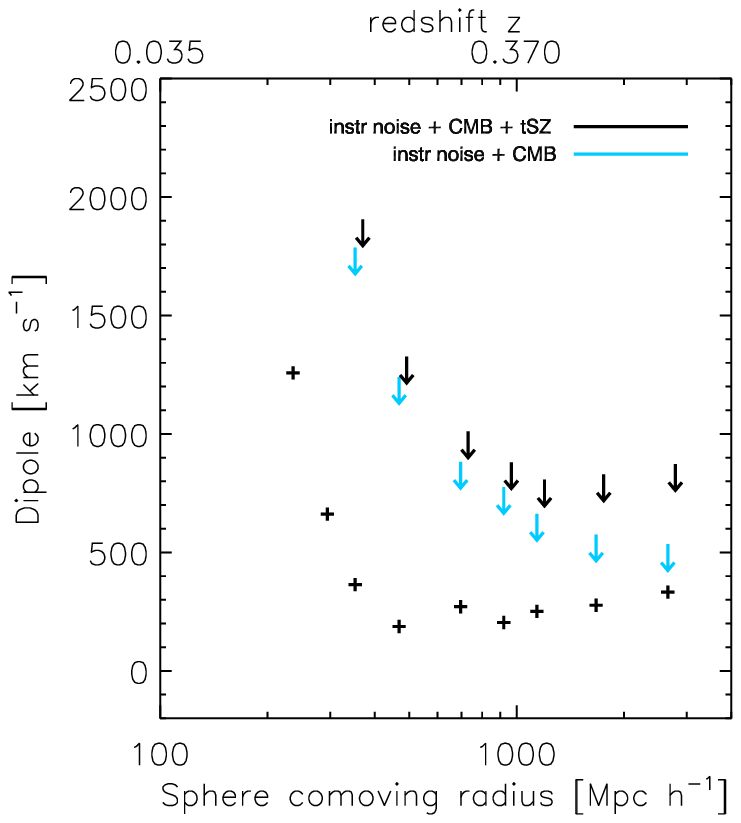}
\caption{Bulk flow amplitude measured in \Planck\ data with the all-sky method,
  after subtraction (vectorially) of the Galactic contribution (black crosses),
  compared with 95\,\% upper limits derived from simulations containing CMB and
  instrumental noise only (blue arrows) or also including tSZ signal (black
  arrows).  The fact that the crosses are below the arrows at all scales
  shows that there is no significant bulk flow detection.}
\label{fig:dipz}
\end{figure}

By restricting the cluster sample to the objects within a specified distance
from us, it is possible to constrain the bulk flow within spheres of a given
comoving radius. This is what is displayed in Fig.~\ref{fig:dipz}, where the
Galactic component has been subtracted. We notice that, for all distances, the
measured bulk flow is below the 95\,\% confidence level as measured from maps
including only CMB, instrumental noise, and tSZ clusters. The upper limits reach
an approximately constant value above scales around 500\,$h^{-1}$\,Mpc, as a
small fraction of the clusters in this sample are at larger distances. The
95\,\% upper limits at 2400\,$h^{-1}$\,Mpc are 893\,km\,s$^{-1}$ when all
sources of noise are considered, reducing to 543\,km\,s$^{-1}$ when CMB plus
instrumental noise are taken into account.

The results reported in Fig.~\ref{fig:dipz} refer to the nominal mask, while in
Table~\ref{t:dipoles} we also quote results for the more restrictive mask. The
two sets of results are very similar, however.

In this analysis, we also fit for the direction of the measured bulk
flow.  Even although the detection is not significant, it might still be
instructive to compare the best fit direction to other potentially relevant
directions.
Results for various cluster configurations and \Planck\ data are displayed
in Fig.~\ref{fig:clunoise}, together with the CMB dipole and the claimed dipole
direction of \citet{kashlinsky08}. We notice that the direction we determine
from \Planck\ data and MCXC clusters is quite different from
both the CMB dipole and the result of \citet{kashlinsky08}. It aligns better
with the direction of the collection of clusters in the map, which happen
to be in a low instrumental noise area of the sky, as one would expect from a
noise--induced measurement.  Indeed, simulations show that the directions of
bulk flows of the magnitude seen in the data cannot be recovered with great
precision. Errors are of the order of tens of degrees, depending on the bulk
flow direction \citep{mak11}.

Finally, we notice that the upper limits to the bulk flow that we find with this
method are above those found in the previous section. This is not surprising,
as we are fitting here for both the velocity direction and amplitude, and we
compute errors in a different way.  The upper limits obtained with this approach
should be considered as more conservative. Nevertheless they are about a factor
of five better than what was found using {\it WMAP\/} data.

\begin{table*}[mbp] 
\begingroup 
\newdimen\tblskip \tblskip=5pt
\caption{Estimated dipole coefficients (columns 2--4) and velocity magnitude (column
  5) using the all-sky method. The values in parentheses are determinations
  using the more restrictive mask which includes 1321 clusters.
  The ``HFI''
  row reports the results obtained for the actual data; as discussed in the
  text, this estimate is significantly contaminated by Galactic foregrounds
  and tSZ. ``PSM diffuse''
  reports the contribution from the diffuse Galactic component found in the
  PSM simulation. In the rows corresponding to thermal SZ (tSZ), instrumental
  noise, and CMB, the first three columns report the mean and 68\,\% confidence
  region error bars, while the last column indicates the 95\,\% upper limit.
  The bottom part of the table refers to results found using ``shifted
  positions'' for each cluster. These are randomly selected between $30\arcm$
  and $1\deg$ from each of the MCXC clusters outside the mask region.
  The notation
  is as in the noise simulations, but it is relative to the distribution found
  for different choices of shifted positions. The velocity magnitude (column 5)
  represents the mean and 68\,\% error in the distribution.  These rows
  indicate the size of the apparent dipole that one could find using this
  method, even without a cosmological dipole existing.}
\label{tab:dipole}
\vskip -3mm
\footnotesize 
\setbox\tablebox=\vbox{ %
\newdimen\digitwidth 
\setbox0=\hbox{\rm 0}
\digitwidth=\wd0
\catcode`*=\active
\def*{\kern\digitwidth}
\newdimen\signwidth
\setbox0=\hbox{+}
\signwidth=\wd0
\catcode`!=\active
\def!{\kern\signwidth}
\newdimen\gtwidth
\setbox0=\hbox{<}
\gtwidth=\wd0
\catcode`?=\active
\def?{\kern\signwidth}

\halign{#\hfil\tabskip=0.2cm& \hfil#\hfil\tabskip=0.2cm&
 \hfil#\hfil\tabskip=0.2cm& \hfil#\hfil\tabskip=0.2cm&
 \hfil#\hfil\tabskip=0.2cm& \hfil#\hfil\tabskip=0cm\cr
\noalign{\doubleline}
\noalign{\vskip -2pt}
\omit& $v_x$& $v_y$& $v_z$& $v$\cr
Maps& [km\,s$^{-1}$]& [km\,s$^{-1}$]& [km\,s$^{-1}$]& [km\,s$^{-1}$]\cr
\noalign{\vskip 3pt\hrule\vskip 3pt}
\multispan5\hfil MCXC  positions\hfil\cr
\noalign{\vskip 3pt\hrule\vskip 2pt}
HFI&         $-188$\quad (147)& !384\quad (!414)& !441\quad (!494)&
 $\,?\,614$\quad $(\,?\,662)$& \cr 
PSM diffuse& $!116$\quad (307)& !436\quad (!327)& !276\quad (!295)&
 $\,\phantom{<}\,529$\quad $(\,\phantom{<}\,537)$& \cr
tSZ& $!238\pm*37$\quad ($232\pm*42$)& $-302\pm*37$\quad ($-319\pm*41$)&
 $-239\pm*26$\quad ($-253\pm*27$)& $<531$\quad $(<549)$& \cr
instr. noise + CMB& $!**0\pm189$\quad ($**0\pm206$)&
 $**-3\pm195$\quad ($**-1\pm217$)&
 $!**6\pm140$\quad ($!**6\pm143$)& $<543$\quad $(<577)$& \cr
instr. noise + CMB + tSZ& $!232\pm187$\quad ($229\pm207$)&
 $-303\pm185$\quad ($-318\pm207$)& $-234\pm142$\quad ($-248\pm145$)&
 $<893$\quad $(<929)$& \cr
\noalign{\vskip 3pt\hrule\vskip 3pt}
\multispan5\hfil Shifted positions\hfil\cr
\noalign{\vskip 3pt\hrule\vskip 3pt}
HFI& $-112\pm214$\quad ($171\pm225$)& $348\pm274$\quad ($304\pm285$)&
 $290\pm136$\quad ($338\pm103$)& $552\pm221$\quad ($591\pm161$)& \cr
PSM diffuse& $*!73\pm154$\quad ($229\pm156$)& $470\pm189$\quad ($367\pm170$)&
 $199\pm*77$\quad ($191\pm*73$)& $553\pm164$\quad ($520\pm112$)& \cr
\noalign{\vskip 3pt\hrule\vskip 3pt}}}
\label{t:dipoles}
\endPlancktable 
\endgroup
\end{table*}

\subsubsection{Revisiting the \citet{kashlinsky10} filter}
\label{sec:kash}

The idea of constraining the local bulk
flow of matter by looking at the dipolar pattern of the kSZ in the galaxy
cluster population was first discussed by \citet{haehnelt95} and further
developed by \citet{kashlinsky00}.  The method
was applied by
\citet{kashlinsky08,kashlinsky09} to {\it WMAP\/} data, analyses that have been
followed by more recent studies \citep{kashlinsky10,kashlinsky11}. In this
section, we perform a direct application of their filter to both
{\it WMAP\/} and \Planck\ data, and interpret it at the light of the
results already outlined in this work. 

\begin{figure*}
\centering
\plotancho{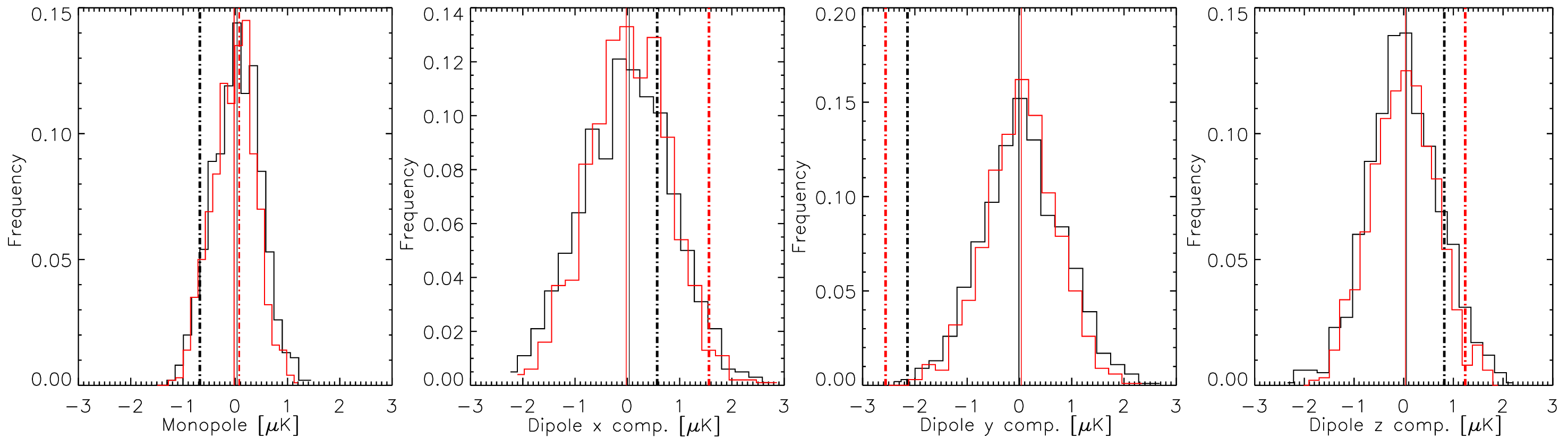}
\caption[fig:histograms_kabke]{Monopole and dipole component estimates after
  applying the spatial filter of \citet{kashlinsky08} on \textit{WMAP}-7 W-band
  data (black lines) and on \Planck\ 2D-ILC maps (red lines). Estimates from
  real MCXC clusters are displayed by vertical, dot-dashed lines. Histograms
  are obtained after repeating the analysis on 1000 random cluster
  configurations, with averages indicated by the vertical, solid lines.  The 
  $y$-component appears discrepant here, but compare
  Fig.~\ref{fig:all_hists}.}
\label{fig:histograms_kabke}
\end{figure*}

We first implement the filter of \citet{kashlinsky10} on the MCXC cluster sample
and the {\it WMAP}-7 data. After using the extended temperature KQ75 mask, we
obtain filtered maps from the cleaned Q, V and W band {\it WMAP\/} data.
Since the filtered maps
for the four W-band Differencing Assemblies (DAs) used by those authors are
publicly available\footnote{The data were downloaded from the URL site
  \tt{http://www.kashlinsky.info}. }, a direct comparison of the filtered maps
can be performed:
for instance, for the filtered maps corresponding to the fourth W-band
DA, the temperature rms outside the joint mask in our filtered map is
74\,$\mu$K, very close to the 77\,$\mu$K obtained from the map used by
\citet{kashlinsky10}. The rms of the difference map amounts to 35\,$\mu$K, and a
visual inspection shows the similarity between both maps. Each cluster is
assigned a radius of $25\arcm$, and the {\tt remove\_dipole} routine from
HEALPix
is used when computing the monopole and dipole in the subset of pixels
surrounding the clusters. The monopole and dipole components obtained for the
\textit{WMAP\/} W band are displayed by the black, vertical dot-dashed lines in
Fig.~\ref{fig:histograms_kabke}. These are in very good agreement with the
results obtained by \citet{kashlinsky10}.

We next distribute the same number of
clusters surviving the mask {\em randomly\/} on the unmasked sky 1000 times,
assign them a circle of radius $25^\prime$ and repeat the monopole and dipole
computation. For each of the 1000 cluster configurations, we separately compute
the monopole and dipole for each of the DAs. This permits us to obtain the rms
for each component and DA, in such a way that a combined estimate of the
monopole and dipole can be extracted from all DAs by inverse-variance weighting
the estimate for each DA. This is carried out
for the real cluster configuration on the
sky and for the 1000 mock (random) configurations. From the latter, we obtain
the histograms shown in Fig.~\ref{fig:histograms_kabke}. The average quantities
out of the 1000 simulations are displayed by the solid, vertical lines. Black
lines refer to \textit{WMAP\/} data, and our results show that the $y$-component
of the dipole is peculiar,
in the sense that it falls far in the negative tail of the distribution.

When repeating these analyses with the 2D-ILC map, we obtain the results
displayed by the red lines in the same figure.
In this case, the dipole components from the real data fall further outside
the distribution provided by the histograms, as
none of the 1000 mock cluster configurations provides a dipole of larger
amplitude than the one measured from the real MCXC sample. These results suggest
that the dipole measured at the MCXC cluster positions is indeed peculiar if
compared to dipole estimates from randomized cluster positions.

Nevertheless, there is one aspect to be studied more closely, namely the angular
distribution of clusters on the sky. In what follows, the filtered map built
upon the 2D-ILC data will be used. So far our Monte Carlo simulations assumed
that clusters were placed randomly on the sky, i.e., the clustering of our
sources has been neglected. We next perform tests in which the angular
configuration of our MCXC cluster sample is preserved. The first test consists
of repeating the filtering and subsequent dipole computation on 1000 CMB mock
skies following the {\it WMAP}-7 best-fit model. These mock CMB maps contain no
kSZ and hence should give rise to no significant kSZ dipole. Out of this
ensemble of mock skies, we compute the dipole using the positions
of MCXC clusters (as described above) and obtain a histogram from the
recovered dipole amplitude. This permits us to judge how peculiar our
measurement is with respect to
the simulation outputs. In a second test, we rotate the clusters' angular
positions around the Galactic $z$-axis {\em on the real filtered map obtained
from the 2D-ILC data}. We conduct 360 rotations of one degree step size,
in Galactic longitude, while preserving Galactic latitude,
and the relative angular configuration of MCXC clusters on
the sky. Since the mask mostly discards pixels at low Galactic latitude
(close to the Galactic plane), most clusters that are originally outside the
mask remain outside the mask after rotating. For each rotation a new value of
the dipole is recorded, and information on dipole statistics is then built
up using
outputs obtained from the real map with the real rotated cluster configuration
on the sky. This rotation test, unlike the one based upon CMB mock skies, 
accounts for the impact of noise, foregrounds and other systematic
signals that may be present in the filtered map.

\begin{figure}
\centering
\includegraphics[width=8.cm]{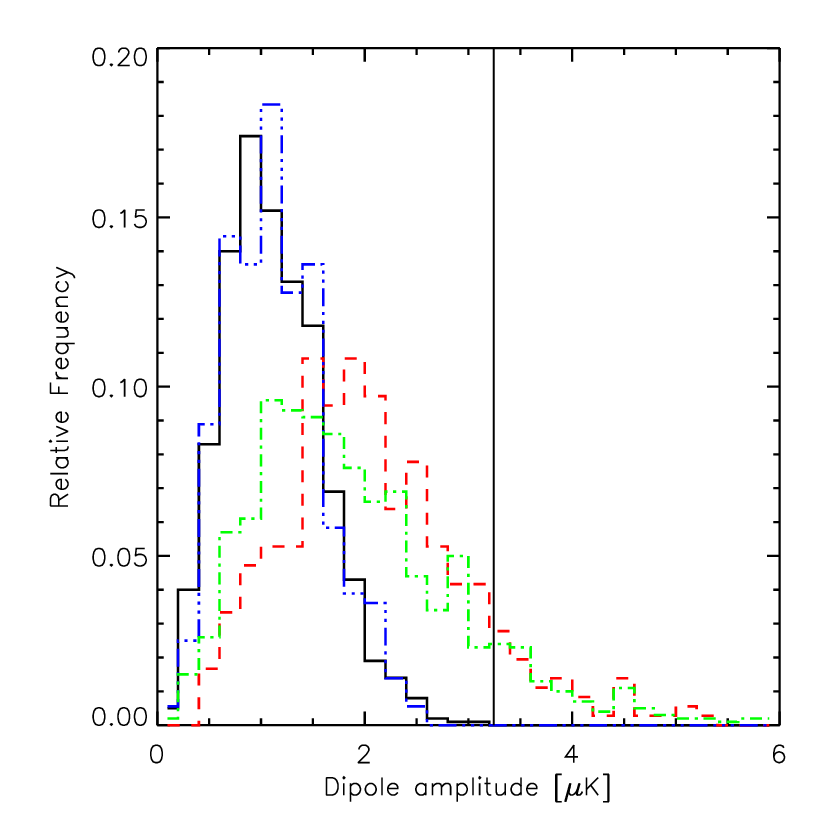}
\caption[fig:all_hists]{Histograms of the recovered cluster dipole amplitude:
  (a) from the 1000 Monte Carlo random cluster configurations on the sky (black,
  solid line); (b) from rotating one random cluster configuration in Galactic
  latitude on the real filtered map (blue, triple dot-dashed line); (c) from
  rotating the real MCXC cluster configuration around the Galactic $z$ axis on
  the real filtered map (red, dashed line); and (d) from applying the filter on
  the position of our MCXC cluster sample in 1000 Monte Carlo CMB simulations
  following the {\it WMAP}-7 best-fit model (green, dot-dashed lines). The
  dipole amplitude recovered at the real MCXC cluster positions on the real
  filtered map is shown by the vertical, solid line.  It is not significantly
  detected, provided one is careful to simulate the most important noise
  contributions.}
\label{fig:all_hists}
\end{figure}

The results are shown in Fig.~\ref{fig:all_hists}. The black histogram
reflects the statistics of the recovered dipole amplitudes after drawing 1000
random cluster configurations on the real filtered map, just as done for
Fig.~\ref{fig:histograms_kabke}. The blue, triple dot-dashed histogram
corresponds to the dipole outputs obtained after rotating in Galactic longitude
{\em one\/} single random cluster configuration applied to
the filtered map obtained from
2D-ILC data. Clearly, this rotation gives rise to a histogram that is very close
to the one obtained from the 1000 random cluster configurations. On the other
hand, the green, dot-dashed histogram reflects the statistics of the recovered
dipole amplitudes obtained from
the 1000 Monte Carlo CMB simulations. Again, this
histogram is fairly close to the one obtained after rotating the real
cluster sample in Galactic longitude on the real filtered map (red dashed
histogram). The recovered dipole amplitude from the real cluster positions on
the real filtered map is displayed by a vertical black line.

\begin{figure}
\centering
\includegraphics[width=8.cm]{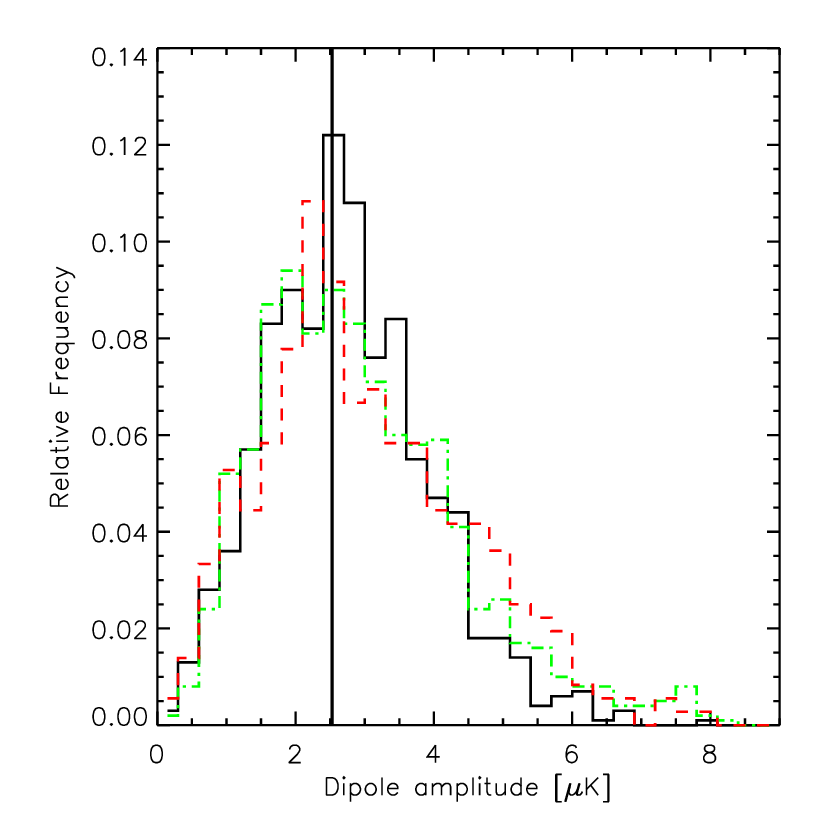}
\caption[fig:all_hists_200]{Same as Fig.~\ref{fig:all_hists}, but restricting
the analysis to the 200 most massive clusters outside the mask. The colour and
line-style coding is identical to that figure. In this case, {\em all\/}
histograms contain the measurement on the real positions of clusters
(displayed by the vertical solid line).
}

\label{fig:all_hists_200}
\end{figure}

While the measured dipole falls in the far positive tail for the simulations
using {\em random\/} cluster configurations
(black and blue histograms), it is
however quite unremarkable when compared to the simulations
accounting for the real configuration
of clusters on the sky (red and green histograms): 
about 11\,\% of cases in both CMB simulations and rotations yield dipoles 
larger in amplitude than the one measured on the real data.  The green histogram
shows that the apparent dipole can be explained by chance alignments of
random, uncorrelated CMB skies.  The impact of instrumental
noise and other component only shifts the histogram slightly (as a comparison of
the red and green histograms suggests). These results show that the dipole
measured for the real MCXC cluster positions is not peculiar when compared to
other dipole computations, either on mock CMB skies or on the filtered 2D-ILC
map for a set of positions in which the angular
clustering of the MCXC sample is
preserved. When repeating this analysis on a sub-sample of MCXC clusters
containing the 200 most massive objects, wider histograms from both rotations
and CMB mock skies are obtained.
The area under the histograms above the apparent dipole
obtained from real data at zero-lag rotation
amounts to about $56\,$\% of the total, see Fig.~\ref{fig:all_hists_200}. 
Unlike for the entire MCXC cluster sample, the histograms obtained after
randomly distributing this sub-sample of massive clusters on the un-masked
filtered map are very similar to those obtained after running CMB mocks or
rotating the clusters in Galactic longitude. This is in better agreement with
\citet{atrioerrors}, who found no significant difference between the histograms
obtained from CMB mocks and from randomly distributing clusters on the filtered
map. This is likely due to the absence of any significant intrinsic dipole in
the angular distribution of this (smaller) cluster sub-sample. 

Finally, we perform a direct comparison of our results with
\citet{kashlinsky11}. For this purpose, we use the {\it WMAP\/} filtered
maps and the sky mask for galaxy clusters at $z<0.25$ used by
\citet{kashlinsky08}; these data are presented as
supplementary materials for \citet{kashlinsky11}. For the filtered maps
corresponding to the four DAs of the {\it WMAP\/} W band data, we apply the
rotation
test in Galactic longitude. We find that, although the $y$-component of the
dipole at no rotation is marginally peculiar (at the
roughly 1--3\,\% level), the amplitude of the dipole is not, since
around 14\,\% of the rotations yield higher amplitude dipoles. This is in
good agreement with the \Planck\ results outlined above. Hence, according
to our estimations of the dipole uncertainty, we conclude that the
roughly $3\,\mu$K dipole measured by \citet{kashlinsky08}
should not be assigned to the clusters' peculiar motion, but rather to
residuals (mostly of CMB origin) in the filtered map.

\subsection{Constraints on inhomogeneous cosmological models}
\label{sec:inhomo}

The sensitivity of the kSZ effect to peculiar velocities and bulk flows makes it
an excellent probe of nonstandard inhomogeneous cosmological models
\citep{goodman95}.  In particular, models in which we are located near the
centre of a spherically symmetric Hubble-scale void have been examined
extensively in recent years as alternatives to standard accelerating models on
Friedmann-Lema\^itre-Robertson-Walker (FLRW) backgrounds \citep[see, e.g.,][and
  references therein]{clarkson12}.  Such void models can easily reproduce the
Type Ia supernova (SN) luminosity distance-redshift data without dark energy or
modified gravity.  However, these models generically predict very strong
outwards-directed bulk flows, due to the greater expansion rate within the void,
and so are expected to produce a large kSZ monopole signal (superimposed, of
course, on the usual kSZ signal from structure).  Constraints on such models
using kSZ upper limits from nine clusters
\citep{holzapfeletal97,bensonetal03,kitayamaetal04} show that the largest voids
are at odds with these early data, assuming purely adiabatic initial conditions
\citep{gbh08b,yns10b}.  Tight constraints have also been imposed using upper
limits on the kSZ power from small-scale CMB experiments
\citep{zhangandstebbins11,zm11}.  
However, the results based on small-scale kSZ power are uncertain, due to our
inability to properly perform perturbation theory in void models and our lack
of knowledge about the small-scale matter power spectrum and baryonic physics
\citep{zm11}.  The very tight \Planck\ constraints on the kSZ monopole
presented above are therefore expected to provide extremely stringent limits
on any such large-scale features, in a manner that is free of the
uncertainties due to small-scale structure.

We first briefly describe our void models and calculation methods; full details
can be found in \citet{mzs10}.  Growing-mode void models are characterized by a
single radial function, e.g.\ the matter density profile. Models with
significant decaying modes are ruled out by their extremely large kSZ and CMB
spectral distortions \citep{zibin11,bcf12} and so will be ignored here. In this
study we consider a family of smooth void profiles \citep[taken from][]{mzs10}
parameterized by a width, $L$, and a depth, $\delta_0 < 0$.  Explicitly, we
superpose (at early times) on a spatially flat background the total matter
density contrast profile
\begin{equation}
\delta(r) = \left\{\begin{array}{ll}
            \displaystyle \delta_0\left[1 - 3\left(\frac{r}{L}\right)^2
               + 2\left(\frac{r}{L}\right)^3\right] & \quad r \le L\,,\\
            \displaystyle 0 & \quad r > L\,,
\label{eqn:polyvoid}
\end{array}\right.
\end{equation}
for comoving radial coordinate $r$ centred on us.  In order to express our
constraints in terms of more directly observable quantities, in place of $L$ we
use the corresponding redshift, $z_{\rm L}$, at which we observe an object at $r
= L$.  In place of the depth, $\delta_0$, we use the local matter density
parameter at the origin today, $\Omega^{\rm loc}_{\rm M} \equiv 8\pi G\rho_{\rm
  M,0}/(3H_0^2)$, where $\rho_{\rm M,0}$ is the current total matter energy
density at the centre. Thus $\Omega^{\rm loc}_{\rm M}$ generalizes the familiar
density parameter of an FLRW cosmology, and the deepest voids have the smallest
values of $\Omega^{\rm loc}_{\rm M}$.

These models require a relativistic treatment at late times, since they are not
well described by small perturbations from an FLRW background.  We use the
Lema\^itre-Tolman-Bondi (LTB) exact solution to Einstein's equations to
calculate the radial velocity, $v_{\parallel,\rm LTB}(z)$, between a comoving
scatterer (i.e., a cluster) and the local CMB frame, using the method of
\citet{mzs10}.

We compare our LTB void models with the AP \Planck\ radial velocity estimates
for 1405 clusters by calculating the likelihood while varying the width and
depth of the void profile.  The likelihood, $\mathcal{L}$, is calculated
using the full error
distributions from the 100 displaced positions for each cluster.  Importantly,
for the deepest and widest void models, the typical velocities
$v_{\parallel,\rm LTB}$
are outside of the range of velocities for the displaced positions.
This means that we cannot calculate the actual likelihoods for these models
(because they are so small).
Instead, we conservatively assign such models likelihoods based on the outermost
sampled regions of the error distributions.  This will almost certainly result
in a large overestimate of the likelihood for these models, and hence we will
{\em under}estimate the confidence at which they are ruled out.

\begin{figure}
\centering
\includegraphics[width=7.cm]{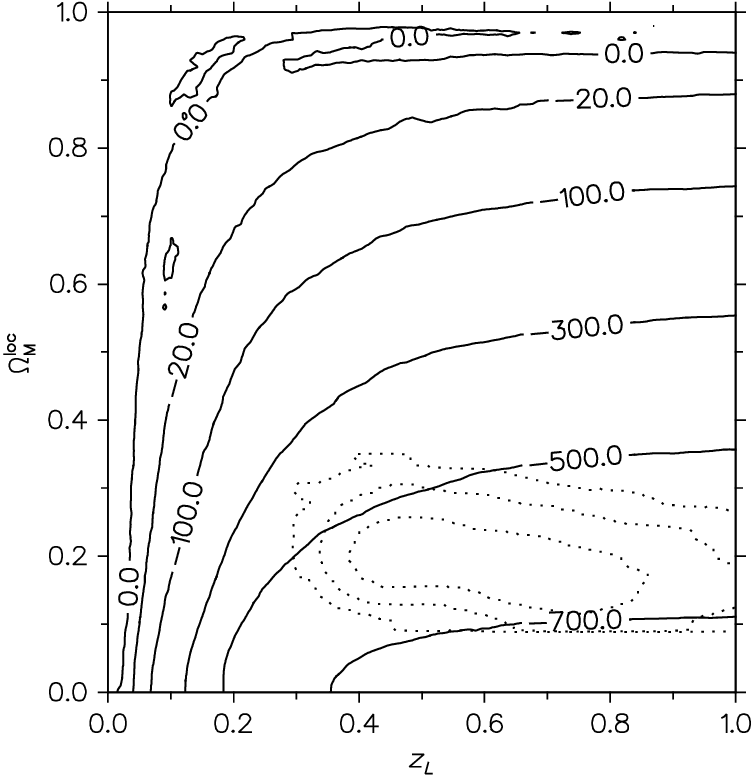}
\caption{Solid contours indicate $\log_{10}(\mathcal{L}/\mathcal{L}_{\rm hom})$
  for the AP \Planck\ frequency maps as a function of central matter
  density parameter, $\Omega^{\rm loc}_{\rm M}$, and the width of the void
  in redshift, $z_{\rm L}$.  The
  deepest voids have the smallest values of $\Omega^{\rm loc}_{\rm M}$.  Dotted
  contours are the 1, 2, and $3\sigma$ confidence levels from \citet{zm11},
  using the Union2 SN data.  Void models which fit the SN data are ruled out at
  very high confidence by the kSZ data.}
\label{voids_kSZ_fig}
\end{figure}

In Fig.~\ref{voids_kSZ_fig} we plot contours for the quantity
$\log_{10}(\mathcal{L}/\mathcal{L}_{\rm hom})$ in the width-depth plane.  
Here $\mathcal{L}_{\rm hom}$ is the
likelihood for the exactly homogeneous model, i.e., the model for which
$\delta_0 = 0$ (which implies $\Omega^{\rm loc}_{\rm M} = 1$ and
$v_{\parallel,\rm LTB}(z) = 0$).  Also shown on the plot are the confidence
levels for the same void models, but using the Union2 compilation of Type Ia
SNe, taken from \citet{zm11}.  The SN data demand deep (i.e., low $\Omega^{\rm
  loc}_{\rm M}$) but wide void profiles, while the \Planck\ kSZ data rules out
all but the very shallowest (i.e., $\Omega^{\rm loc}_{\rm M} \simeq 1$)
or narrowest (i.e., small $z_{\rm L}$)
profiles.  Adiabatic void models are
thus ruled out at extremely high confidence.  It is easy to understand the
strength of this result: voids fitting the SNe have $v_{\parallel,\rm LTB}(z)
\sim 10^4\kms$ at $z \simeq 0.5$ \citep{mzs10}, which places them a few standard
deviations into the tails of roughly $1000$ cluster measurements.

\section{Robustness of the results}
\label{sec:robust}

Returning to our main goal of determining the kSZ peculiar velocity constraints,
we now address the sensitivity of our results to uncertainties in the density
profile adopted for clusters and the errors in estimates of the optical
depth. We also quantify the impact of non-CMB noise sources on our error bars.

\subsection{Impact of changes in the density profile}

\begin{figure}
\centering
\includegraphics[width=8.cm]{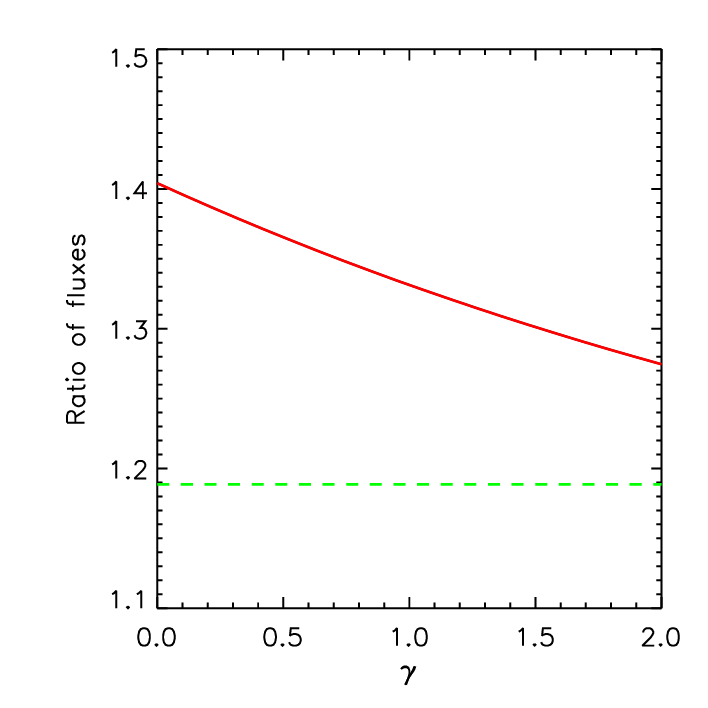}
\caption[fig:ratioss_flux]{The solid, red line displays the ratio of kSZ fluxes
  computed in circles centred on clusters with radii $\sqrt{2}\,\theta_{\rm
    500}$ and $\theta_{\rm 500}$ for different scalings of the entropy with
  radius, $K(r)\propto r^{\gamma}$. The green, dashed line observes the
  isothermal case, in which density exactly traces the universal pressure
  profile adopted in this work.}
\label{fig:ratios_flux}
\end{figure}

As mentioned in Sect.~\ref{sec:clusters}, the adopted density profile at radii
below $R_{\rm 500}$ is a fit from REXCESS observations, but at larger radii
($x>1$) the density is expressed in terms of the gas pressure \citep[which
  follows the universal profile of][]{arnaudetal10} and the entropy $K(r)=k
T(r)/n_{\rm e}^{2/3}(r)$. As well as being physically motivated, the reason for
this is the existence of some constraints on the scaling of the entropy for
$r>R_{\rm 500}$. As long as the shock front is beyond $5\times R_{\rm {500}}$
(i.e., beyond our upper radial integration limit), the entropy should increase
with radius with a power law of the type $K(r) \propto r^{\gamma}$, with
$\gamma=1.1$ for the adiabatic case, as predicted by
\citet{voitetal05}. Observations at $r<R_{\rm 500}$ are reasonably close to this
prediction, but at larger radii, the scaling should become shallower \citep[as
  suggested by the observations of ][]{walkeretal12}. Our computations for the
AP filter adopted the fixed value $\gamma = 0.5$ for $r>R_{\rm 500}$, but we
explored the impact of different scalings when estimating the ratio of fluxes
inside the inner and outer circles of the AP filter. The red solid line in
Fig.~\ref{fig:ratios_flux} displays the ratio of the kSZ flux computed in
circles of radii $\sqrt{2}\,\theta_{\rm 500}$ and $\theta_{\rm 500}$ for
different scalings of the entropy, $K(r)\propto r^{\gamma}$.  This ratio depends
on the profile, but not on the cluster mass or redshift. The horizontal, dashed
green line displays the isothermal case at {\em all\/} radii,
including for $x<1$. When
translating these flux ratios into velocity constraints, we find that the
$\gamma=0.5$ profile introduces a boost in the peculiar velocity amplitude
estimate of about 28\,\% with respect to the case where clusters are assumed to
be isothermal. For the extreme case of $\gamma=0$ this error amounts to 36\,\%,
while for $\gamma = 1.1$ it goes down to 20\,\%. With respect to our reference
model of $\gamma=0.5$, we expect that variations of $\gamma$ in the range
$[-1.1, 0]$ will introduce changes in the velocity constraints
at the level of $\pm
6$\,\%. If the pressure profile is changed to use the parameters that best fit
the external profiles measured out to $3\times R_{\rm {500}}$
\citep{planck2012-V}, then the change in the
velocity constraints are small enough to be considered negligible.

Regarding the uMMF, only the isothermal profile was implemented in the
filter. In order to assess the impact of this approximation, we conduct a
numerical experiment, consisting of assigning a 1000\,km\,s$^{-1}$ amplitude
dipole to the cluster set of our simulations, and extracting this dipole after
injecting both the isothermal and non-isothermal profiles to the clusters, but
considering only the isothermal profile in the uMMF definition. We recover
amplitude values of $1016\pm 106$\,km\,s$^{-1}$ and $966\pm 106$\,km\,s$^{-1}$
for the isothermal and non-isothermal profiles, respectively. The errors do not
change, which is expected, since the profile coded in the uMMF is isothermal in
both cases. The difference in the recovered values, however, suggest
a potential bias in
the amplitude estimate of about 50\,km\,s$^{-1}$, which amounts to around 5\,\%
of the signal amplitude. This test indicates that the uMMF is less
sensitive to the assumed density profile than the AP, probably because the
former assigns more weight to the central regions of the cluster where the noise
(mostly induced by the CMB) is lower and where the decreasing radial pattern of
temperature is less dramatic.
  
Hence, we conclude that uncertainties in the radial gas distribution in galaxy
clusters should introduce errors in the limits imposed on peculiar velocities at
the 5--30\,\% level.

\subsection{Other sources of error}

\begin{table}[tmb] 
\begingroup 
\newdimen\tblskip \tblskip=5pt
\caption{Impact of uncertainties in estimates of the optical depth of clusters
  on the constraints imposed on the kSZ monopole, rms, and dipole from \Planck\
  data. Different $\tau$ error amplitudes (given by $\sigma_{\epsilon}$) and two
  cluster (sub-)samples are considered here. The percent levels correspond to
  the fractional changes found in the kSZ monopole error bar (third column), the
  95\,\% confidence limit for the kSZ-induced rms-excess (fourth column), and the
  kSZ dipole error bar (fifth column).
}
\label{tab:error_table}
\vskip -3mm
\footnotesize 
\setbox\tablebox=\vbox{ %
\newdimen\digitwidth 
\setbox0=\hbox{\rm 0}
\digitwidth=\wd0
\catcode`*=\active
\def*{\kern\digitwidth}
\newdimen\signwidth
\setbox0=\hbox{+}
\signwidth=\wd0
\catcode`!=\active
\def!{\kern\signwidth}
\halign{\hfil#\hfil\tabskip=0.2cm& \hfil#\hfil\tabskip=0.2cm&
 \hfil#\hfil\tabskip=0.2cm& \hfil#\hfil\tabskip=0.2cm&
 \hfil#\hfil\tabskip=0.2cm& \hfil#\hfil\tabskip=0pt\cr
\noalign{\doubleline}
\noalign{\vskip -2pt}
Error on $\tau$& No. clusters& $\Delta[\langle v\rangle]$&
 $\Delta[\langle v^2\rangle]$& $\Delta[\langle {\rm dipole}\rangle]$\cr
 $\sigma_{\epsilon}$& & [\%]& [\%]& [\%]\cr
\noalign{\vskip 3pt\hrule\vskip 3pt}
\multispan5\hfil AP\hfil\cr
\noalign{\vskip 3pt\hrule\vskip 3pt}
 0.2& 1405& 1& 1& 3\cr
 0.4& 1405& 2& 2& 5\cr
 0.2& *100& 2& 2& 3\cr
 0.4& *100& 5& 5& 6\cr
\noalign{\vskip 3pt\hrule\vskip 3pt}
\multispan5\hfil uMMF \hfil\cr
\noalign{\vskip 3pt\hrule\vskip 3pt}
 0.2& 1405& 1& 1& 2\cr
 0.4& 1405& 2& 3& 5\cr
 0.2& *100& 2& 2& 3\cr
 0.4& *100& 5& 5& 6\cr
\noalign{\vskip 3pt\hrule\vskip 3pt}}}
\endPlancktable 
\endgroup
\end{table}

After repeating the analysis of the AP and MF filters on a pure CMB simulated
map, we found that errors decreased to about 70\,\% of their amplitude on HFI
raw maps. This shows that the main limiting factor when estimating kSZ
velocities is the intrinsic CMB component.  The presence of point sources,
instrumental noise and other foregrounds should be included in the remaining
roughly 30\,\%, and after properly accounting for the spectral response of HFI
detectors, there seems not to be any significant tSZ leakage biasing the
peculiar velocity estimates.

The AP and uMMF/MF velocity estimates, however, rely on an accurate knowledge of
the cluster optical depths within a given radius. For each cluster, the
integrated optical depth is estimated from the adopted radial density profile,
which itself relies on the $Y_{500}$--$M_{500}$ and $T$--$M_{500}$
relations. Apart from the uncertainties in the shape of the profile (addressed
above), the intrinsic scatter in these scaling relations could have an impact on
the peculiar velocity estimates. In order to test this we conduct a
Monte Carlo analysis consisting of introducing {\em un-correlated\/} errors to the
real estimates of the cluster optical depths. We adopt a log-normal model for
the errors on the $\tau_{500,\,j}$ estimate for the $j$th cluster:
\begin{equation}
\tilde{\tau}_{500,\,j} = \tau_{500,\,j} \, \exp{(\epsilon_j)}, 
\label{eq:err_tausMC}
\end{equation}
with $\epsilon_j$ being a normally distributed variable of zero mean. The symbol
$\tilde{\tau}_{500,\,j} $ denotes the Monte Carlo estimate of the $j$th
cluster's optical depth obtained from the real estimate $\tau_{500,\,j}$. For
each of the 100 Monte Carlo simulations, we simulate values of $\tau$ for each
cluster and then repeat the full analysis,
setting constraints on the kSZ monopole,
kSZ-induced rms excess, and kSZ dipole, as outlined in previous sections.

These analyses show that our constraints on the kSZ monopole, velocity rms, and
dipole uncertainty change by less than 10\,\% when considering errors in the
optical depth estimation of the order 20--40\,\%. Typical changes are at the
level of a few percent (see Table~\ref{tab:error_table}); since we are
constraining ensemble quantities obtained from sub-samples of the cluster
catalogue, errors tend to average out if they are independent from cluster to
cluster (as we expect them to be). Provided that the relative uncertainty in
cluster luminosities is about 40\,\% for the $L$--$M$ scaling relation
\citep{prat}, and combining it with the approximate
scalings $L \propto M^{4/3}$ and $M
\propto R^3$, one deduces that the relative uncertainty in $R$ should be at the
level of about 10\,\%. For the spherical estimates of
$\tau_{\rm{sph,\,500}}$ this translates into a roughly 30\,\% uncertainty,
decreasing to about 20\,\%
for the cylindrical optical depth estimate ($\tau_{\rm cyl} \propto R^2$). From
the results of our Monte Carlo approach above, we conclude that errors in the
optical depth estimates should not significantly bias our kSZ constraints.

\section{Conclusions}
\label{sec:discussion}

The MCXC cluster sample has been used to search for signatures of peculiar
velocities in the \Planck\ CMB data. For this purpose, two different filters were
applied: aperture photometry; and the unbiased Multi-frequency Matched
Filter. The former is a simple, quick and robust tool for providing estimates of
kSZ-induced temperature anisotropies, and although it detects the tSZ-induced
monopole and rms excess at the cluster positions, it fails to detect the kSZ
effect, setting a constraint on the kSZ-induced radial velocity rms at the level
of 1200\,km\,s$^{-1}$ (95\,\% confidence level) for a massive and distant MCXC
sub-sample of 1000 sources. By effectively removing the tSZ signal, matched
filters are able to place stronger constraints, reaching the level of
800\,km\,s$^{-1}$ (95\,\% C.L.) for a subsample of 100 massive clusters. All
these values, however, lie a factor of 3--5 above $\Lambda$CDM
expectations for clusters of typical mass $2\times 10^{14}\, \Msolar$ at
$z\simeq 0.15$. Thus, while our constraints are fully consistent with $\Lambda$CDM expectations, a detection of the radial velocity rms would require significant improvements over the present analysis.

Both methods also provide measurements of the clusters' average velocity that
are compatible with zero: these are at the level of 120--160\,km\,s$^{-1}$
(95\,\% confidence level) for the uMMF and AP filters. The fact that this
constraint applies to a cluster sample whose mean redshift is $z\sim 0.18$
provides very strong evidence that the CMB is {\it mostly} at rest with respect
to those observers (as opposed to the relative motion of our local CMB to those
sources, which is of the order of $c z \sim 54\,000$\,km\,s$^{-1}$). By itself,
this measurement constitutes an unprecedented and valuable confirmation of a
prediction of the standard cosmological scenario, and has strong implications in
discussions of the homogeneity of our Universe.

In this context, the large number and redshift distribution of the
\Planck\ cluster kSZ measurements are ideal for constraining void models, which
attempt to explain the apparent acceleration without dark energy or modified
gravity.  Indeed, void models which fit the Union2 SN data are ruled out at
extremely high confidence.  In principle it may be possible to cancel the kSZ
effect generated at cluster positions in these models with a large (order unity)
isocurvature mode at last scattering \citep{yns10b}, but this would almost
certainly entail substantial fine tuning of the isocurvature mode.  Therefore
the \Planck\ kSZ data strongly support the conclusions of previous studies which
found that void models generically predict very low $H_0$
\citep[e.g.,][]{zms08,bcf12} and too large kSZ power on small scales
\citep{zhangandstebbins11}.

\Planck's constraints on the amplitude of the local bulk flow provide an
independent view of a long ongoing debate. Unfortunately, our results are
not sensitive to the local volumes where many claims for bulk flows have been
raised; the limit of 390\,km\,s$^{-1}$ within spheres of 350\,$h^{-1}$\,Mpc at
95\,\% C.L. does not permit us to confront claims at the level of
400--700\,km\,s$^{-1}$ within radii of 50--120\,$h^{-1}$\,Mpc, and on the
convergence of the measured CMB dipole within these cosmological volumes
\citep{hudson04,watkinsetal09,feldmanetal10,nusserdavis11,nusserbranchinianddavis11,branchinidavisandnusser12,Courtoisetal2012,mascott12}.
The number of galaxy clusters present in those spheres is too low
(78 entries in the MCXC catalogue within 80$\,h^{-1}$\,Mpc) to decrease the
statistical noise significantly. It would thus be required that future CMB
experiments have sufficient angular resolution and sensitivity for
galaxy groups and clusters in the neighbourhood of the Local
Group in order to provide a kSZ view of the local dipole.
However, on larger scales,
\Planck\ is able to set strict constraints on the amplitude of bulk flows
(below 254\,km\,s$^{-1}$ at 95\,\% C.L. for a radius of 2\,$h^{-1}$\,Gpc),
in clear contradiction
with some previous claims \citep{kashlinsky08,kashlinsky10,abatefeldman11}.
It is worth remarking that the conclusions derived from our analysis are
practically insensitive to few-percent changes in the set of cosmological
parameters.

The linear continuity equation states that peculiar velocity surveys are
sensitive
to fluctuations in the distribution of matter and energy on scales larger than
density or galaxy surveys. The fact that \Planck\ is able to set such strong
constraints on peculiar velocities in a cluster population at
$\langle z \rangle \sim 0.18$ (and extending out to $z\sim 1$),
translates into correspondingly
strong constraints on the amplitude of primordial fluctuations at Gpc scales. If
the Universe were inhomogeneous on scales larger than the size of the volume
containing our cluster catalogue, these clusters would show a significant
dipolar pattern in their kSZ velocities. We conclude that \Planck\ constraints
on peculiar velocities are compatible with $\Lambda$CDM expectations, and
constitute an unprecedented piece of evidence for the local homogeneity of the
Universe in the super-Gpc regime.

\begin{acknowledgements}
The development of \Planck\ has been supported by: ESA; CNES and
CNRS/INSU-IN2P3-INP (France); ASI, CNR, and INAF (Italy); NASA and DoE (USA);
STFC and UKSA (UK); CSIC, MICINN, JA and RES (Spain); Tekes, AoF and CSC
(Finland); DLR and MPG (Germany); CSA (Canada); DTU Space (Denmark); SER/SSO
(Switzerland); RCN (Norway); SFI (Ireland); FCT/MCTES (Portugal); and PRACE
(EU).  A description of the Planck Collaboration and a list of its members,
including the technical or scientific activities in which they have been
involved, can be found at \url{http://www.rssd.esa.int/Planck}.
The authors from the consortia funded principally by CNES, CNRS, ASI, NASA, and
Danish Natural Research Council acknowledge the use of the pipeline running
infrastructures Magique3 at Institut d'Astrophysique de Paris (France), CPAC at
Cambridge (UK), and USPDC at IPAC (USA).
We acknowledge the use of the HEALPix
package, \textit{WMAP\/} data and the LAMBDA archive
(\url{http://lambda.gsfc.nasa.gov}).
\end{acknowledgements}

\bibliographystyle{aa}
\bibliography{Planck_bib,pip13}

\end{document}